\documentclass[referee,a4paper,12pt,traditabstract]{jswsc} 

\usepackage{graphicx}
\usepackage{txfonts}
\usepackage{caption}
\usepackage{subfigure}
\usepackage{epstopdf}
\usepackage[displaymath,mathlines]{lineno}
\usepackage[authoryear,round]{natbib}
\usepackage[backref]{hyperref}
\usepackage{url}
\usepackage{pgffor}
\usepackage{multirow}

\hypersetup{colorlinks=true,citecolor=cyan,urlcolor=cyan,linkcolor=blue}

\begin{document}


   \title{The Variation of Radiation Effective Dose Rates and Single Event Effect Rates at Aviation Altitudes with 
   Magnetospheric Conditions and Geographic Location}

   \titlerunning{Rad. Dose Rate Variation with Mag. Conditions and Location, 2023}

   \authorrunning{Davis et al.}

   \author{C. S. W. Davis\inst{1},
           K. Ryden\inst{1},
           F. Lei\inst{1},
           B. Clewer\inst{1},
           A. Hands\inst{1},
           C. Dyer\inst{1,2}
           }

   \institute{University of Surrey
         \and
             CSDRadConsultancy
             }


 
  \abstract
   {The geographic structure of radiation dose rates at aircraft altitudes in Earth's atmosphere during the irradiation of 
   Earth by proton spectra from incoming solar particle events is examined using the recently developed MAIRE+ software. 
   Conditions are examined under two incoming proton spectra, a low/hard spectral index and a high/soft spectral index spectra, 
   which are representative of some of the solar particle events that have caused reasonably sized Ground-Level Enhancements/Events 
   (GLEs) over the past 70 years. It is found through the use of `cut-throughs' of the atmosphere, that the atmosphere can be divided 
   into three volumes; a high dose rate polar region, a low dose rate equatorial region, and a transition region between the two. The 
   location of these regions as a function of latitude, longitude and altitude is characterised. It is also found that the location of 
   the transition region changes for different magnetospheric disturbance levels, implying that the total radiation dose rate an aircraft 
   will experience if it passes through the transition region will be subject to large systematic uncertainties, particularly during 
   the currently unknown levels of magnetospheric disturbance that a major solar event could cause. The impact that various 
   magnetospheric conditions might have on dose rates that specific flight routes might experience is also discussed.
   }        


   \maketitle

\section{Introduction}
Radiation dose rates in Earth's atmosphere are reasonably 
predictable most of the time. During quiet periods of relative 
solar inactivity, cosmic rays and the secondary particles they 
produce when passing through matter are the main source of 
atmospheric radiation levels. Cosmic rays originate from outside 
the solar system, and bombard Earth at a roughly constant rate 
day-to-day, typically only varying in intensity on the scale of 
years as the solar cycle varies. However, every so often events 
such as Coronal Mass Ejections (CMEs) or solar flares can cause 
the Sun to eject a large cloud of charged particles. If such an 
event is sufficiently high energy, such that it would cause 
significant radiation increases, it is known as a Solar Particle 
Event (SPE) \citep{reames2013two}, or in more common English, a 
solar storm. If an SPE happens to hit Earth and is comprised of 
particles that are sufficiently high energy such that detectable 
levels of dose-inducing radiation are generated in Earth's 
atmosphere from the event, the SPE can cause what is known as a 
Ground-Level Enhancement (GLE) \citep{poluianov2017gle}, which 
is also sometimes called a Ground-Level Event. 

During a Ground-Level Enhancement (GLE), high energy solar 
particles hitting the Earth's atmosphere generate showers of 
particles, which can hit both electronics and humans and lead 
to radiation effects \citep{dyer2020single, hubert2020impact, 
matthia2015economic}. Major GLEs, such as the GLE of February 
1956 (sometimes labeled as GLE05) \citep{belov2005study}, have 
been known to induce so much radiation in the atmosphere that 
it dwarfs the radiation induced by cosmic rays by orders of 
magnitude. It is unknown exactly how strong these events can 
get, as the radiation dose rate due to GLEs has only been 
detectable since the 1940s, and only 73 GLEs have been directly 
measured to date, as of the time of writing \citep{gleDatabase}. 
However recent research into cosmogenic radionuclides has 
revealed several events in the past that have exceeded events 
such as GLE05 by orders of magnitude \citep{dyer2017extreme}. 
If such an event happened today, it could pose a serious risk 
to the electronics in planes flying at any particularly 
susceptible regions of the atmosphere.

Because GLEs occur on a highly intermittent irregular timescale 
and only last for several days at maximum, there are only a few 
experimental datasets available that directly measured the full radiation dose 
experienced by aircraft during such events 
\citep{beck2009overview} as 
there are only a few regular flights 
that have contained permanently stationed dose rate monitors.
MAIRE+, which is used for the simulations in this paper, was validated
against data during the GLEs of 1989 from CREAM on Concorde 
\citep{Hands2022new,dyer1989measurements}. While there are 
some projects underway to develop sensors that are capable of 
detecting such dose rates 
\citep{clewerCitizen, clewerdemonstration} while being 
permanently stationed on aircraft such that they catch an 
event in action, for the time being it is necessary to rely on 
simulation data to understand what might happen during such 
events. 

There has been a reasonable level of past research that has been 
done into the variation of dose rates during a GLE with location 
and magnetospheric conditions. Such research in the past has 
been generally limited by the computational power available at 
the time, as both trajectory calculations within Earth's 
magnetosphere and dose rate calculations are quite 
computationally complex. It has therefore only been possible to 
perform full-scale calculations of dose rates on reasonable 
timescales and without the aid of high-performance computers in 
recent years.

It has been known for a long time from theoretical calculations, 
simulations and experimental observations from sources such as 
the global neutron monitor network \citep{mishev2020current} 
that dose rates generally increase with latitude, and also 
increase with altitude until about 20~km/65~kft in altitude 
(this altitude is known as the Pfotzer maximum, and is higher in 
altitude than most currently operating commercial aviation) 
\citep{grieder2001cosmic}. The increase in dose rates with 
altitude below the Pfotzer maximum is caused by the complex 
physics underlying the penetration of the atmosphere by incoming 
particles, and the generation of secondary particles at 
different atmospheric levels. The increase in dose rates with 
increasing latitude on the other hand, is caused by Earth's 
magnetic field becoming less effective at shielding Earth from 
incoming particles at regions closer to Earth's magnetic poles. 
This effect is typically quantified by a value at each 
latitude-longitude coordinate known as 
`vertical cut-off rigidity' (also sometimes referred to as 
effective vertical cut-off rigidity) \citep{cooke1991cosmic}, a 
parameter that quantifies how difficult it is for a particle 
traveling towards Earth to penetrate the magnetic shielding of 
Earth.

It is also generally well known that increases in magnetospheric 
disturbance levels caused by solar wind and solar storms also 
generally cause increases in dose-rate across Earth. This is 
because disturbances to Earth's magnetosphere generally cause 
reductions in vertical cut-off rigidity across Earth 
\citep{smart2005review}, allowing more dose-inducing particles 
to penetrate Earth's magnetosphere and hit the atmosphere.

While the general characteristics of how dose rate varies with 
changes in single parameters and at single locations are known, 
there has been little published modern research examining how 
dose rates vary as a complex multidimensional function of all 
geographic location and relevant parameters associated with a 
solar storm. The atmosphere is a 3-dimensional environment, and 
so understanding the full volumetric picture of dose rates in 
the atmosphere is crucial for gaining a qualitative 
understanding of the atmosphere during GLEs. A knowledge of the 
big picture for how dose rates can vary across the atmosphere is 
especially important for airlines and policy makers, who will 
have to provide guidance to pilots on the safe locations to fly, 
and volumes of the atmosphere to avoid, during major solar 
storms.

Additionally, it is well known that during a solar event of the 
sort of scale that has not been seen so far since the invention 
of proper measurement equipment at the dawn of the space age, 
levels of disturbance in Earth's magnetosphere could reach 
levels that have never been seen before and that it is not 
currently possible to simulate with verifiable accuracy. 
Therefore building a qualitative picture of what happens to the 
volumes of the atmosphere that are associated with certain dose 
rates as magnetospheric disturbance increases would be useful 
for qualitatively understanding what might happen during a major 
GLE.

The work presented in this paper has been therefore performed to 
examine the full structure of the radiation dose rate field in 
the atmosphere during the bombardment of Earth by two 
representative simulated spectra of protons that may occur 
during a GLE, a low spectral index spectrum, and a high spectral 
index spectrum. The atmospheric structure of dose rates were 
examined through the use of atmospheric `cut-throughs', and the 
variation of the volumes associated with these dose rates with 
changes in magnetospheric disturbance levels were investigated. 
Only the ambient dose equivalent rate field is investigated 
within this paper, however it should be expected that other 
classifications of radiation dose rate will exhibit the same or 
similar atmospheric structure. Specific dose rates that 
transatlantic flights will experience during different 
magnetospheric conditions are also discussed.

\section{Methodology}

Simulation results were gathered using a variant of 
MAIRE+\citep{Hands2022new} known as MAIRE-S. MAIRE+ is a new software 
that has been developed recently to calculate radiation dose rates at aircraft altitudes
at any time using the current solar and magnetospheric conditions. During quiet solar
conditions, MAIRE+ uses a variant of the ISO model developed by Matthiä et al. \citep{matthia2013ready} to determine the cosmic ray
spectrum. During high energy solar event conditions, MAIRE+ adds a `GLE' dose rate component
onto dose rates by assuming a rigidity power law spectrum and applies an algorithm to
determine the extra GLE proton spectrum occurring at a given time\citep{Hands2022new}.
MAIRE-S is simply a version of MAIRE+ that has been modified to be able to be 
programmatically run across historic and user defined solar and 
magnetospheric conditions. The assumptions that are present in 
MAIRE+ are therefore the assumptions that should also be 
considered with regards to the data presented in this paper. 
Some of the main assumptions of MAIRE+ include:
\begin{itemize}
    \item The incoming particle distribution hitting Earth's magnetosphere is isotropic. This is true for many GLEs, particularly during the tail end of a GLE.
    \item Vertically incident particles dominate radiation doses, such that vertical cut-off rigidities can be used to describe the full effect of Earth's magnetosphere on dose rates.
    \item The assumptions present in the Tsyganenko 1989 
    \citep{tsyganenko1989magnetospheric} magnetic field model 
    and the extension to the Tsyganenko 1989 model made by 
    Boberg et al. \citep{boberg1995geomagnetic}.
\end{itemize}

All results in this paper were generated from two single 
`snapshot' input spectra. The two spectra and corresponding 
atmospheric dose rates, and vertical cut-off rigidities, were 
taken from MAIRE+ simulations of GLE42, which occurred in 
September 1989. The spectra were chosen somewhat arbitrarily in 
order to showcase a relatively hard differential power law 
spectral index, and a relatively soft differential power law 
spectral index; due to the variability in GLE intensity the 
normalisation factor of the spectra were not considered to be 
too significant within the scope of this particular study into 
the geographic structure and variability of the dose rates. 
According to Tylka et al. \citep{tylka2009new}, spectral indices 
for most recorded GLEs range between about 3 and 10 (the 
spectral indices reported by Tylka et al. are reported in terms 
of an integral power law and therefore the number 1 needs to be 
added to it to get the differential spectral indices used 
here).


Only dose rates due to the GLE component of incoming particles 
were considered here. The total dose rate would actually be the 
summation of GLE component + cosmic ray component, however for 
the purposes of this study only the GLE component was 
considered. In any case, during a major GLE of importance, the 
GCR component would be dwarfed by the GLE component. This GLE 
component was described by a simple power law in rigidity, with 
the relevant parameters for the two spectra shown in 
table~\ref{tab:inputtedSpectra}.

\begin{table}[]
    \centering
    \caption{Characteristics defining the spectra used in the simulations described in this paper. GLE proton spectra were modelled as a power law distribution in rigidity (with units of GV).}
    \begin{tabular}{c c c c}
    \hline
     Spectrum label & GLE normalisation factor used & GLE spectral index \\
     & (protons/m²/s/sr/GV) & & \\
     \hline
     `hard' & $1.69\times10^4$ & 2.65 \\
     `soft' & $2.67\times10^5$ & 6.54 \\
    \end{tabular}
    \label{tab:inputtedSpectra}
\end{table}

These two spectra were chosen as representative of periods 
of time during GLEs where the GLE spectra might be soft and hard 
respectively. The spectra given in table~\ref{tab:inputtedSpectra} 
where directly taken from runs of MAIRE-S across GLE42 in 1989, 
so both spectra are 
possible in real world scenarios, and the spectral indicies 
represent different regions of the potential range of GLE spectral indicies 
as aggregated by Tylka et al.\citep{tylka2009new}. The main parameter of interest 
for spectra that may significantly alter the structure of radiation 
dose rates throughout the atmosphere is spectral index, which 
will affect how many particles are able to penetrate the 
atmosphere at different locations due to cut-off rigidity, and 
alter the geographic structure of radiation dose rates. The 
normalisation factors used here are somewhat arbitrary. While 
the normalisations do represent real MAIRE-S-predicted 
normalisation factors during GLE42, they do not affect the 
structure of dose rates throughout Earth's atmosphere, and only 
act to scale overall dose rates. Normalisation factors also 
vary by orders of magnitude between different GLEs.

\begin{figure}
   \centering
   \includegraphics[width = \textwidth]{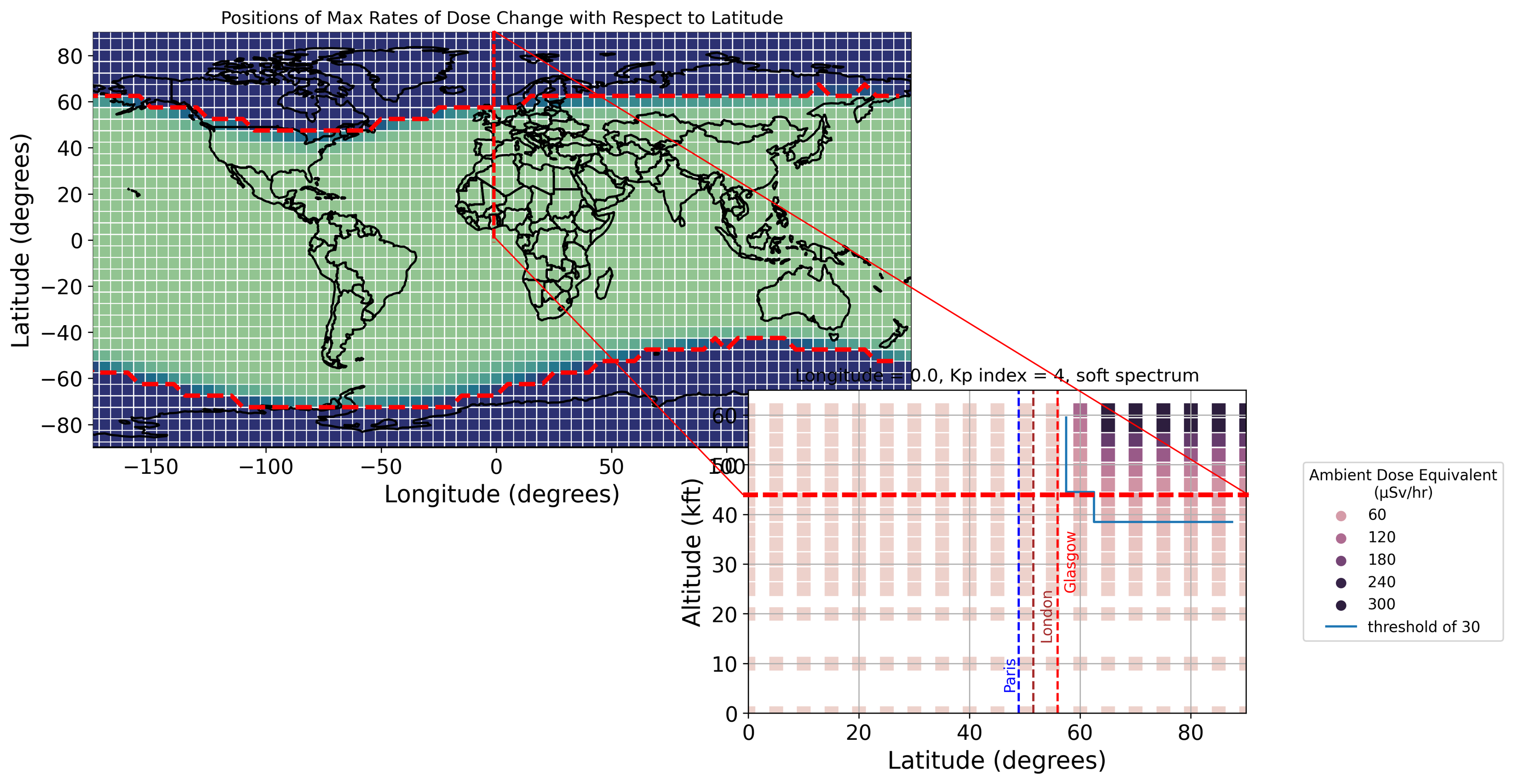}
   \caption{An example of an atmospheric `cut-through' as used in this paper. The cut-throughs are effectively atmospheric cross-sections at a particular latitude. The displayed cut-through shows ambient dose equivalent rates as a function of latitude and altitude at a longitude of 0.0\textdegree.}
   \label{fig:xsectionExplanationDiagram}
\end{figure}

Throughout this paper, dose rate `cut-throughs' have been 
generated using output dose rate maps generated by MAIRE-S. 
These cut-throughs are effectively geometric cross-sections of 
the atmosphere, showing dose-rates as a function of latitude and 
altitude at a given longitude, as displayed in 
figure~\ref{fig:xsectionExplanationDiagram}. The term 
`cut-through' will be used in this paper as opposed to 
cross-section in an attempt to prevent possible confusion with 
the entirely different concept of interaction cross-sections 
which are also used in radiation physics.

\section{The Variation of Dose Rates as Function of Kp Index}

\begin{figure}
   \includegraphics{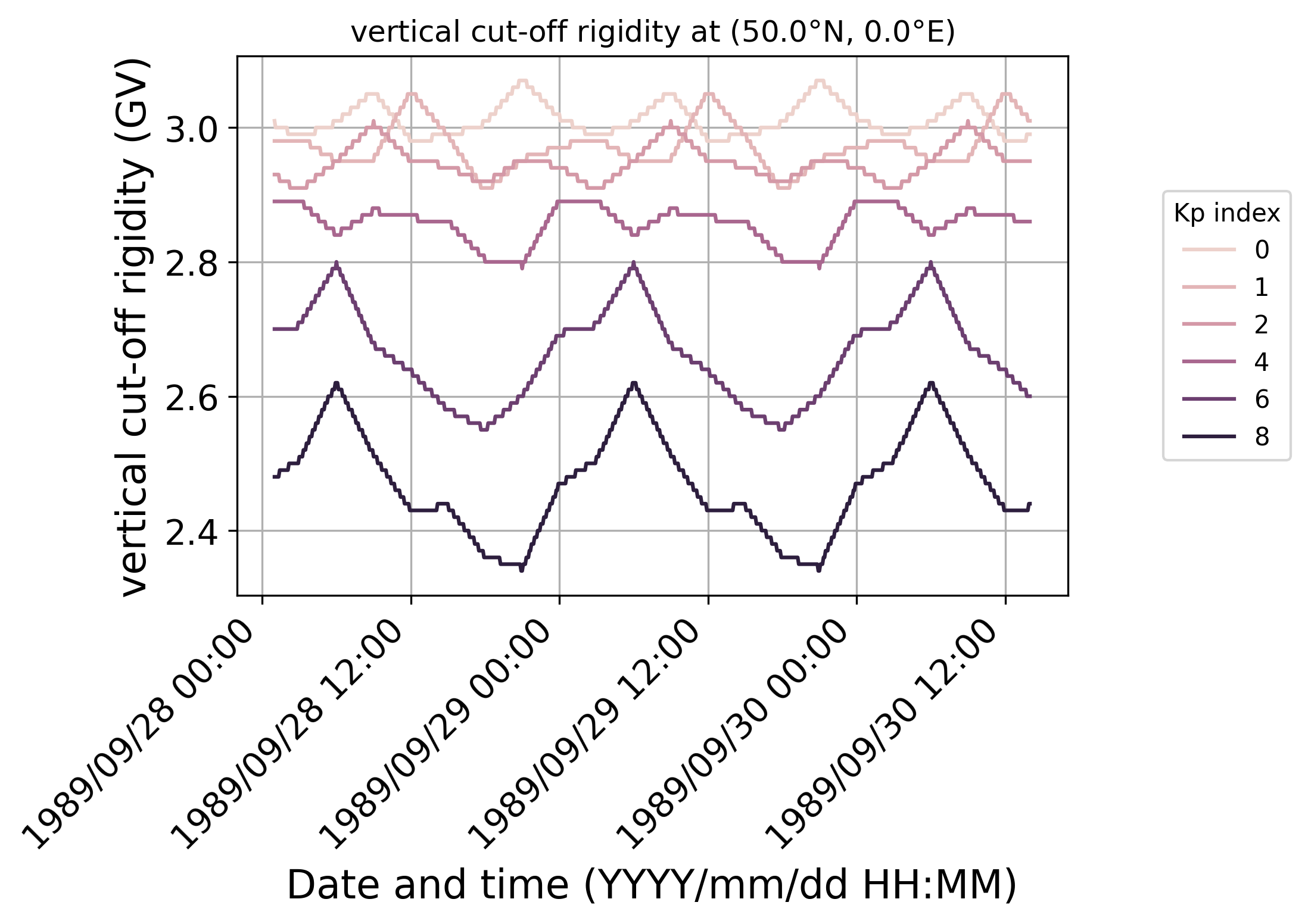}
   \caption{Vertical cut-off rigidities as a function of Kp-index and datetime at (50.0\textdegree N, 0.0\textdegree E). Increases in Kp index decrease the effective cut-off rigidity by up to approximately 0.7 GV at maximum, and the vertical cut-off rigidity varies by up to about 0.25 GV across a full day. The curves are somewhat jagged in this plot due to the relatively granular interpolation with respect to time that MAIRE+ uses to calculate cut-off rigidities within a day.}
   \label{fig:rcVsDatetimeUK}
\end{figure}

The rigidity cut-off at (50.0\textdegree N, 0.0\textdegree E) in each of the Kp index simulations is plotted in figure~\ref{fig:rcVsDatetimeUK} as a function of date and time. The date and time here correspond to the internal geomagnetic field between 28/09/1989 00:00 UTC and 30/09/1989 12:00 UTC. As Kp was held constant for each of these simulations, all variations in the rigidity cut-off are therefore caused by the rotation of the Earth rather than direct magnetospheric phenomena. Similar results for the magnitude of daily variation of cut-off rigidity can be found in \citep{smart2000magnetospheric} albeit for a different time period, and at Durham, USA rather than the (50.0\textdegree N, 0.0\textdegree E) used here.

figure~\ref{fig:rcVsDatetimeUK} indicates that the rigidity cut-off at (50.0\textdegree, 0.0\textdegree) decreases from lows of about 3.0 GV to lows of below 2.4 GV when Kp index is changed from 0 to 8. This lines up with research on rigidity cut-offs that has been performed before by Smart et al. \citep{smart2006geomagnetic, smart2009fifty}, as does the general structure of the variation in cut-off rigidity over time shown in figure~\ref{fig:rcVsDatetimeUK}. One interesting point to note is that the variability of cut-off rigidity over time increases as Kp index increases here (from a range of approximately 0.1 GV at Kp = 0 to a range of approximately 0.25 GV at Kp = 8), indicating that the time of day may become an important variable in dose rate calculations at high Kp indices.

\begin{figure}
   \centering
   \begin{tabular}{c|c}

      \subfigure{\includegraphics[trim={0 0 0 0},clip,width=0.4\columnwidth]{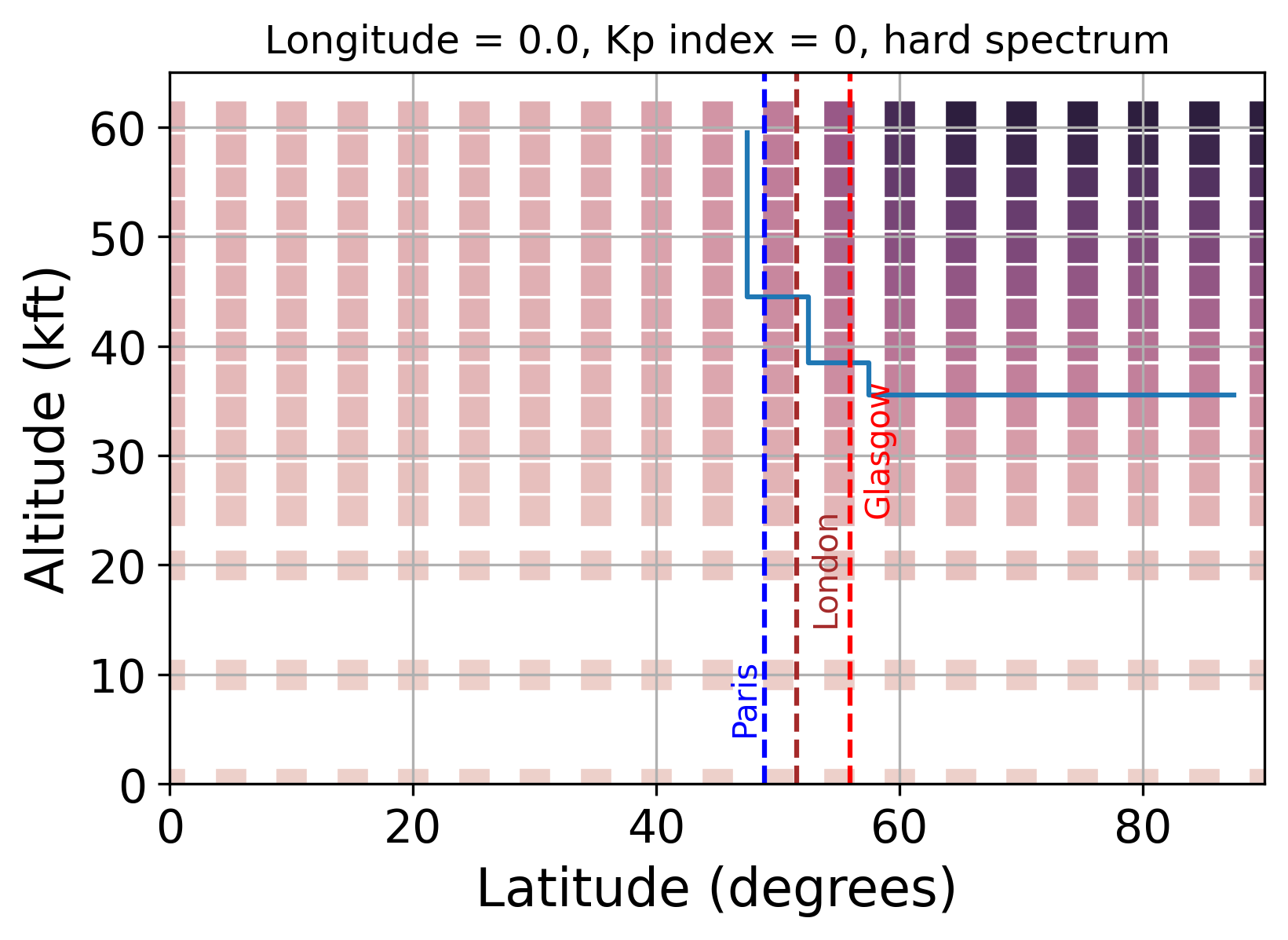}} 
      &
      \subfigure{\includegraphics[trim={0 0 0 0},clip,width=0.4\columnwidth]{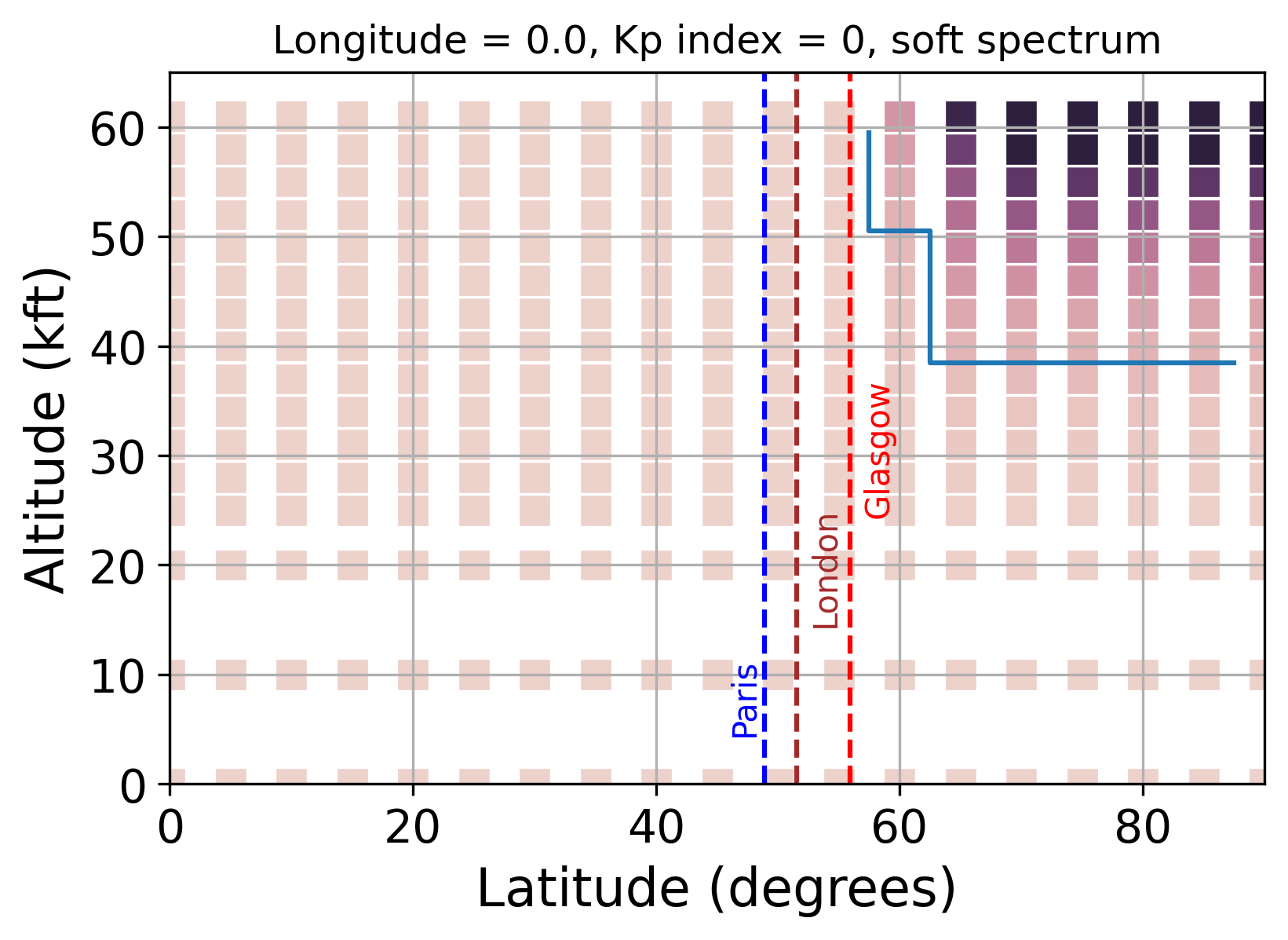}} \\

      \subfigure{\includegraphics[trim={0 0 0 0},clip,width=0.4\columnwidth]{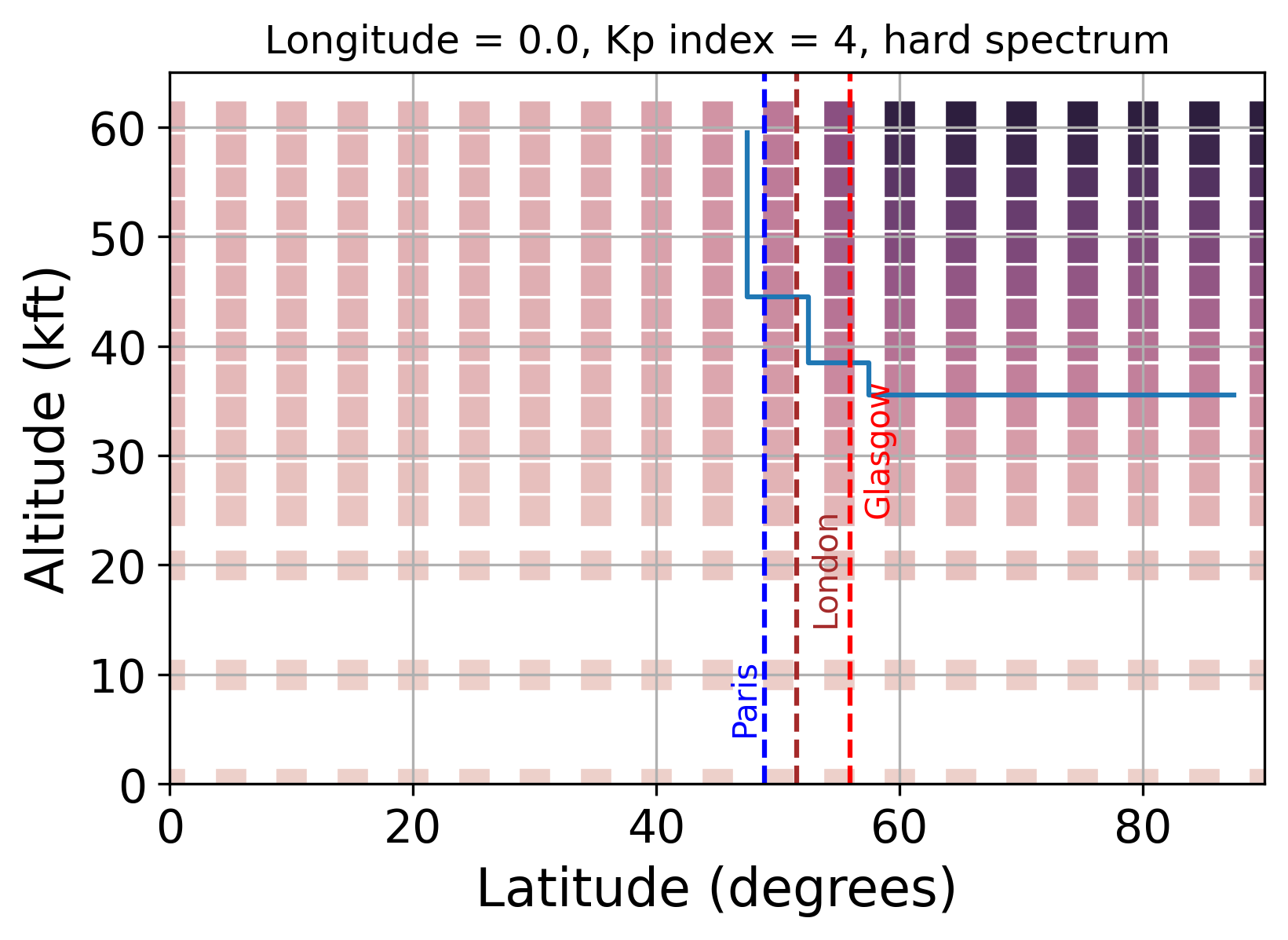}} 
      &
      \subfigure{\includegraphics[trim={0 0 0 0},clip,width=0.4\columnwidth]{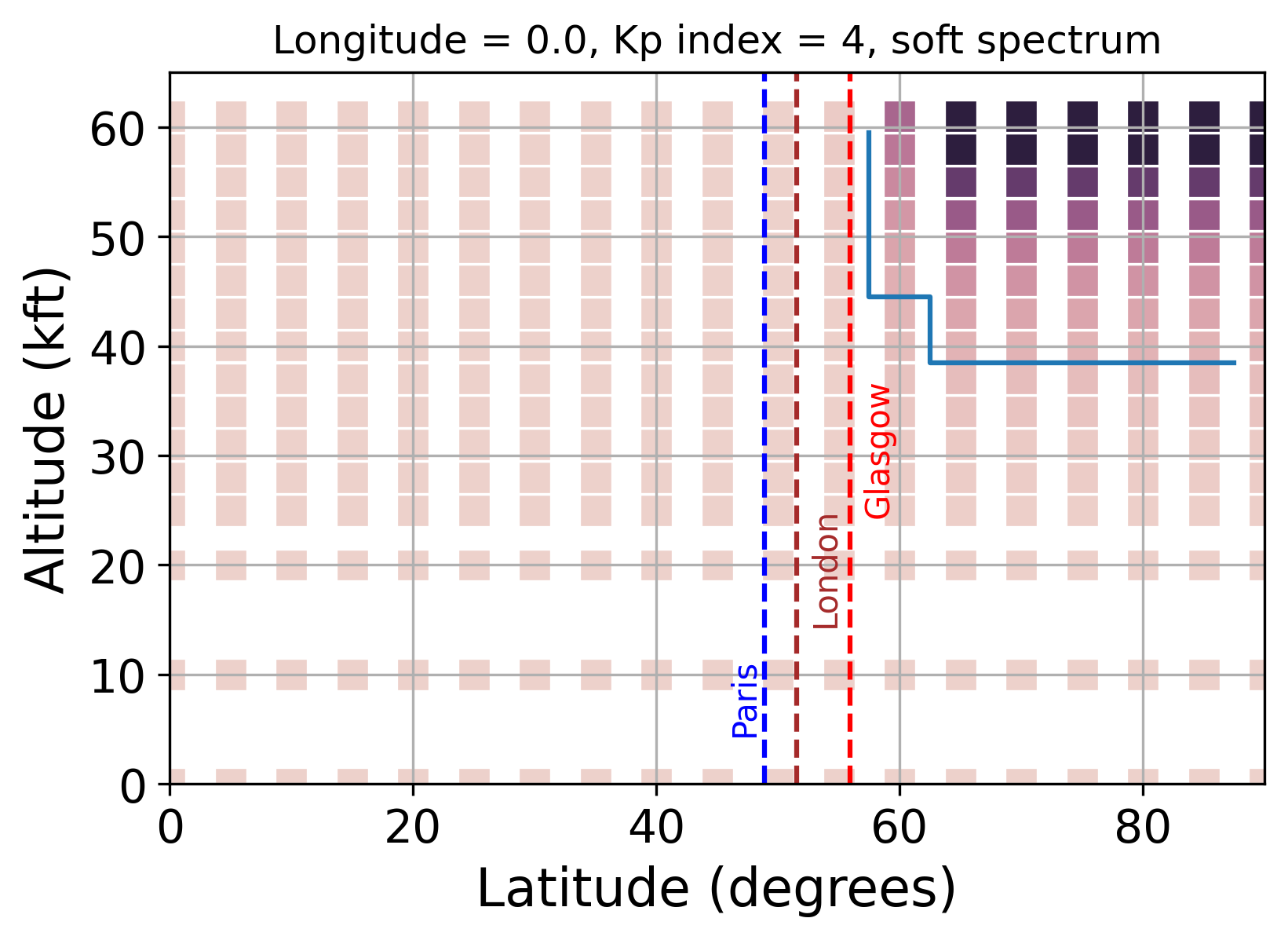}} \\
      
      \subfigure{\includegraphics[trim={0 0 0 0},clip,width=0.4\columnwidth]{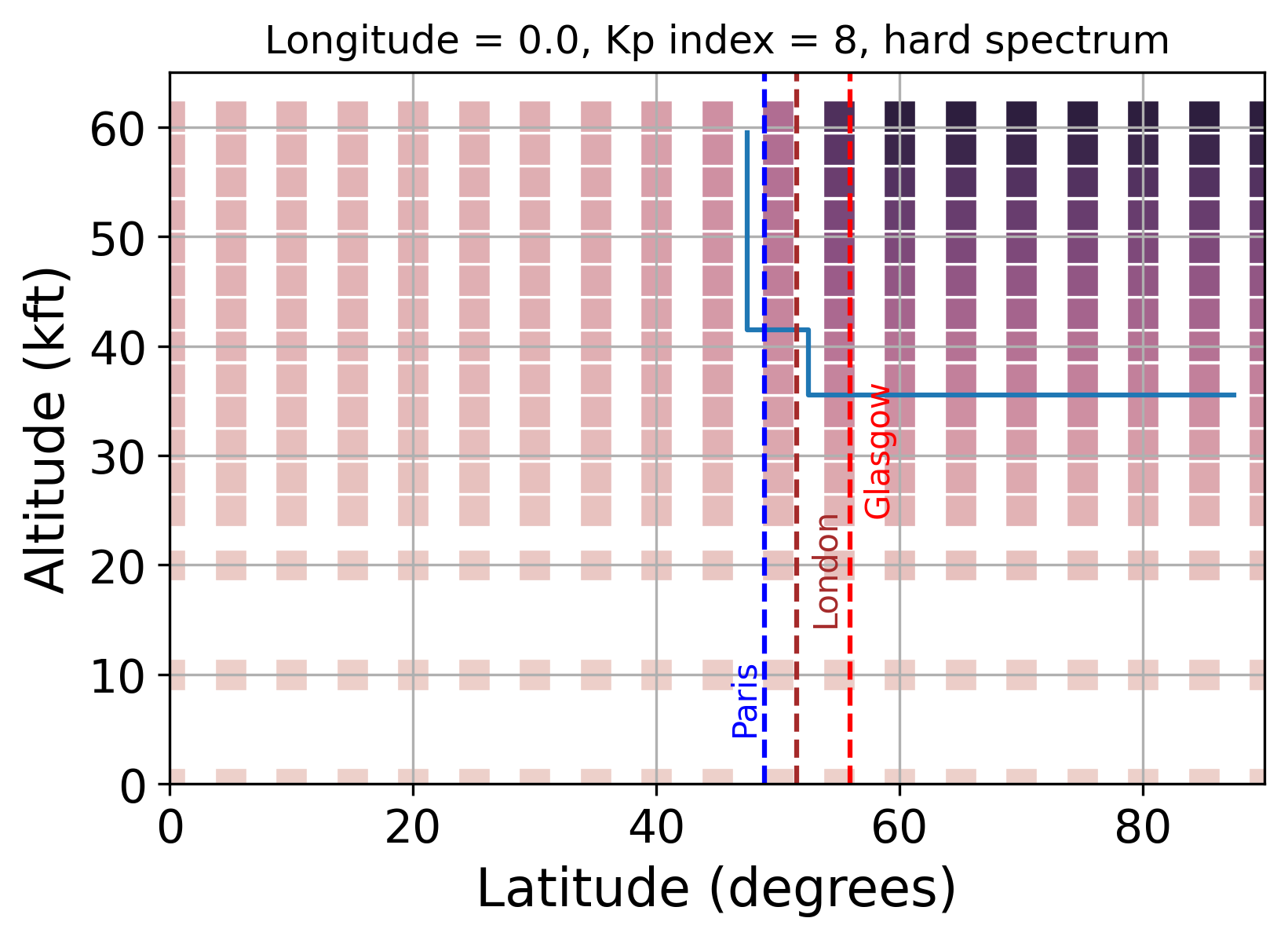}} 
      &
      \subfigure{\includegraphics[trim={0 0 0 0},clip,width=0.4\columnwidth]{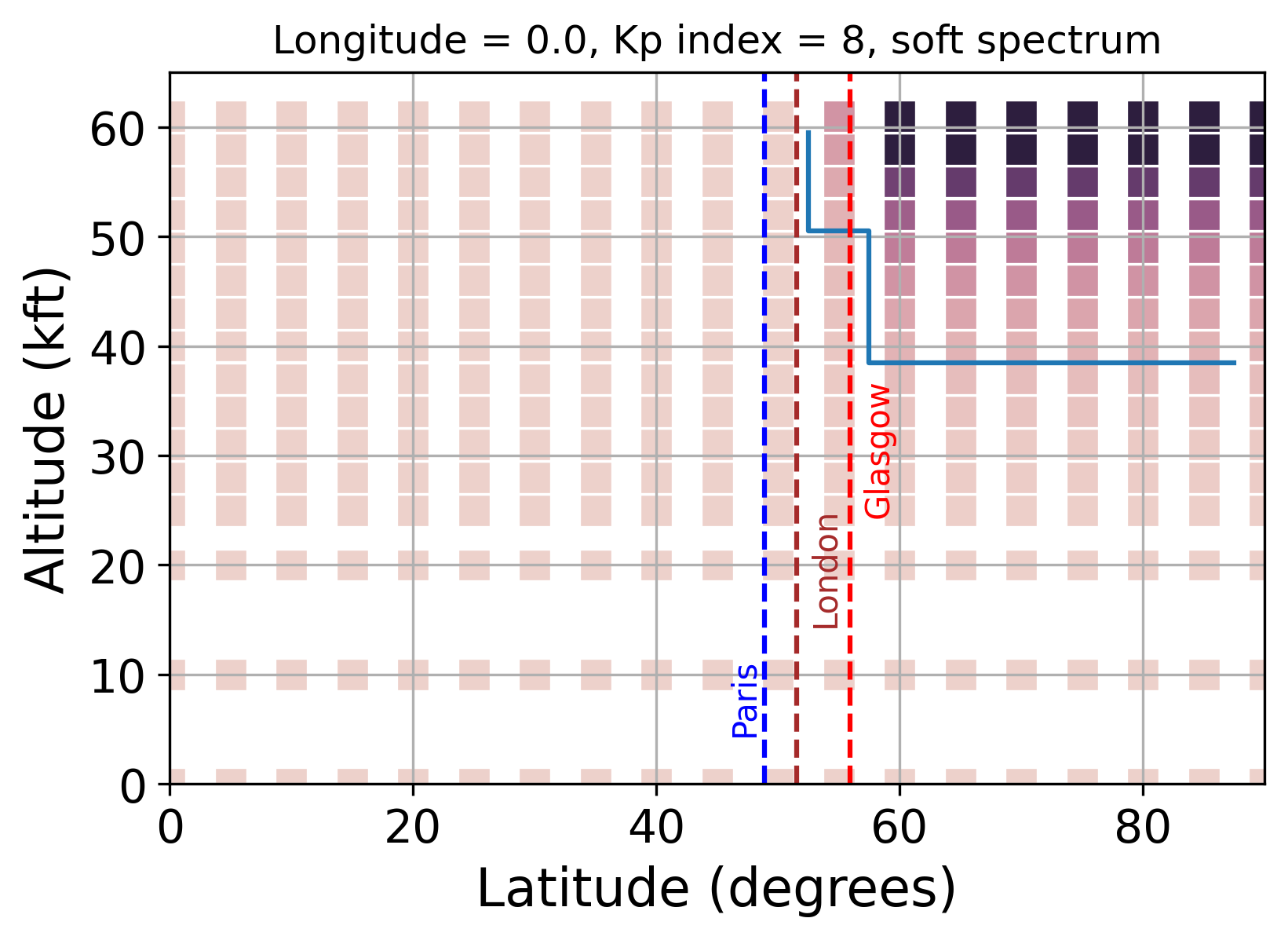}} \\

      \subfigure{\includegraphics[trim={0 0 0 0},clip,width=0.4\columnwidth]{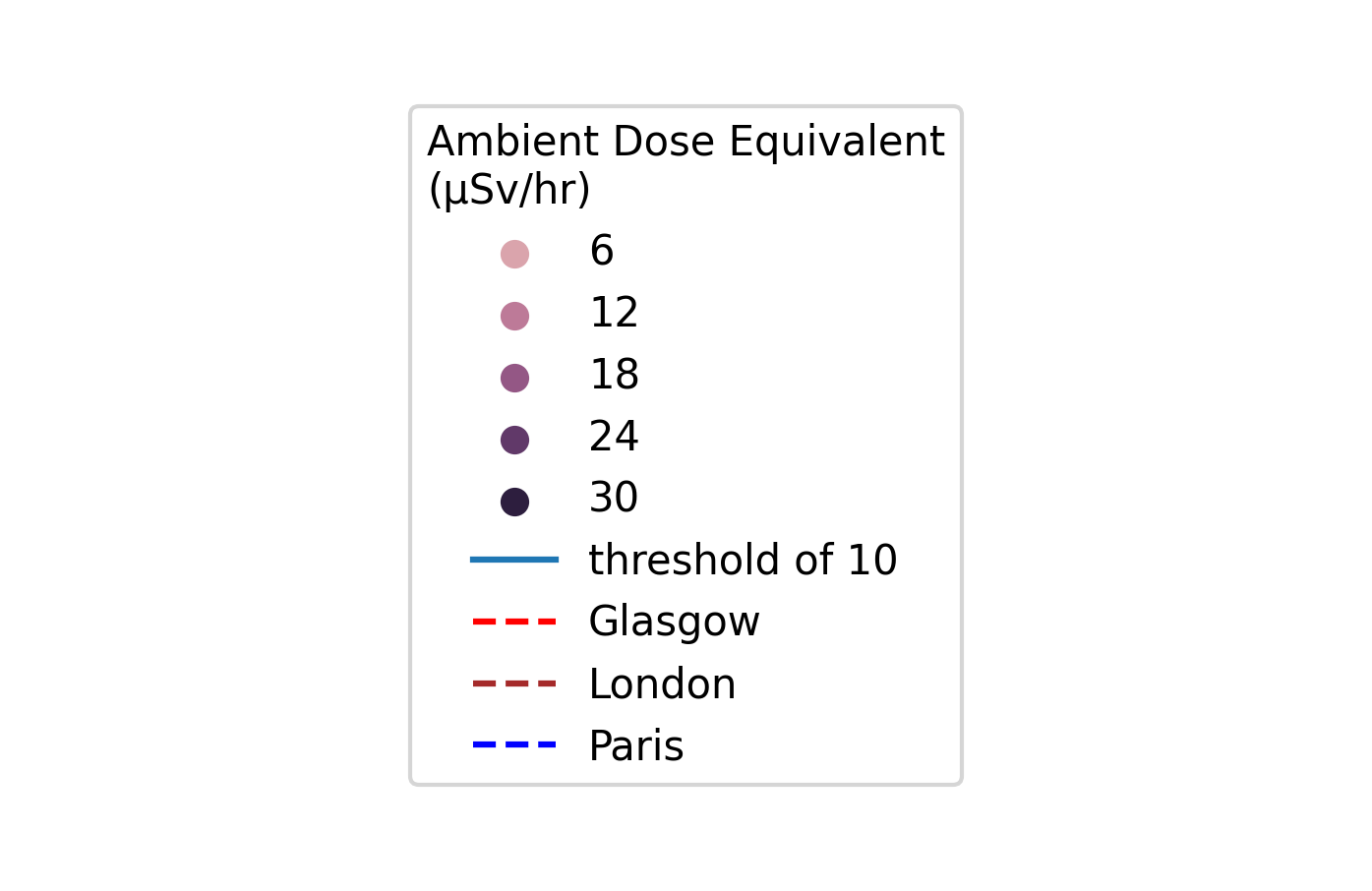}} 
      &
      \subfigure{\includegraphics[trim={0 0 0 0},clip,width=0.4\columnwidth]{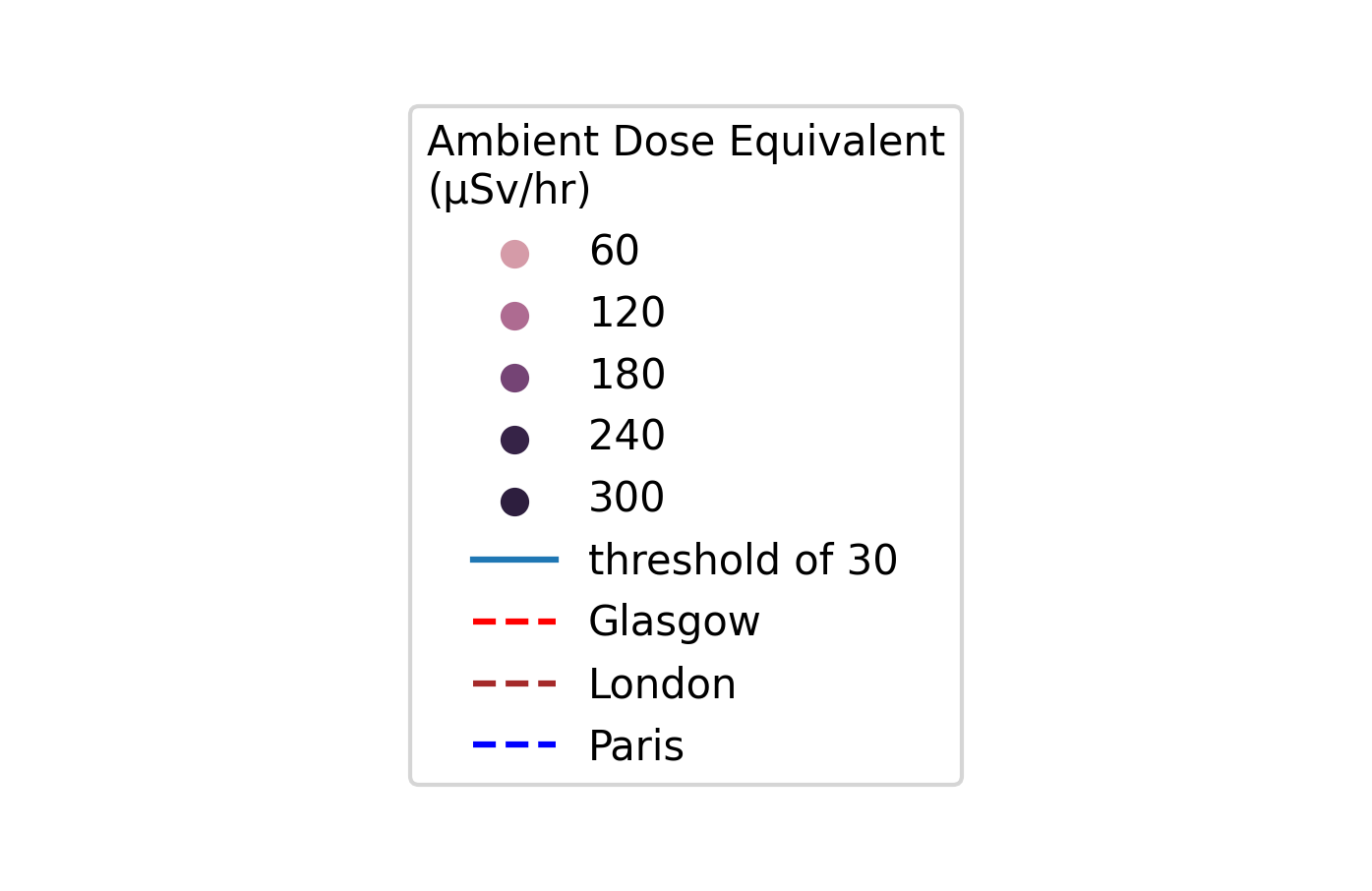}} \\
   \end{tabular}
   \caption{Ambient dose equivalent rate atmospheric cut-throughs 
   for multiple Kp indices. The 30~$\mu Sv / hr$ curve plotted shows 
   the overall shape of the high dose rate polar region, and shows 
   where the region where an aircraft may cross between ICAO thresholds. 
   The main variations in dose rate with Kp index occur at the lower 
   latitude side of the high dose rate polar region, where the region 
   expands to slightly lower latitudes as Kp index increases. Note that 
   the color scales for each spectral case are slightly different.}
   \label{fig:xsectionKpAll}
\end{figure}

The effect of Kp index on the MAIRE-S GLE dose rates can be seen in the atmospheric cut-throughs plotted in figure~\ref{fig:xsectionKpAll}. There is minimal variation in dose rate at most regions of the atmosphere, except for regions that are already slightly south of the transition region between the equatorial region of the sky and the polar region.

Essentially this could be thought of as the transition region 
moving to lower altitudes as Kp index increases. While the 
transition region is relatively narrow with respect to the 
whole atmosphere, the latitude of the transition region is 
positioned at a location that a significant proportion of 
flights cross daily, such as transatlantic flights for instance. 
Additionally, and perhaps more importantly, during a GLE on the 
scale of a 1 in a 100 or 1 in a 1000 year event, magnetospheric 
conditions could reach extreme levels that have never been 
measured before, such that they could vastly exceed the 
magnetospheric upset levels described by Kp = 8, as the physics 
of such an event is not well understood yet. The results given 
in figure~\ref{fig:xsectionKpAll} indicate that in such an 
event, the transition region between high and low dose rates 
would move southwards, perhaps even approaching latitudes of 
50\textdegree N, which roughly corresponds to cities such as 
London, Paris, Brussels and Frankfurt near the simulated 
longitude of 0\textdegree E in figure~\ref{fig:xsectionKpAll}. 
The results given here are consistent with similar findings from 
Shea and Smart (2005)\citep{smart2005review}, where it was found 
that the latitude corresponding to a given vertical cut-off 
rigidity decreases with increasing Kp-index, across the entire 
vertical cut-off rigidity range.

figure~\ref{fig:xsectionPercentIncreases} below displays the 
percentage increase of ambient dose equivalent rate as Kp index 
increases, and further highlights the influence that changing Kp 
has on radiation dose rates in the transition area between 
equatorial and polar regions. It also shows how much more 
significant the effect is between Kp=0 and Kp=8, where dose 
rates within the transition region increase by up to about 
1250\% in the soft spectrum case relative to dose rates in the 
more equatorial region. Dose rates for the hard spectrum case 
increase by up to 40\% near the transition region at 60 kft 
altitudes.

\begin{figure}
   \centering
      \foreach \KpIndex in {4,8}{
         \foreach \specNumber in {1,2}{
            \subfigure{\includegraphics[trim={0 0 0 0},clip,width=0.4\columnwidth]{percentIncreaseKp\KpIndex Kp0Timestamp\specNumber_no_legend.png}}
         }
      }

   \subfigure{\includegraphics[trim={0 0 0 0},clip,width=0.4\columnwidth]{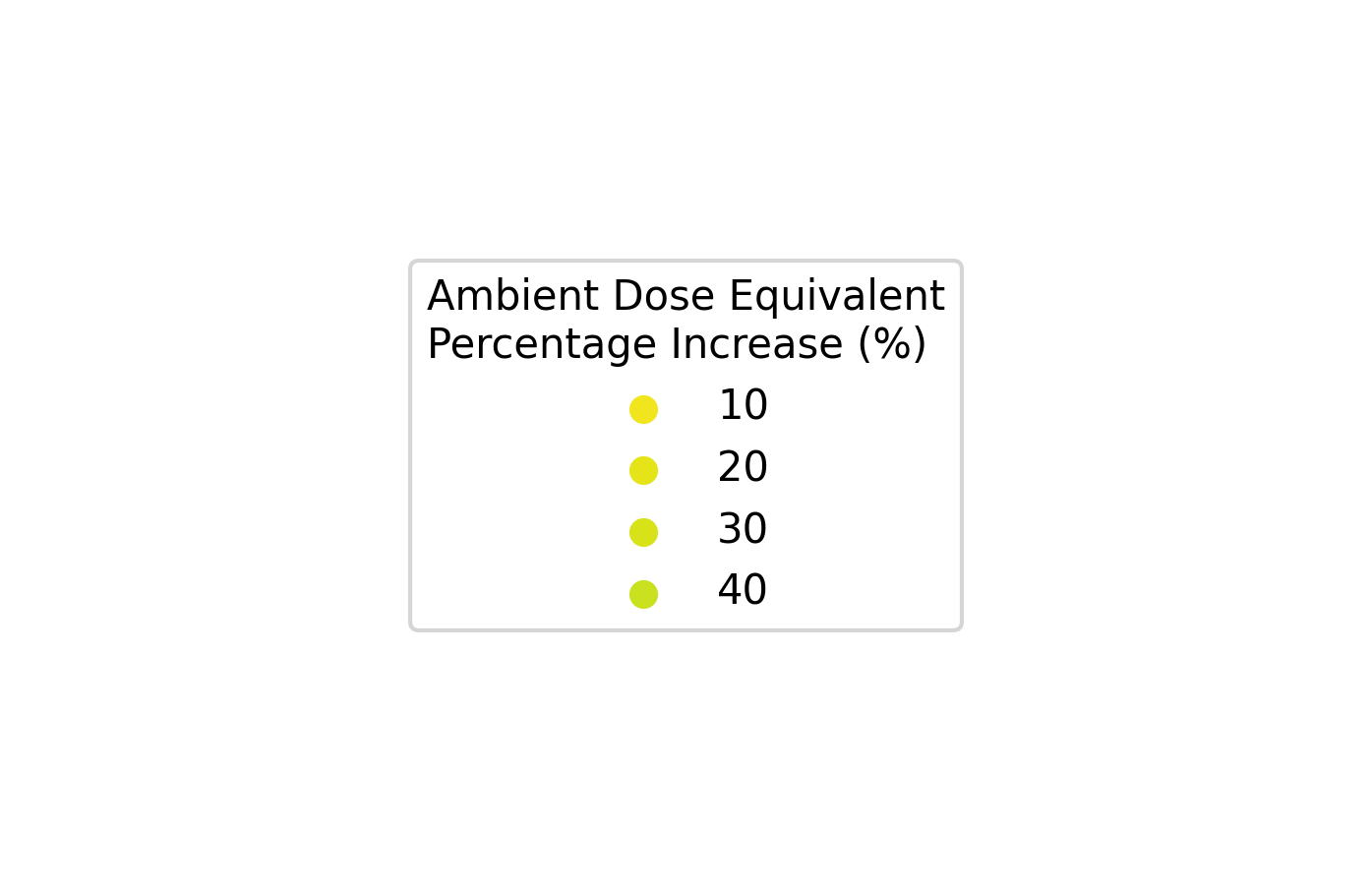}}
   \subfigure{\includegraphics[trim={0 0 0 0},clip,width=0.4\columnwidth]{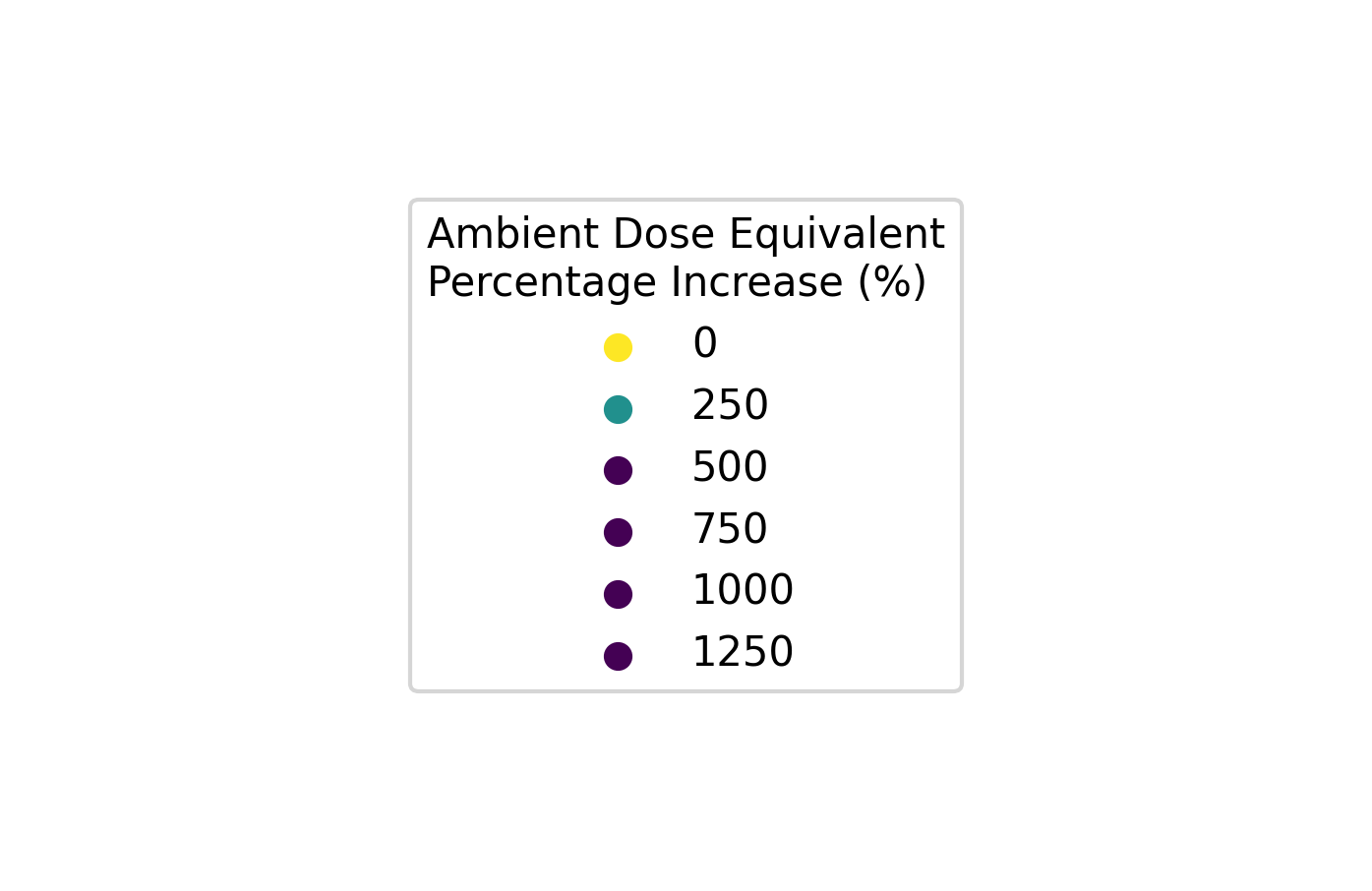}}

   \caption{Percentage increases in total ambient dose equivalent rate (GCR-induced + GLE-induced) 
   between different Kp indices. Changes in Kp index between Kp=0 and Kp=8 show increases of up to 
   1250\% around the transition region between the equatorial and polar areas in the soft spectral 
   index case. In the hard spectral index case, the percentage increase is only up to approximately 
   40\% at maximum.}
   \label{fig:xsectionPercentIncreases}
\end{figure}

\section{The Variation of Dose Rate Atmospheric Cut-through Structures as a Function of Longitude}

\begin{figure}
   \centering
   \begin{tabular}{c|c}

      \subfigure{\includegraphics[trim={0 0 0 0},clip,width=0.34\columnwidth]{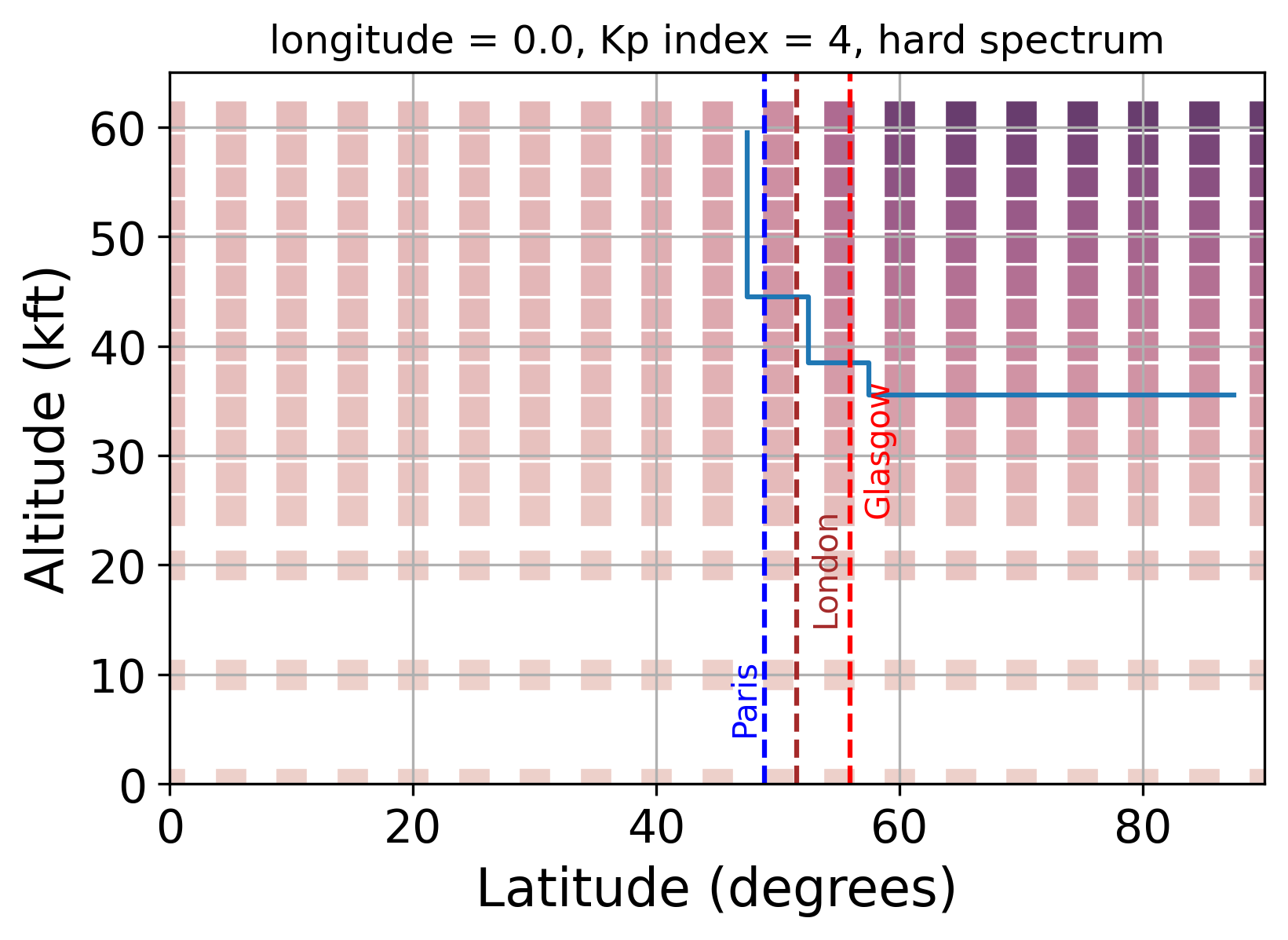}}
      &
      \subfigure{\includegraphics[trim={0 0 0 0},clip,width=0.34\columnwidth]{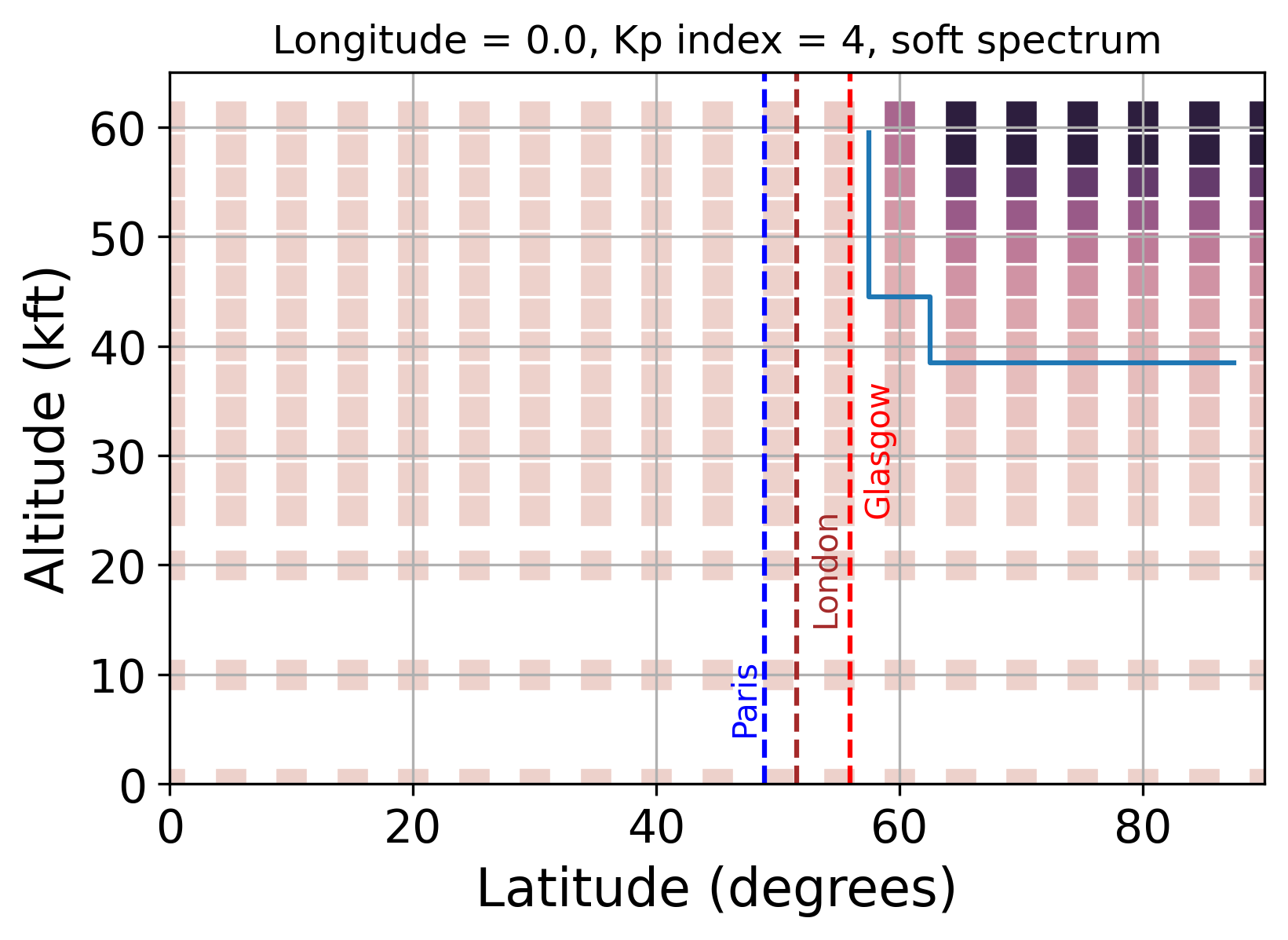}} \\

      \subfigure{\includegraphics[trim={0 0 0 0},clip,width=0.34\columnwidth]{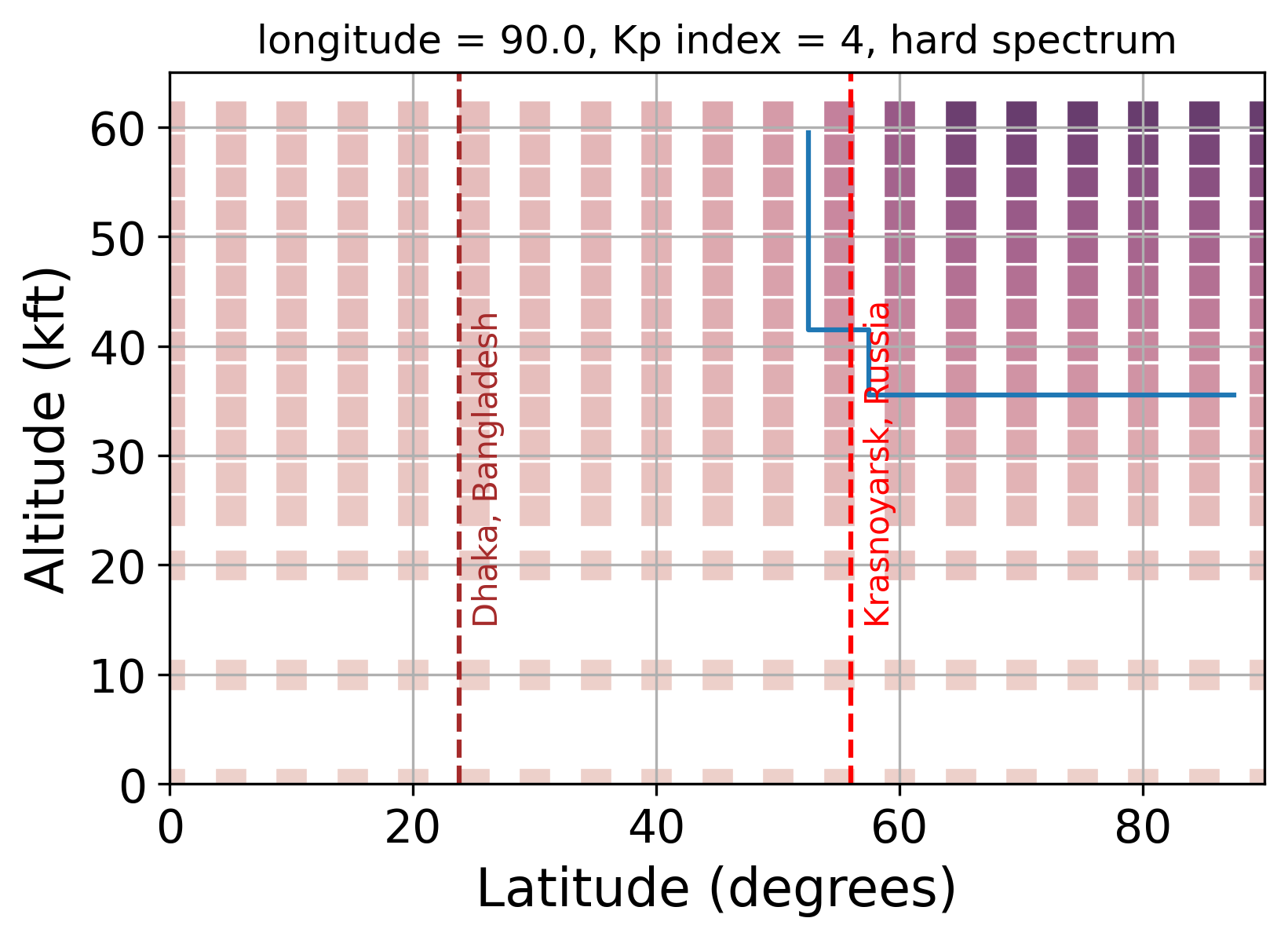}}
      &
      \subfigure{\includegraphics[trim={0 0 0 0},clip,width=0.34\columnwidth]{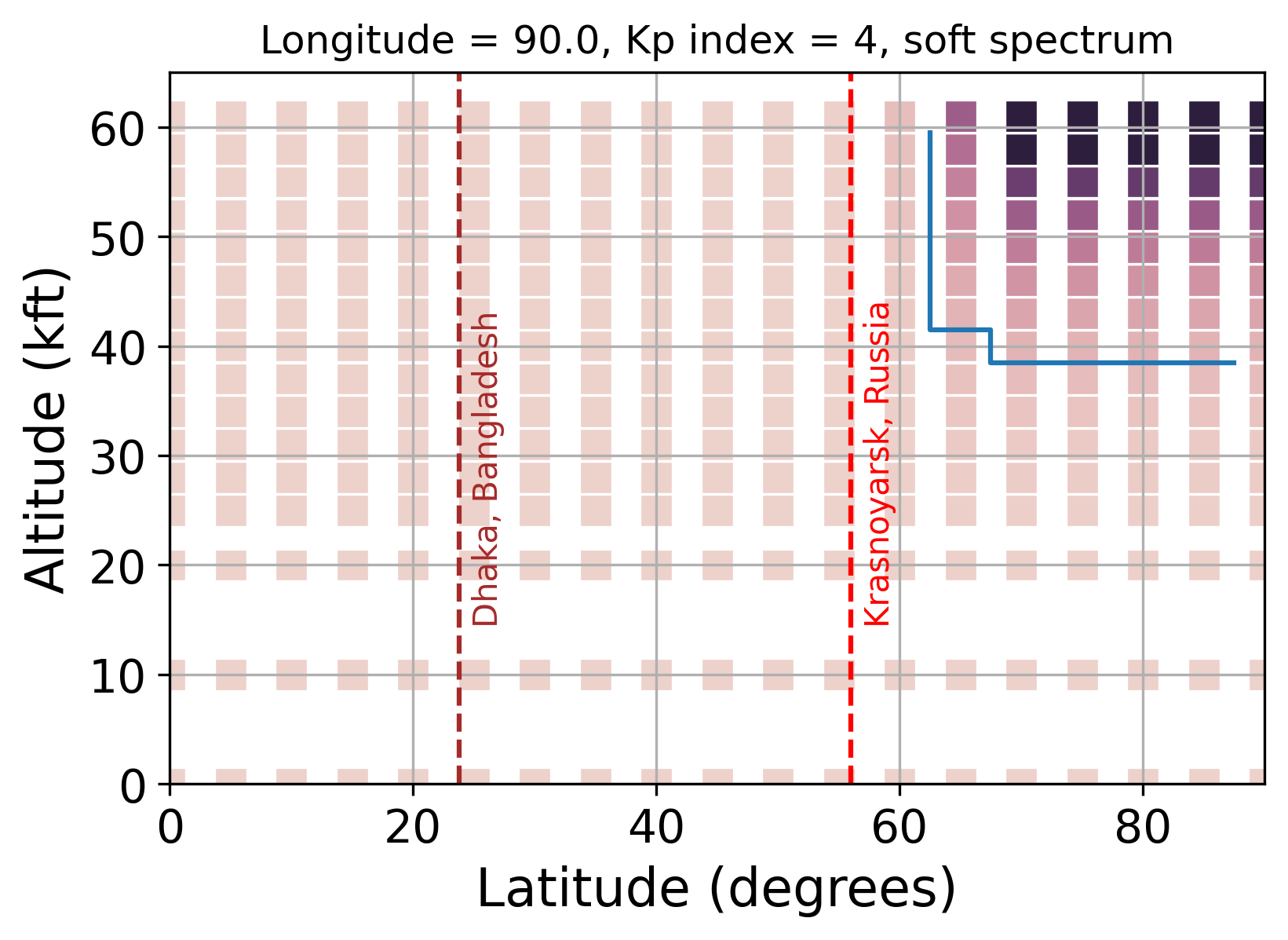}} \\

      \subfigure{\includegraphics[trim={0 0 0 0},clip,width=0.34\columnwidth]{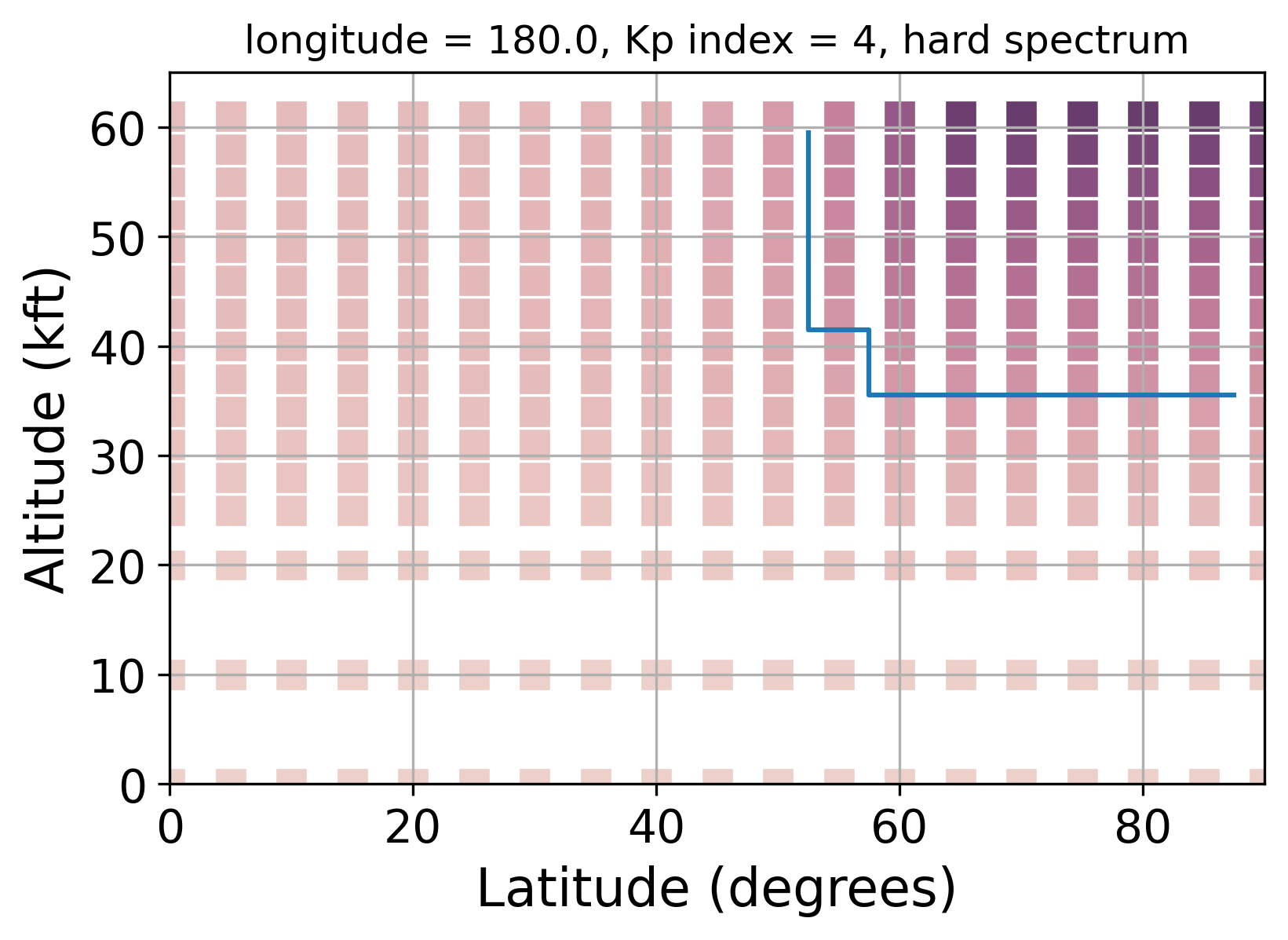}}
      &
      \subfigure{\includegraphics[trim={0 0 0 0},clip,width=0.34\columnwidth]{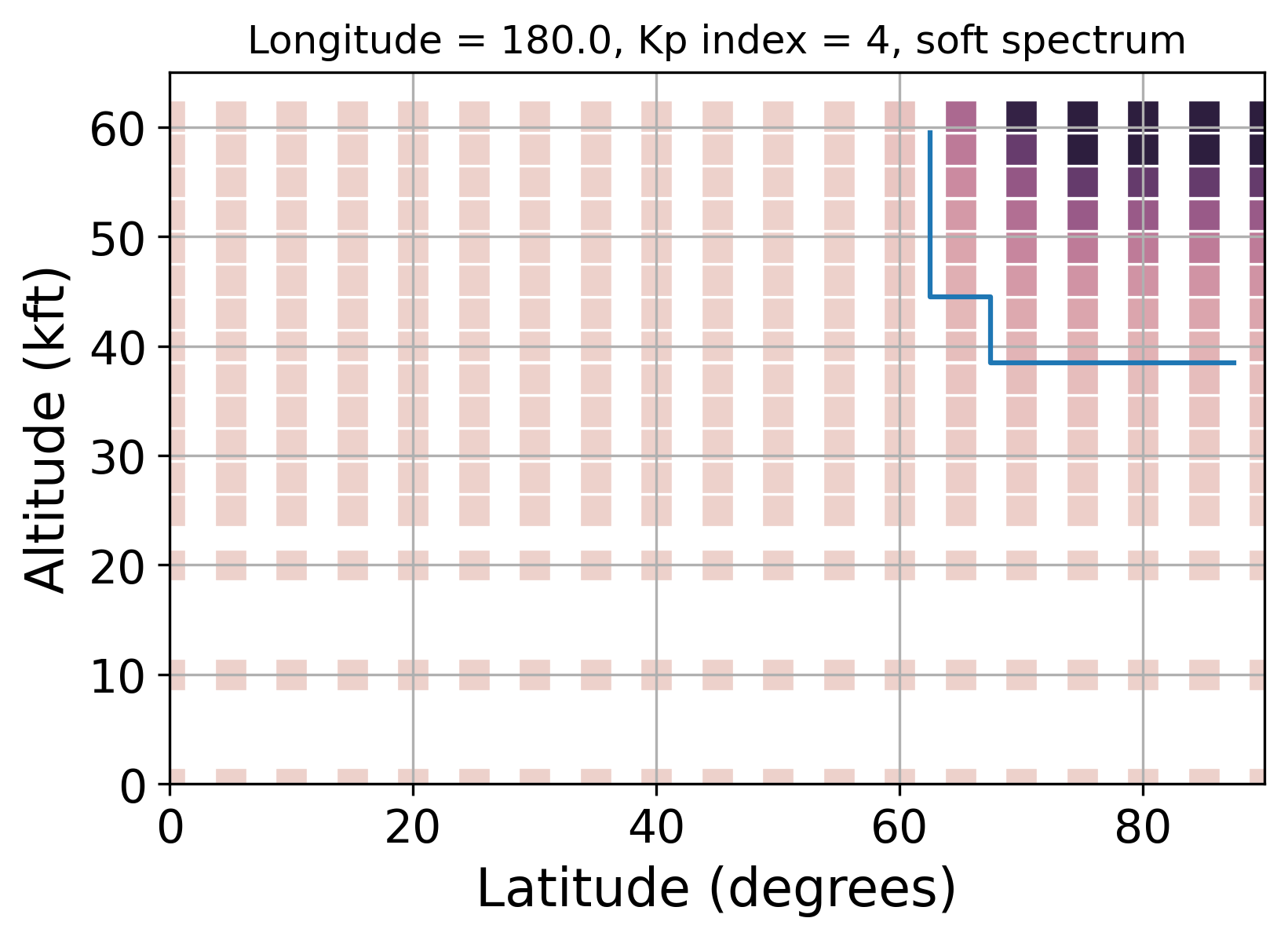}} \\

      \subfigure{\includegraphics[trim={0 0 0 0},clip,width=0.34\columnwidth]{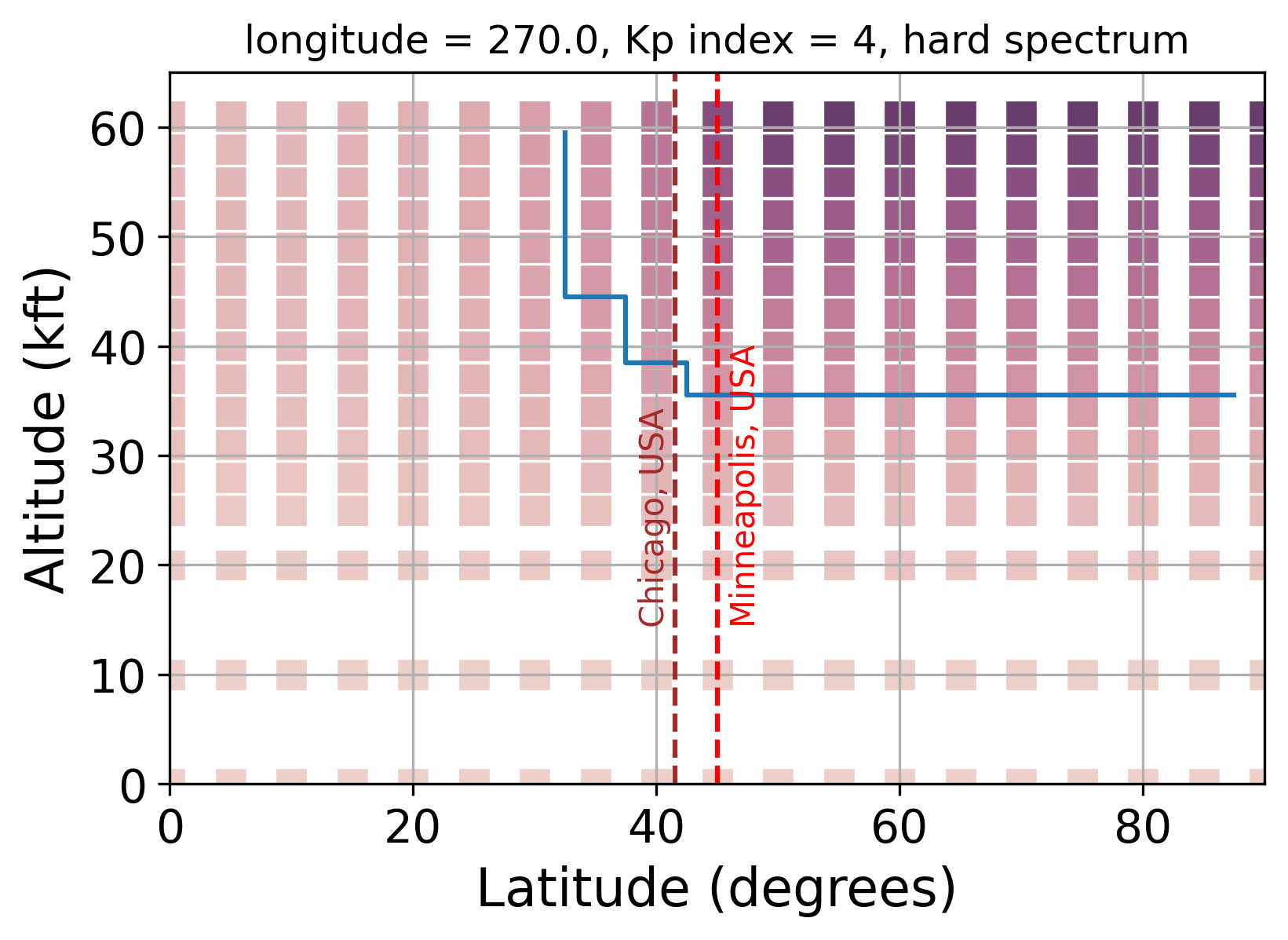}}
      &
      \subfigure{\includegraphics[trim={0 0 0 0},clip,width=0.34\columnwidth]{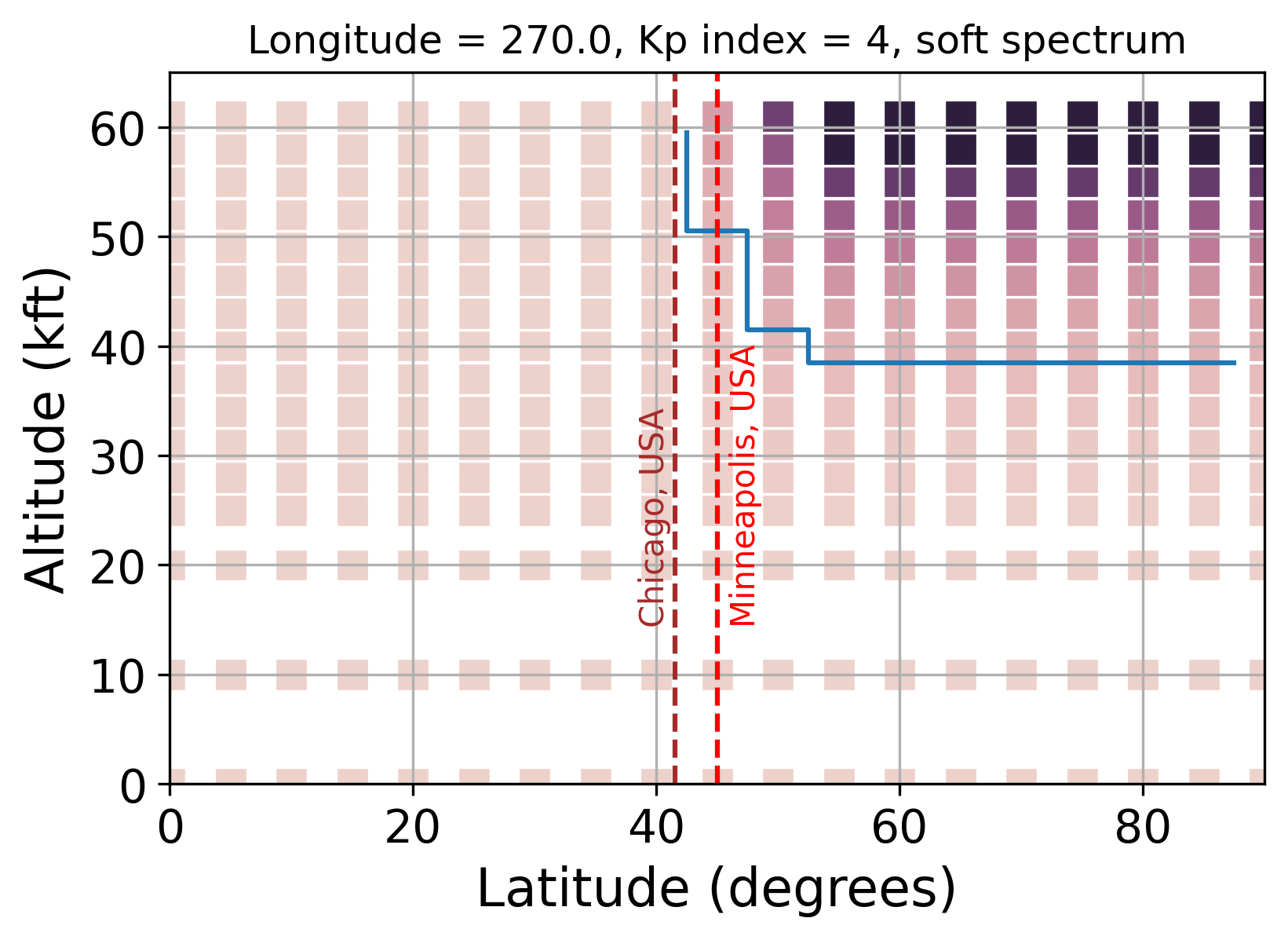}} \\

      \subfigure{\includegraphics[trim={0 1cm 0 1cm},clip,width=0.4\columnwidth]{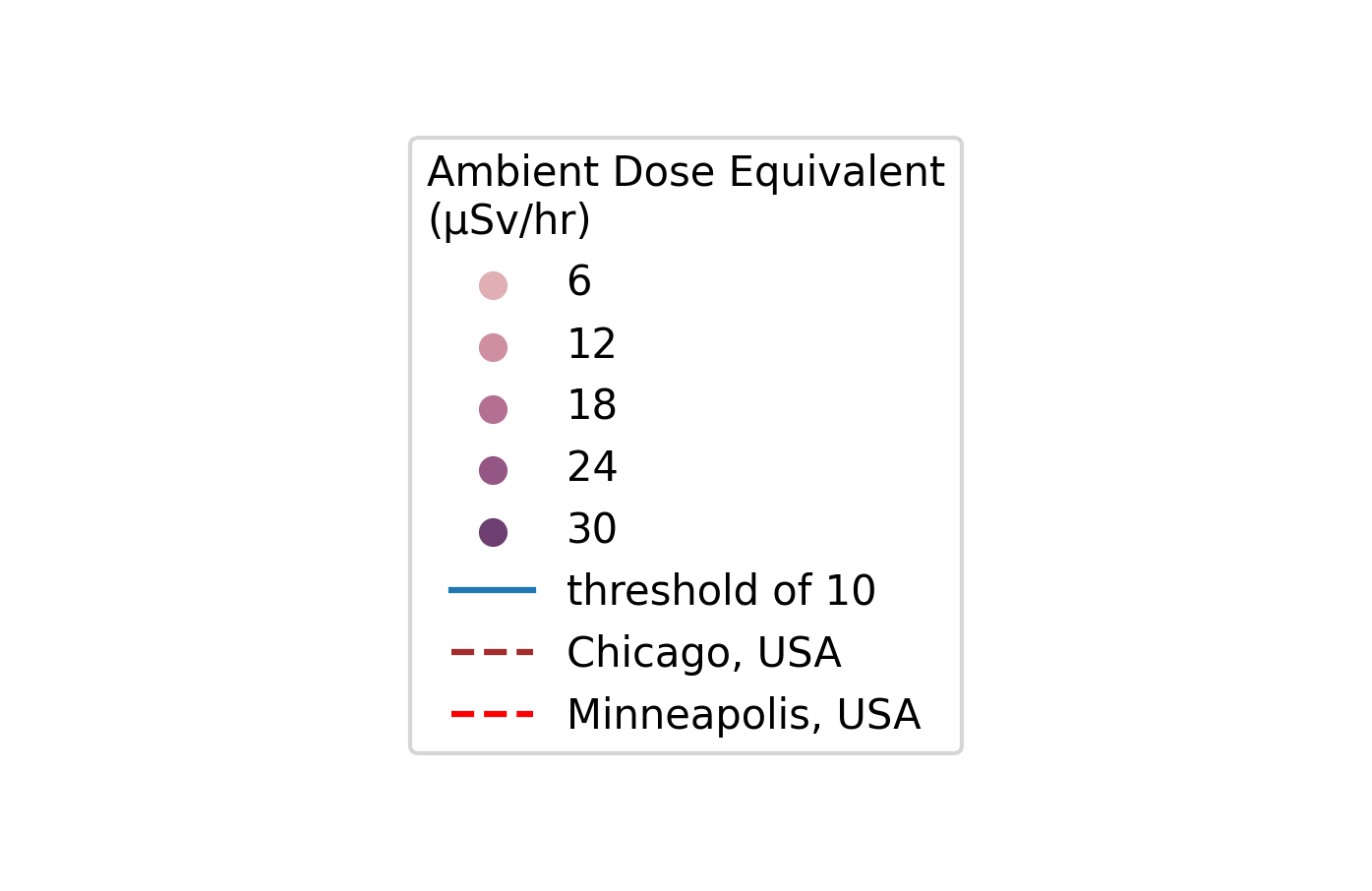}}
      &
      \subfigure{\includegraphics[trim={0 1cm 0 1cm},clip,width=0.4\columnwidth]{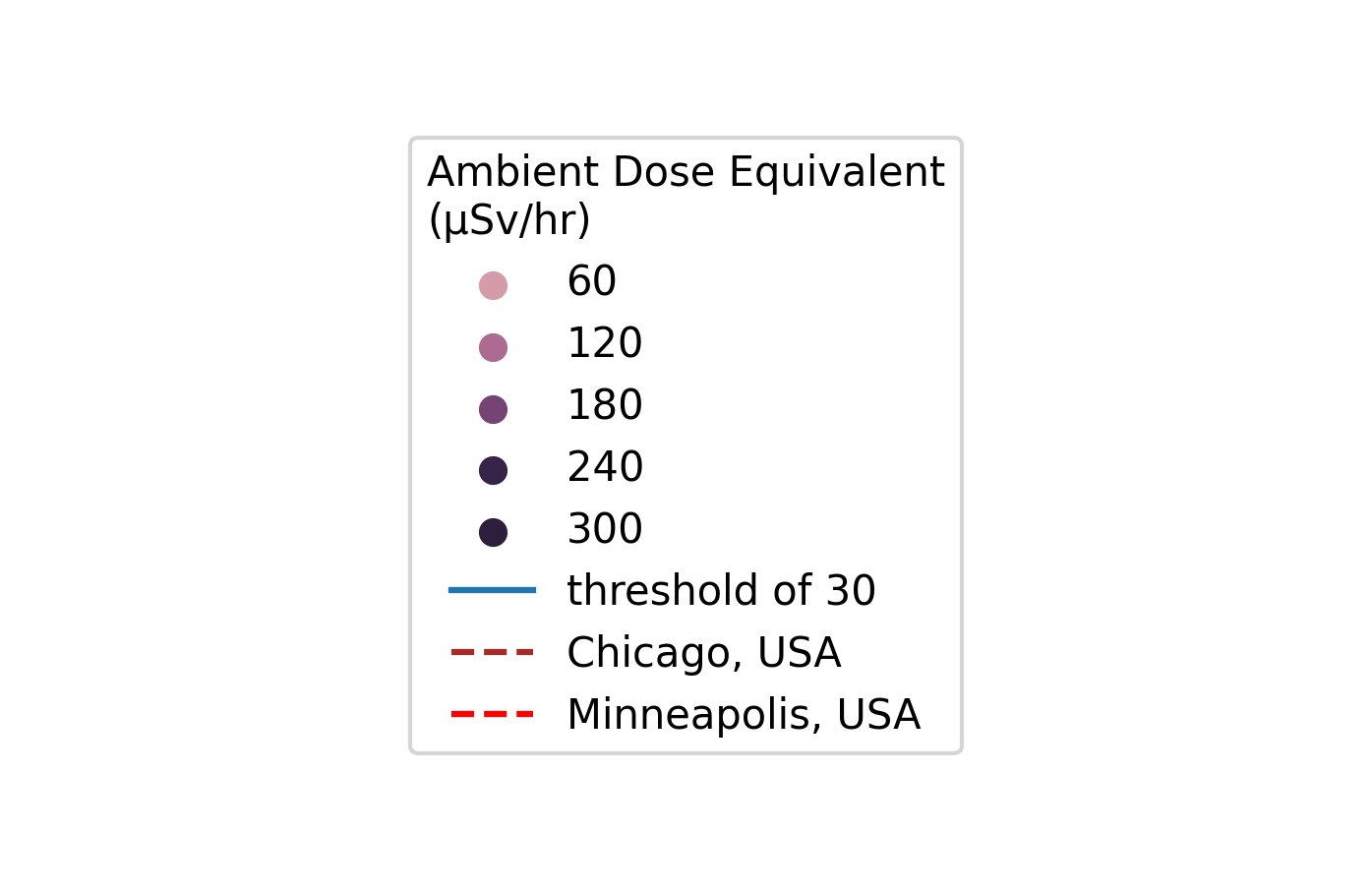}} \\
   \end{tabular}
   \caption{Ambient dose equivalent cut-throughs for 
   different longitudes. Several cities near to the plotted 
   longitudes are also displayed for geographic context. 
   There is minimal change in dose rate with altitude as 
   longitude changes, however the high dose rate polar region 
   stretches and contracts as longitude changes. At 
   longitude=270.0\textdegree E, corresponding to North 
   American airspace, the high dose rate region stretches as far 
   south as approximately 45\textdegree N-50\textdegree N, due 
   to its closer proximity to the North magnetic pole in 1989.}
   \label{fig:mergedLongitudeVariationPlots}
\end{figure}

In addition to examining how changes in Kp indices affect radiation dose rates, it is also useful to examine the impact that longitude has on atmospheric cut-throughs. Cut-throughs of the atmosphere are displayed in figure~\ref{fig:mergedLongitudeVariationPlots} and illustrate that the main impact of longitude appears to be to move the cut-off threshold position towards lower latitudes at certain locations, while not significantly altering the dose rate profile as a function of changing altitude. This effect is caused by the location of the magnetic north pole differing somewhat from the location of the geographic north pole, which was positioned over northern Canada in 1989 (and still is today).

\begin{figure}
   \centering

   \subfigure{\includegraphics[trim={0 0 0 0},clip,width=0.3\columnwidth]{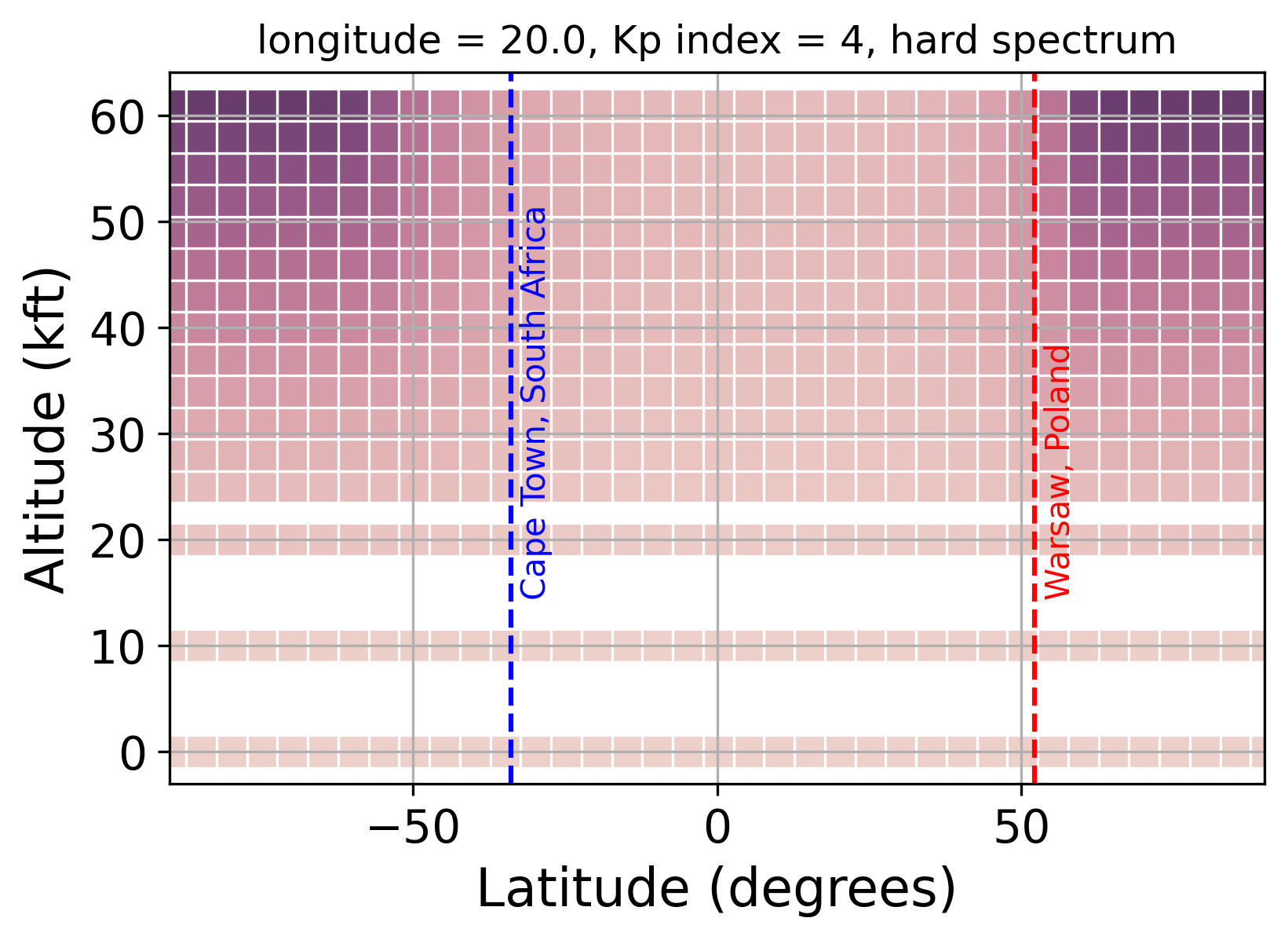}}
   \subfigure{\includegraphics[trim={0 0 0 0},clip,width=0.3\columnwidth]{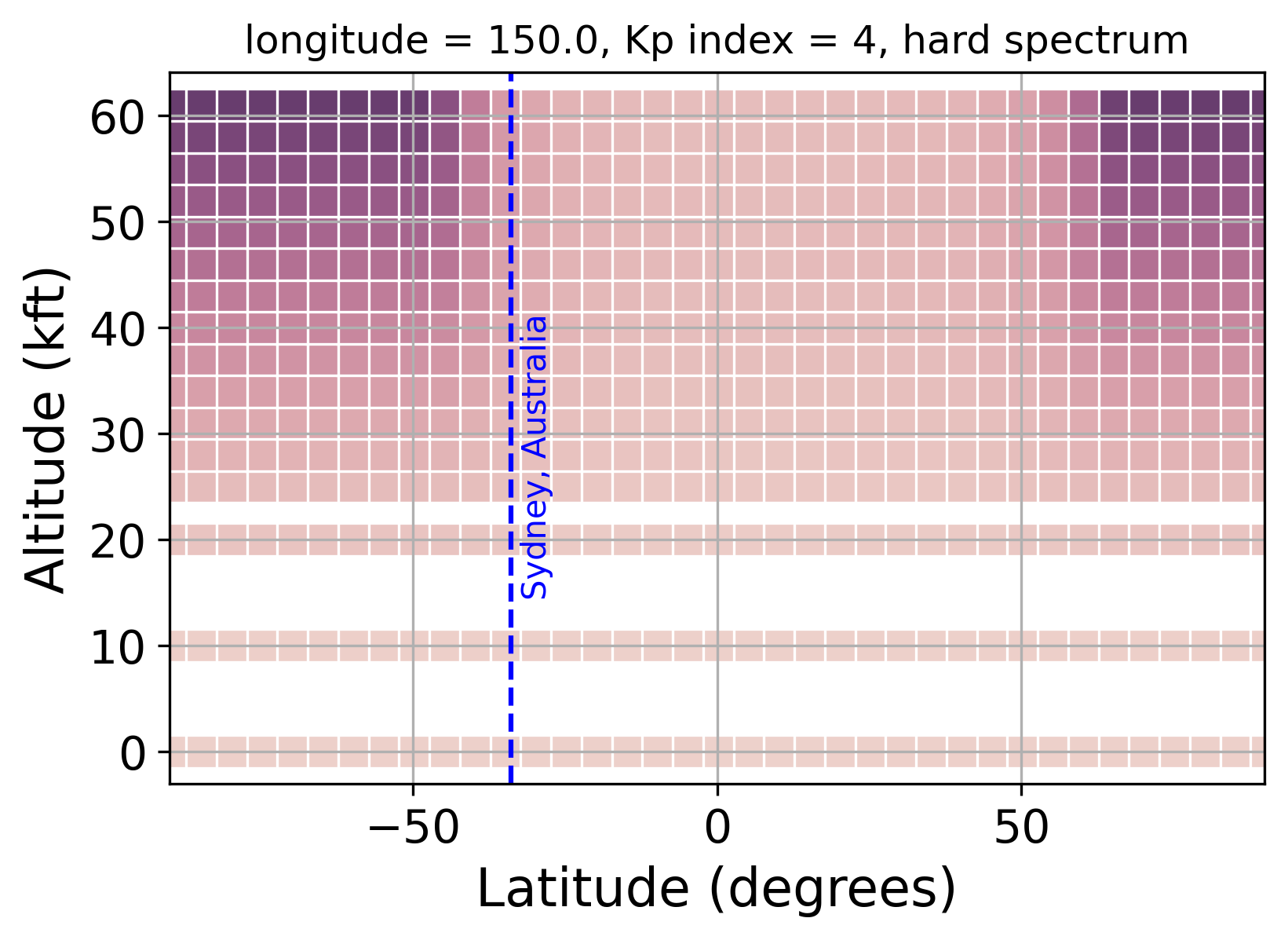}}
   \subfigure{\includegraphics[trim={0 0 0 0},clip,width=0.3\columnwidth]{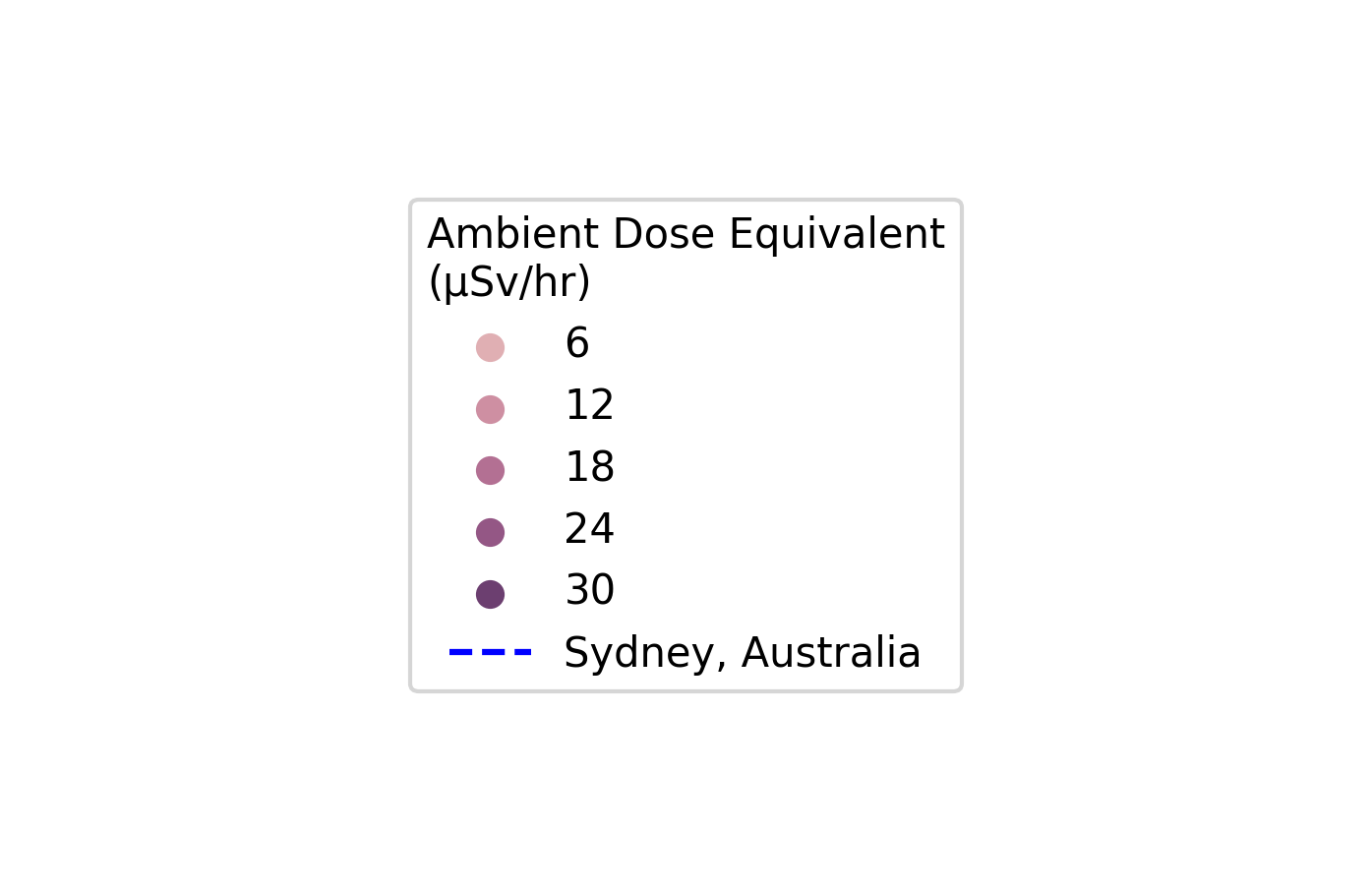}}

   \caption{Cut-throughs of the full range of latitudes above 
   Earth's atmosphere during hard spectral index conditions at 
   longitudes corresponding to Cape Town, South Africa and 
   Sydney, Australia.}
   \label{fig:mergedFullLatRange}
\end{figure}

Plots showing the full range of Earth's latitudes are also 
displayed in figure~\ref{fig:mergedFullLatRange}, so that the 
qualitative full dose rate structure in Earth's atmosphere can 
be seen, as well as dose rates above several southern 
hemispheric locations. 

figure~\ref{fig:mergedLongitudeVariationPlots}, together with figure~\ref{fig:xsectionKpAll} highlights the fact that almost the entirety of the geometry for the event can be described by a narrow transition region which varies with longitude and magnetospheric conditions under the approximations and spectra given earlier in table~\ref{tab:inputtedSpectra}. The low latitude/`equatorial' side of this transition region has relatively low dose rates at all altitudes. In contrast, high latitude/`polar' sides of the transition region exhibit high dose rates but with a reasonably consistent profile at constant altitude. It should be noted that this high latitude dose rate region stretches to much lower latitudes than the auroral region. 

While the transition region between these two areas is reasonably narrow with respect to the total range of possible latitudes, it still wide enough, on the order of several degrees latitude, such that some aircraft may spend a reasonable time in the transition region during a flight. While an aircraft will experience a lower overall dose rate in this region than if it flew exclusively over polar regions, flying over the transition region would lead to a significant uncertainty in dose rates, given how much dose rates in this region appear to vary with magnetospheric conditions and with flight latitude. It should also be noted that some of the busiest flight routes in the world pass through the latitudes associated with this transition region, including most transatlantic Europe to USA flights for instance. Therefore it is useful to examine the structure and properties of this transition region in more detail.

\section{The Sensitivity of Dose Rate to Latitude}

The transition region described above can be thought of as the region of atmospheric cut-throughs that are most sensitive to changes in latitude. The appropriate quantity that can be used to describe the transition region is therefore the rate of change of dose rate with respect to latitude, $\frac{\partial H^*(10)}{\partial \theta}$, where $H^*(10)$ is ambient dose equivalent, and $\theta$ is latitude.

\begin{figure}
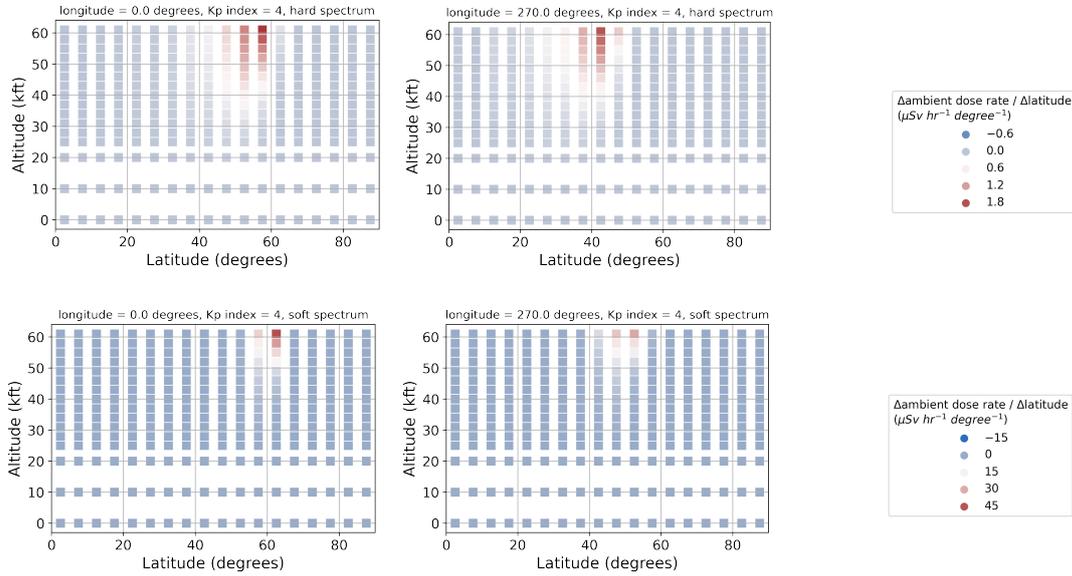

   \centering
   \foreach \spectrum in {Hard,Soft}{
      \foreach \longitude in {0,270}{
         \subfigure{\includegraphics[trim={0 0 0 0},clip,width=0.3\columnwidth]{diffxsection-longitude_\longitude.0\spectrum.png}}
         }
      \subfigure{\includegraphics[trim={0 0 0 0},clip,width=0.3\columnwidth]{diffxsection-longitude_0withLegend\spectrum_legend.png}}
      }
   \caption{The sensitivity of dose rates to changing latitude, 
            expressed in terms of the derivative of ambient dose 
            equivalent to latitude, $\frac{\partial H^*(10)}{\partial \theta}$. 
            The area of maximum sensitivity corresponds to the transition area 
            between the equatorial and polar dose rate regions. As was previously 
            indicated in figure~\ref{fig:mergedLongitudeVariationPlots}, the transition 
            region for longitude=270.0\textdegree is at a lower latitude than for 
            longitude=0.0\textdegree.}
   \label{fig:latitudeSensitivityMaps}
\end{figure}

\begin{figure}
   \centering
   \subfigure{\includegraphics[trim={0 0 0 0},clip,width=0.4\columnwidth]{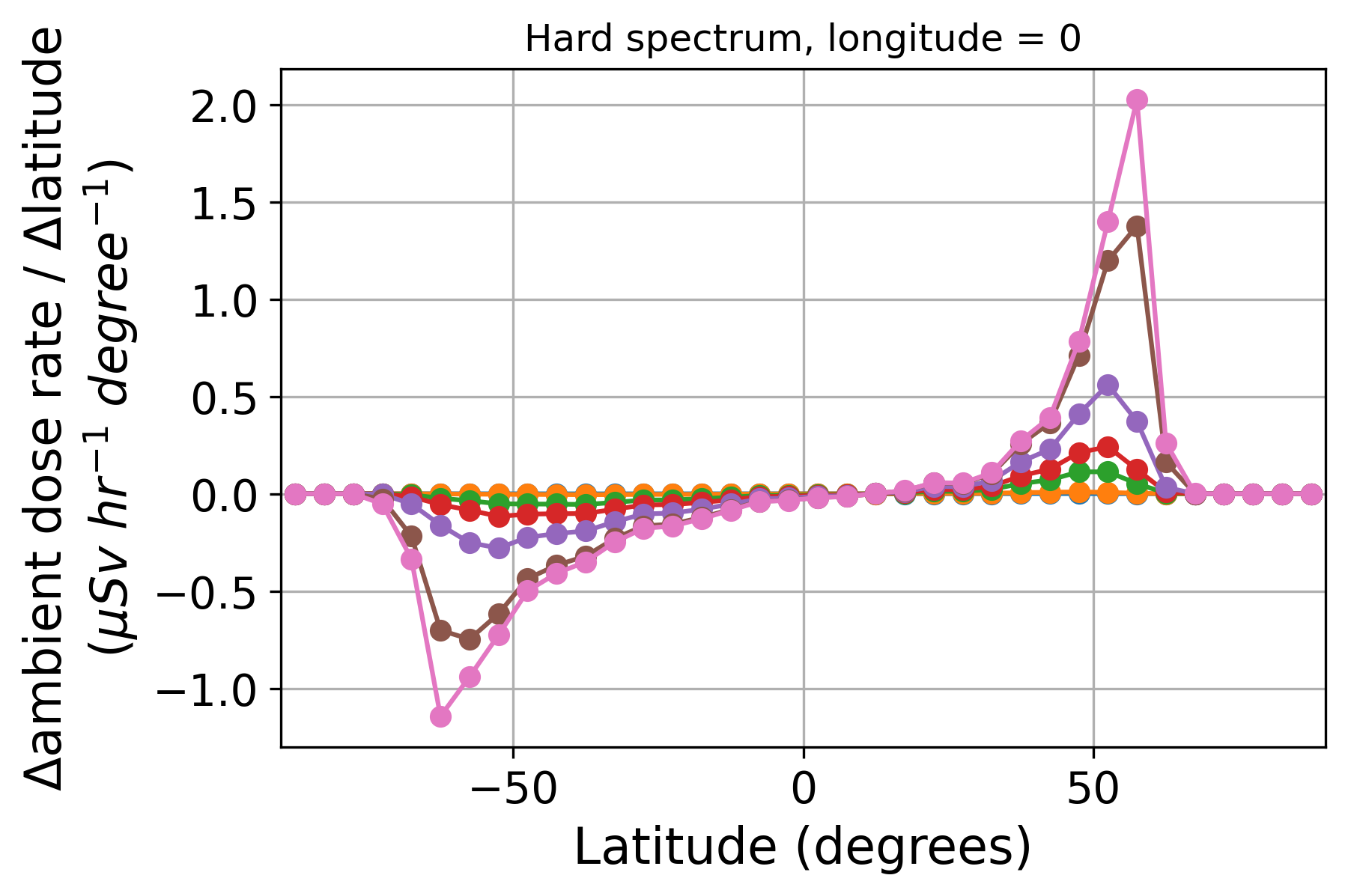}}
   \subfigure{\includegraphics[trim={0 0 0 0},clip,width=0.4\columnwidth]{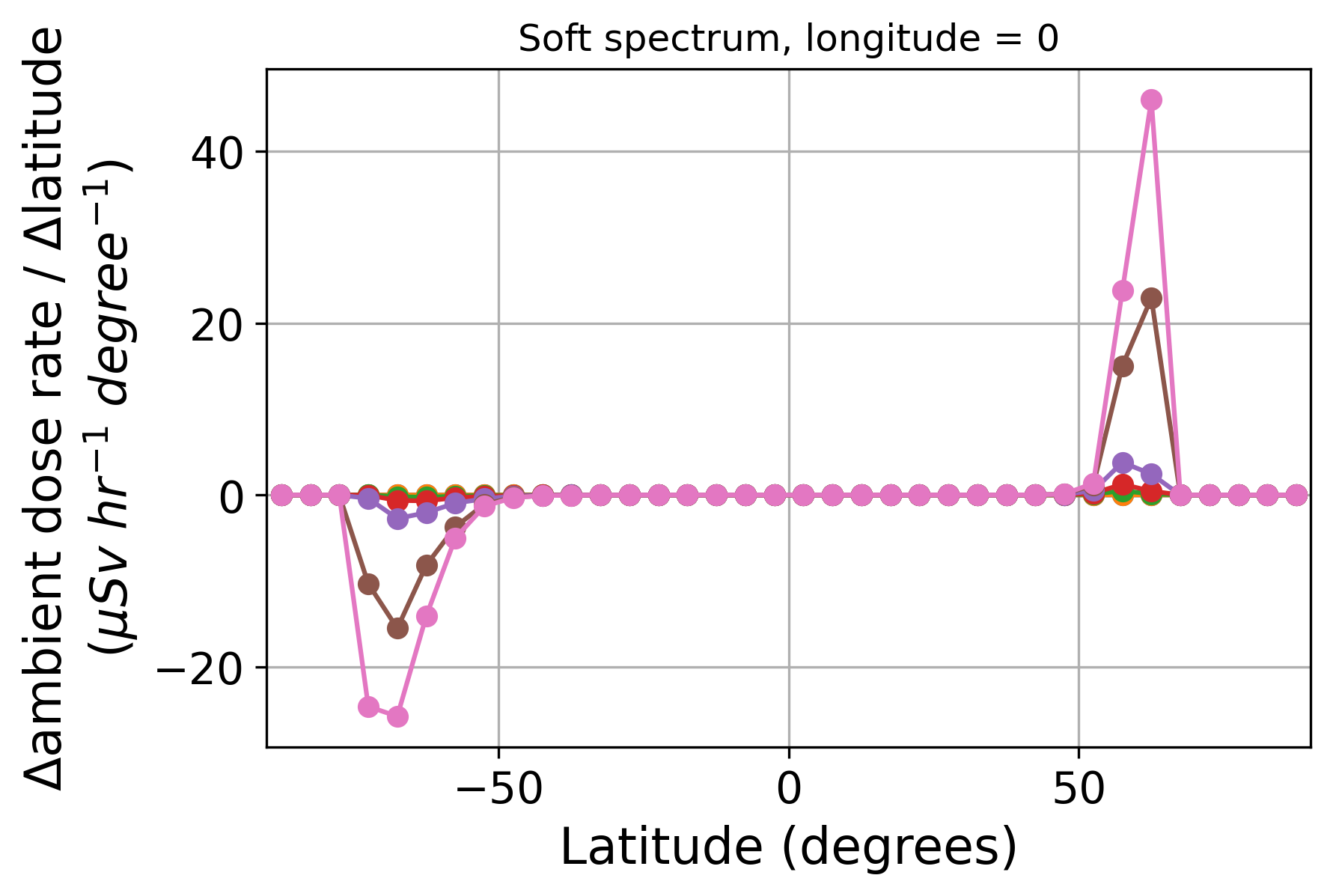}}
   \subfigure{\includegraphics[trim={0 0 0 0},clip,width=0.4\columnwidth]{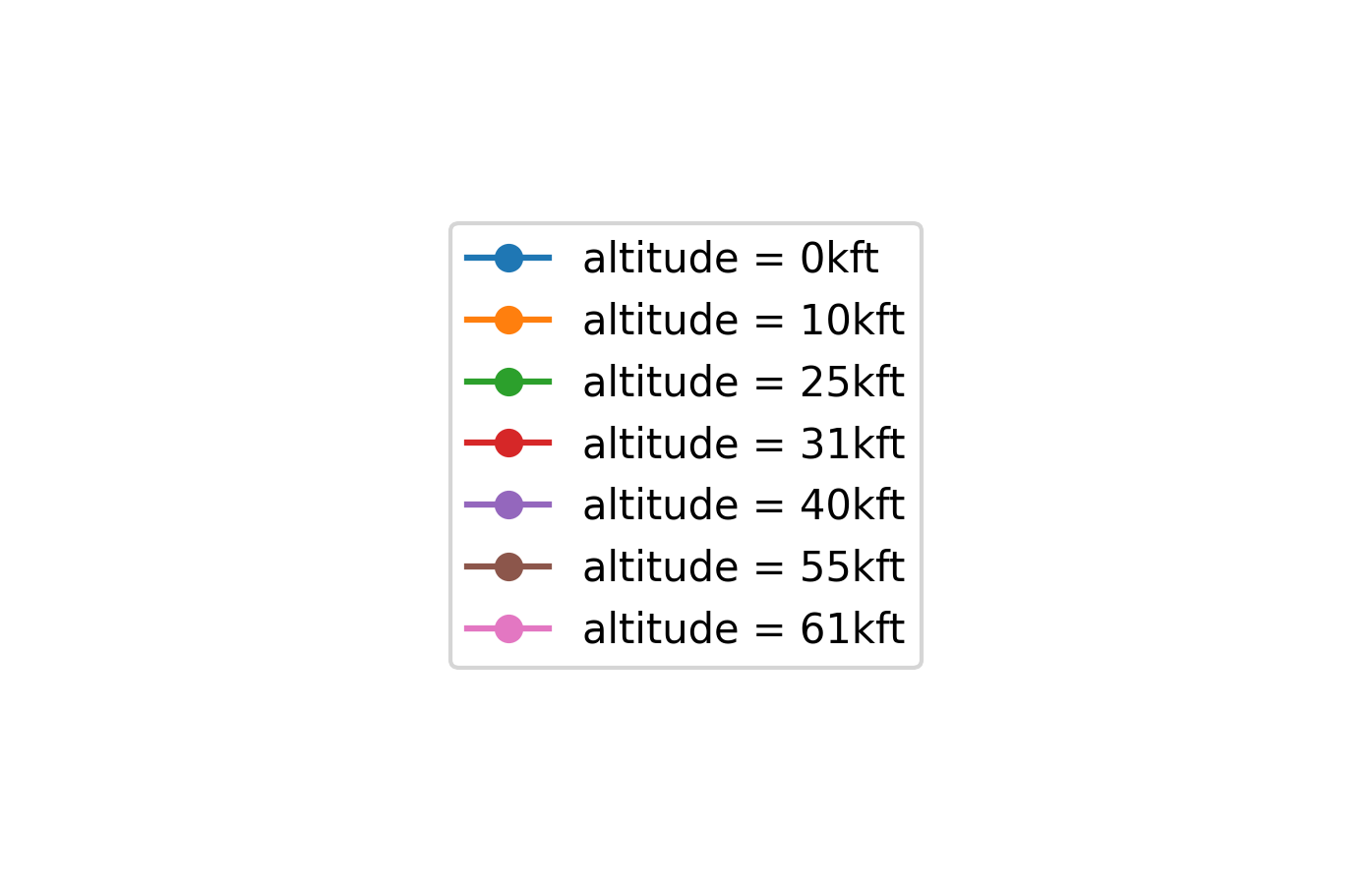}}

   \caption{The sensitivity of ambient dose equivalent rate to 
   latitude, $\frac{\partial H}{\partial \theta}$, plotted as a 
   function of latitude and altitude for longitude=0.0\textdegree E. 
   Spectrum 1, which was had a lower spectral index, shows a much 
   wider transition region than spectrum 2, which had a higher 
   spectral index.}
   \label{fig:latitudeSensitivity}
\end{figure}

$\frac{\partial H^*(10)}{\partial \theta}$ could be determined 
using the data for the atmospheric cut-throughs in this paper 
through taking the difference between dose rates as latitude 
changes, i.e. $\frac{\partial H^*(10)}{\partial \theta} \approx \frac{\Delta H^*(10)}{\Delta \theta}$. 
The difference in latitudes corresponding to these changes was 
therefore always 5\textdegree, the resolution of the outputted 
data here. $\frac{\Delta H^*(10)}{\Delta \theta}$ is plotted as 
a function of latitude and altitude in 
figure~\ref{fig:latitudeSensitivityMaps} and 
figure~\ref{fig:latitudeSensitivity}.

figure~\ref{fig:latitudeSensitivityMaps} shows the feature that 
has been discussed previously; that there is a reasonably thin 
transition region which is positioned at a specific latitude 
that varies with longitude. 
figure~\ref{fig:latitudeSensitivity} shows the same data but 
displayed as a line plot rather than as a cut-through. It is 
straightforward to see the location of the transition region as 
a peak in $\frac{\Delta H^*(10)}{\Delta \theta}$ on 
figure~\ref{fig:latitudeSensitivity}. 
It is interesting to note that the transition region is 
significantly wider and less sharp in `hard spectrum' case than 
in the `soft spectrum' case. This could reflect how the spectrum 
is broader in the case of the harder spectrum, and is therefore 
susceptible to a broader range of cut-off rigidities than in the 
softer spectrum case.

In both spectral cases, the latitude sensitivity of dose rates 
increases significantly with altitude, as might be expected 
given that overall dose rates also increase significantly with 
increasing altitude.

\begin{figure}
   \centering
   \subfigure{\includegraphics[trim={0 0 0 0},clip,width=\columnwidth]{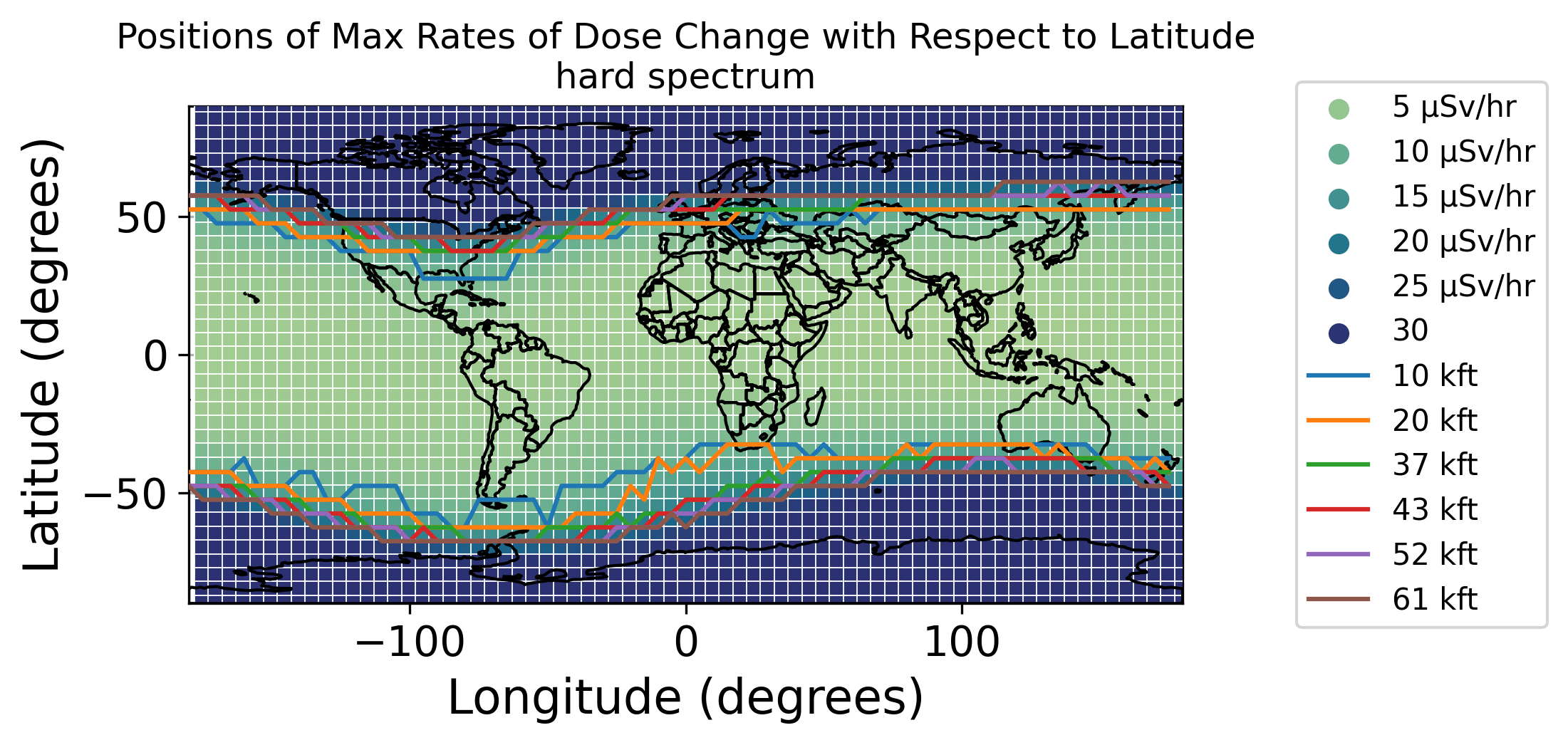}}
   \subfigure{\includegraphics[trim={0 0 0 0},clip,width=\columnwidth]{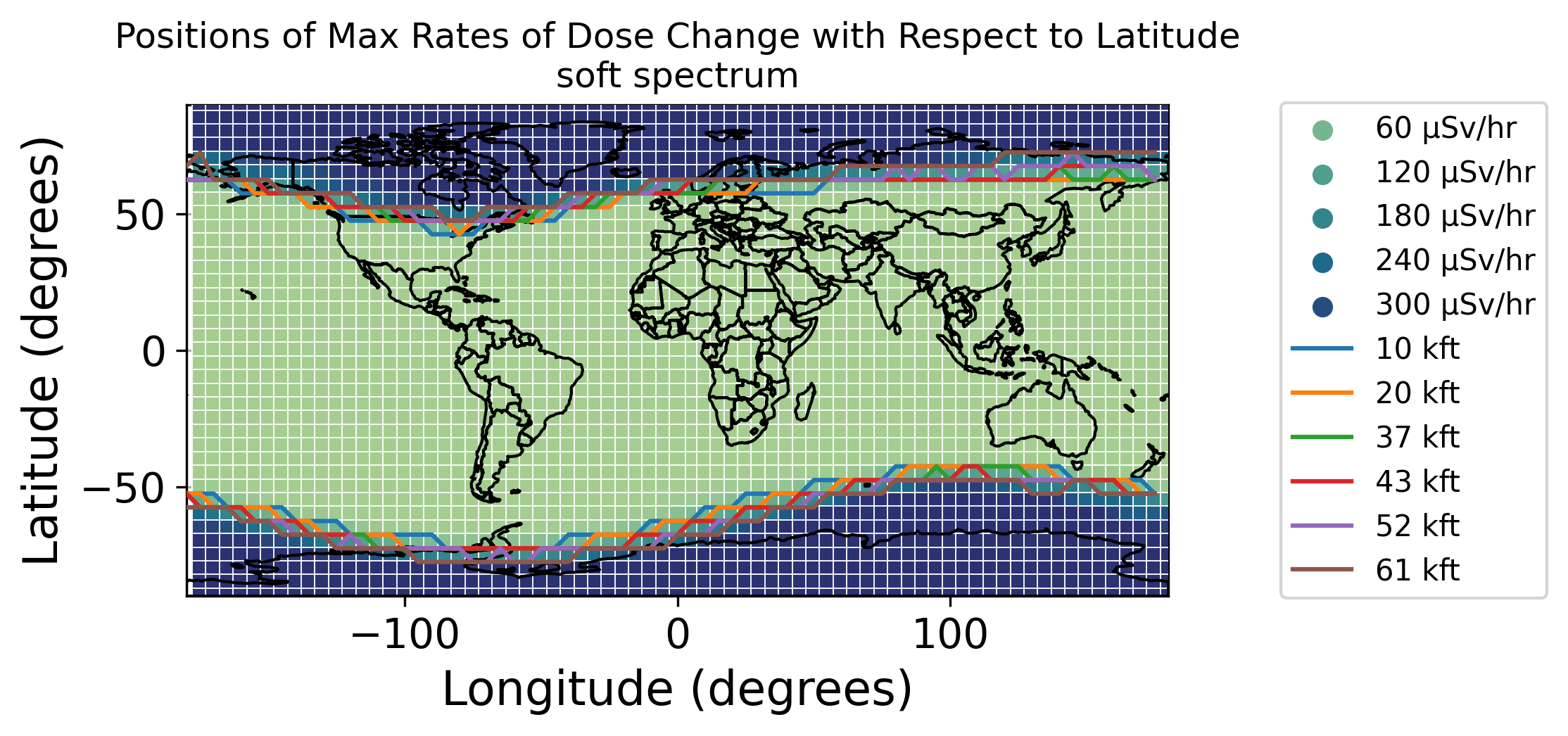}}
   \caption{Positions of maximum sensitivity of ambient dose 
   equivalent to latitude, $\frac{\Delta H^*(10)}{\Delta \theta}|_{max}$, 
   plotted on a dose rate map of Earth's atmosphere at 61,000 ft. The 
   positions of maximum sensitivity correspond well with the clear 
   transition regions in dose rate between polar and equatorial 
   dose rate areas that can be observed on the dose rate map. The 
   transition region moves slightly towards the poles as altitude 
   is increased.}
   \label{fig:maxDoseRatePosAndMap}
\end{figure}

Its possible to use the data in 
figure~\ref{fig:latitudeSensitivity} to characterise the 
location of the transition region by taking the location of the 
transition region as the latitude at which the latitude 
sensitivity $\frac{\Delta H^*(10)}{\Delta \theta}$ is at a 
maximum. Determining the location of the maximum latitude 
sensitivity for each atmospheric cut-through gives a curve when 
plotted on a map as in figure~\ref{fig:maxDoseRatePosAndMap}. 
Ambient dose equivalent rates at 37 kft are also plotted on 
figure~\ref{fig:maxDoseRatePosAndMap} for comparison.

figure~\ref{fig:maxDoseRatePosAndMap} shows how the location of 
maximum latitude sensitivity divides the dose rate map into two 
regions quite well. Its also interesting to note that the curves 
of maximal latitude sensitivity move further from the equator as 
altitude is increased. This is likely to be due to dose rates at 
lower altitudes being less affected by lower energy particles 
than at higher altitudes, and therefore being slightly less 
susceptible to lower vertical cut-off rigidities.

\section{Implications for Aircraft, Air Crew, and Airlines}

Radiation in general has two consequences for the aviation 
industry. The first and most well-known about impact is the 
effect of radiation on crew and passengers. Pilots and cabin 
crew consistently flying through routes that experience high 
dose rates, such as any flights traveling across the North 
Atlantic (most transatlantic flights from Europe to North 
America) are at higher risk of developing cancer than the general 
public. This is true during quiet solar conditions, however 
during a major solar particle event this effect is even more 
dangerous to crew, as radiation doses can increase by factors of 
1000 or more. In instances like that, air crew might breach the 
recommended general public exposure limit within a single flight 
\citep{dyer2007solar,cannon2013extreme,tobiska2015advances,dyer2017extreme}, 
and could even go well above this during a solar particle event 
on the largest scales that research using tree rings and ice 
cores has indicated have occurred during the last few millennia 
\citep{miyake2012signature,miyake2013another,dyer2017extreme, mekhaldi2015multiradionuclide, 
kovaltsov2014fluence}.

The second impact is often less discussed but has a potentially 
higher impact than crew exposure, and deadlier. Electronics 
exposed to radiation can experience device errors 
\citep{olsen1993neutron,hands2009seu,cannon2013extreme,dyer2017extreme}, including issues known 
as Single Event Upsets (SEUs) and Single Event Latch-ups (SELs). 
SEUs are single errors introduced in electronics, often memory 
devices, as a particle passes through it, flipping individual 
bits. 

If a Single Event Upset (SEU) happens in a particularly important 
part of a device, or if so many upsets are happening at once that 
the aircraft electronics are overwhelmed, aircraft electronics 
could respond in unknown and dangerous ways. Alternatively, 
electronic devices may shut down, potentially leaving pilots to 
deal with multiple electronics failures simultaneously during 
extreme radiation conditions.

Single Event Latch-ups (SELs) on the other hand, represent more 
permanent damage to devices, and cannot be fixed by on-board 
systems. SELs could therefore not only cause electronics to crash 
or give false readings, but could actually cause devices to 
completely fail.

Each of these issues represent a potential immediate threat to 
aircraft and passengers aboard as soon as the aircraft enters a 
region with high enough radiation levels that electronics 
failures could occur. Dyer et al. \citep{dyer2020single} found 
that modern electronics devices at ground-level may have a 
significant probability of failure during an event ten times 
greater than the GLE that occurred in February 1956 (21\% of 
silicon power metal–oxide–semiconductor field-effect transistors 
and 14\% of insulated gate bipolar transistors if the devices are 
unrated). As has been discussed in this paper, the single-event 
effect rate is orders of magnitude greater at aviation altitudes.

MAIRE+ is capable of estimating static RAM (SRAM) SEU and SEL rates at a particular location by multiplying its 
internally calculated $>$10 MeV neutron flux by cross-sections of $10^{-13}$ upsets/cm²/bit and $10^{-8}$ 
latch-ups/cm²/device respectively. These cross-sections are representative values that are within the range 
of cross-sections that devices have been shown to exhibit \citep{dyer2020single}. It should be noted that individual 
devices have highly variable cross-sections, although devices with these cross-sections have been used in avionics. 
Figure~\ref{fig:SEU} and figure~\ref{fig:SEL} below shows SEU and SEL rates plotted in the same format as the `cut-through' plots above 
and for the incoming particle spectra that have been investigated in this paper.

\begin{figure}
   \centering
   \subfigure{\includegraphics[trim={0 0 0 0},clip,width=\columnwidth]{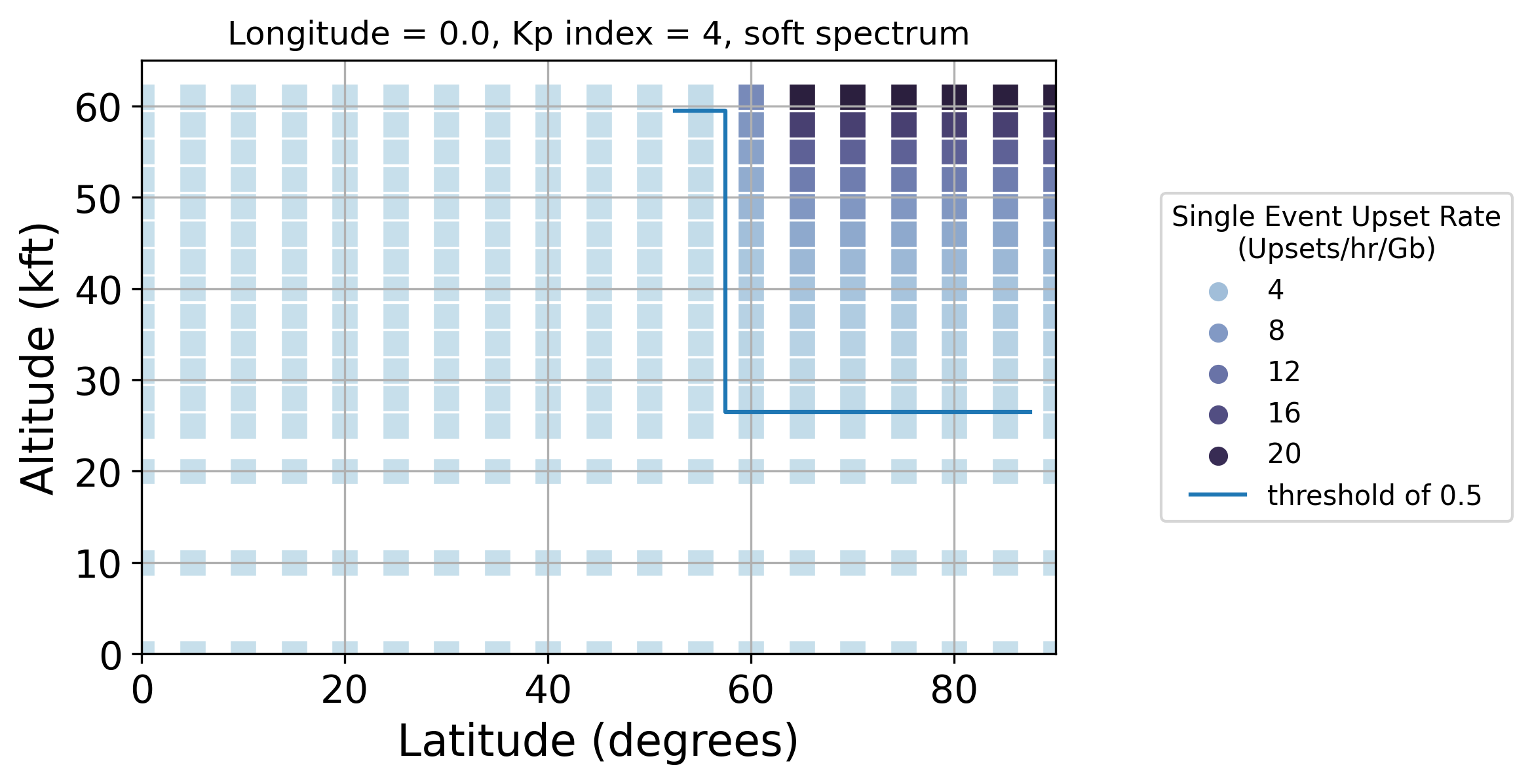}}
   \subfigure{\includegraphics[trim={0 0 0 0},clip,width=\columnwidth]{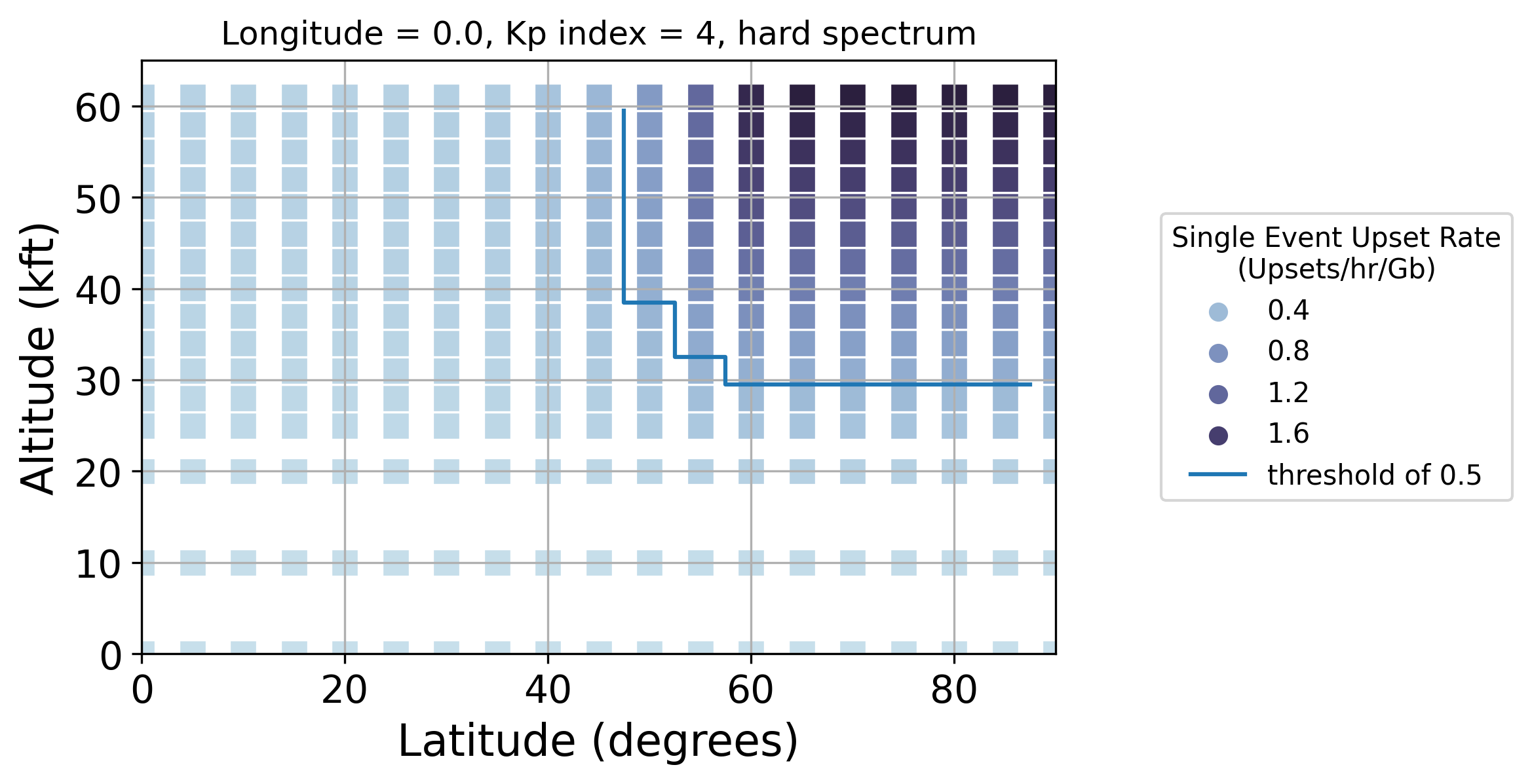}}
   \caption{The MAIRE+ predicted Single Event Upset (SEU) rates as 
   a function of latitude and altitude at longitude = 0.0°N during 
   the spectral conditions. The threshold of 0.5 upsets/hr/Gb is 
   primarily a guide to the eye in this plot rather than a hard 
   boundary between some safe level and some dangerous level.}
   \label{fig:SEU}
\end{figure}

\begin{figure}
   \centering
   \subfigure{\includegraphics[trim={0 0 0 0},clip,width=\columnwidth]{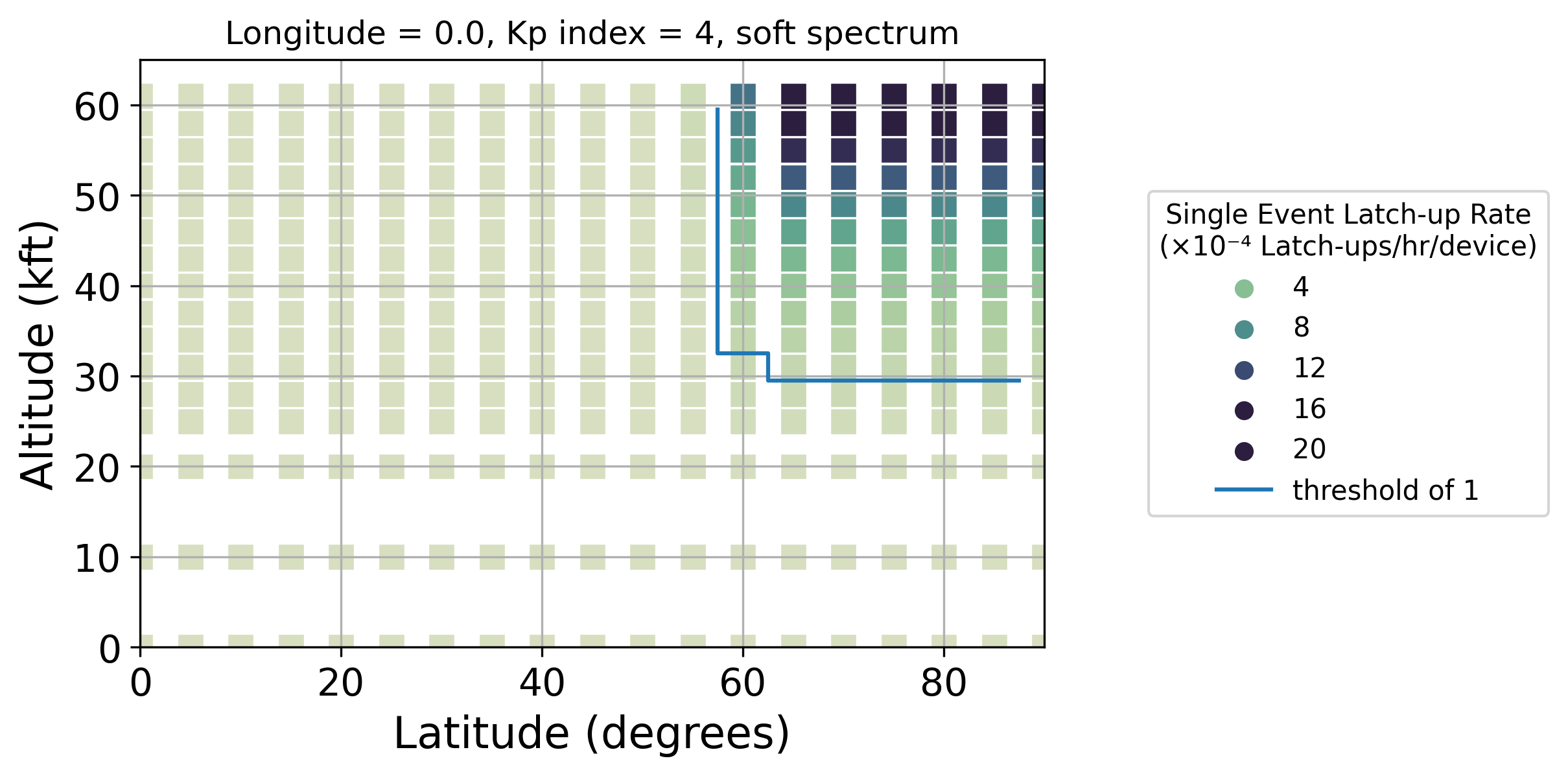}}
   \subfigure{\includegraphics[trim={0 0 0 0},clip,width=\columnwidth]{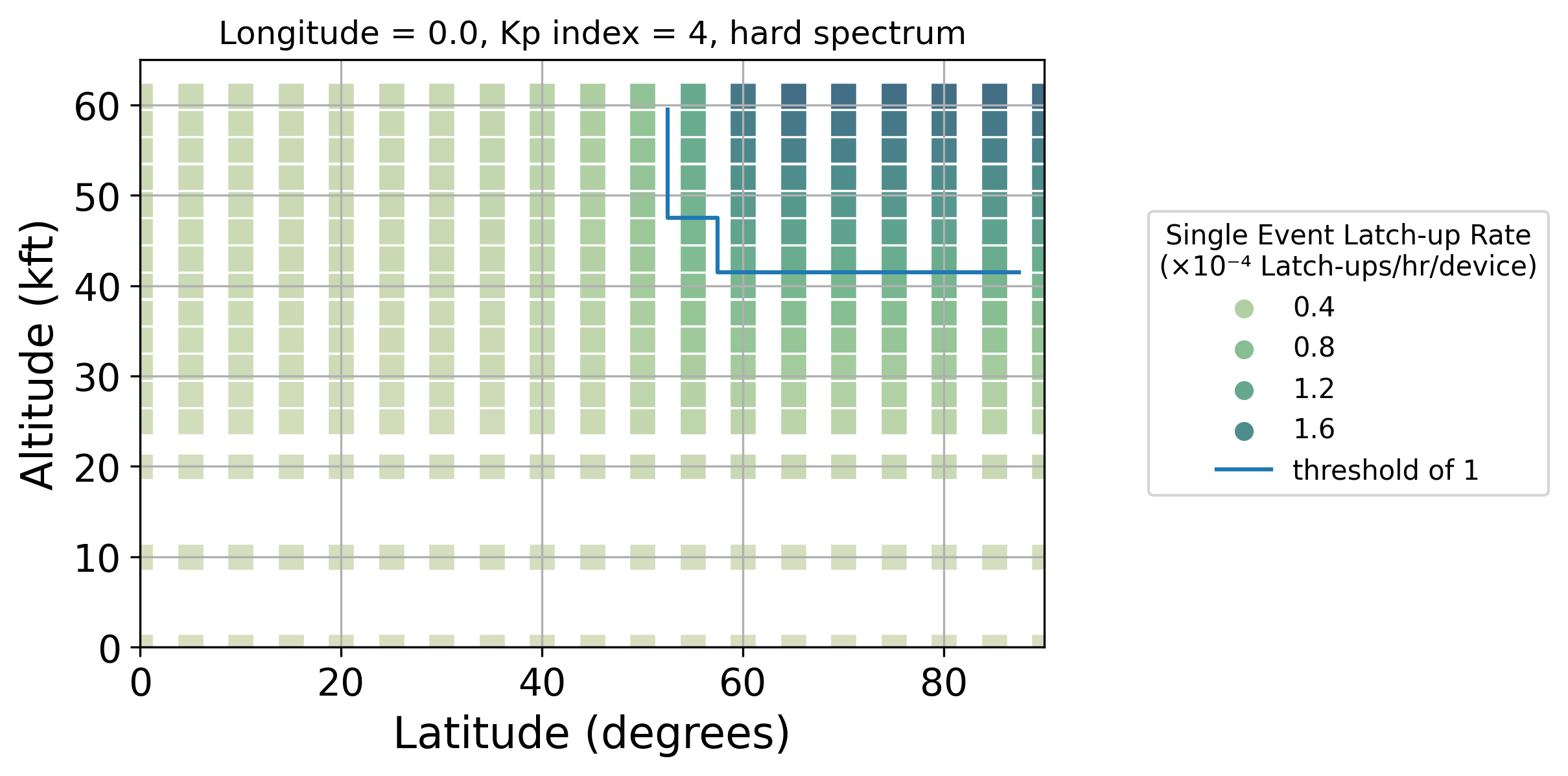}}
   \caption{The MAIRE+ predicted Single Event Latch-up (SEL) rates 
   as a function of latitude and altitude at longitude = 0.0°N 
   during the spectral conditions. The threshold of 1$\times10^{-4}$ 
   latch-ups/hr/device is primarily a guide to the eye in this plot 
   rather than a hard boundary between some safe level and some 
   dangerous level.}
   \label{fig:SEL}
\end{figure}

Figure~\ref{fig:SEU} and figure~\ref{fig:SEL} show a very similar relationship 
between SEU/SEL rates and latitude as were shown in 
figure~\ref{fig:xsectionKpAll} and figure~\ref{fig:mergedLongitudeVariationPlots} 
previously. The actual rate of SEUs that aircraft electronics 
can cope with at 40,000 feet during an event similar to 
something like the GLE that occurred in February 1956 is 
unknown, as most devices are not tested for these conditions. 
It is possible that aircraft electronics are indeed safe under 
the conditions of a 1 in 100 year intensity radiation storm, but 
it is not yet possible to rule out a worst-case scenario where 
aircraft to the north of approximately 65\textdegree North in 
latitude (and approximately 50\textdegree North above the USA) 
all suffer multiple electronics failures simultaneously.

The message that airlines should take from this and the earlier 
plots displaying radiation dose rates is that during a major 
solar storm they should endeavour to reduce aircraft altitude, 
and also possibly reduce their latitude, depending on what route 
the flight is taking. This is already somewhat known to 
airlines, and has been suggested by other previous papers 
\citep{matthia2015economic}. ICAO advice does suggest that 
planes in high latitude regions could attempt to reduce both 
altitude and geomagnetic latitude during radiation storms 
\citep{ICAOinternational2018manual}. However, this advice is 
usually discussed from the perspective of generally reducing the 
longer term accumulated doses received by the aircraft crew, 
rather than addressing the potential immediate threat that 
SEU/SEL errors pose to the aircraft. The existence of threats 
due to high SEU/SEL rates means that speed and urgency may be 
paramount once an alert has been issued for a large GLE. This 
necessary rapidity would also be currently nearly impossible if 
this large solar event occurs at the same time as a HF radio 
blackout, which would render communication with many aircraft 
nearly impossible. This is not to mention the difficulties that 
would occur with such an event happening at the same time as 
other solar storm related effects on society.

What the research presented in this paper adds to this 
discussion is some clarity over which strategies could be chosen 
by an aircraft during a large GLE. For the incoming radiation 
distribution discussed in this paper, radiation dose rates were 
found not to vary significantly with latitude except at the 
boundary between the low dose rate, equatorial, region and the 
high dose rate, polar, region. Under these conditions most 
aircraft situated well within the polar region would not benefit 
much from changing latitude, and reducing their altitude might 
be preferable. It should be noted that flying at lower altitudes 
can take a significant amount of fuel, so it may not always be 
possible for an aircraft to fly at low altitudes underneath a 
region of high dose rates for long, but an aircraft may be able 
to at least do so for long enough to miss the peak of an event.

In contrast, an aircraft traveling approximately along the 
boundary/transition between the two dose rate regions (which 
happens to align approximately with many commonly used 
transatlantic routes) might be able to significantly to reduce 
their radiation dose and SEU/SEL rate by traveling just a few 
degrees further south - providing the location of the transition 
region could be accurately determined in real time, which would 
require an accurate real-time determination of magnetospheric 
disturbance levels and incoming particle spectral index. Flight 
routes are constantly updated according to atmospheric factors 
such as zones of high turbulence, and zones of high radiation 
dose could simply be considered another factor for such flights 
in what route they should take.

\begin{figure}
   \centering
   \subfigure{\includegraphics[trim={0 0 0 0},clip,width=0.4\columnwidth]{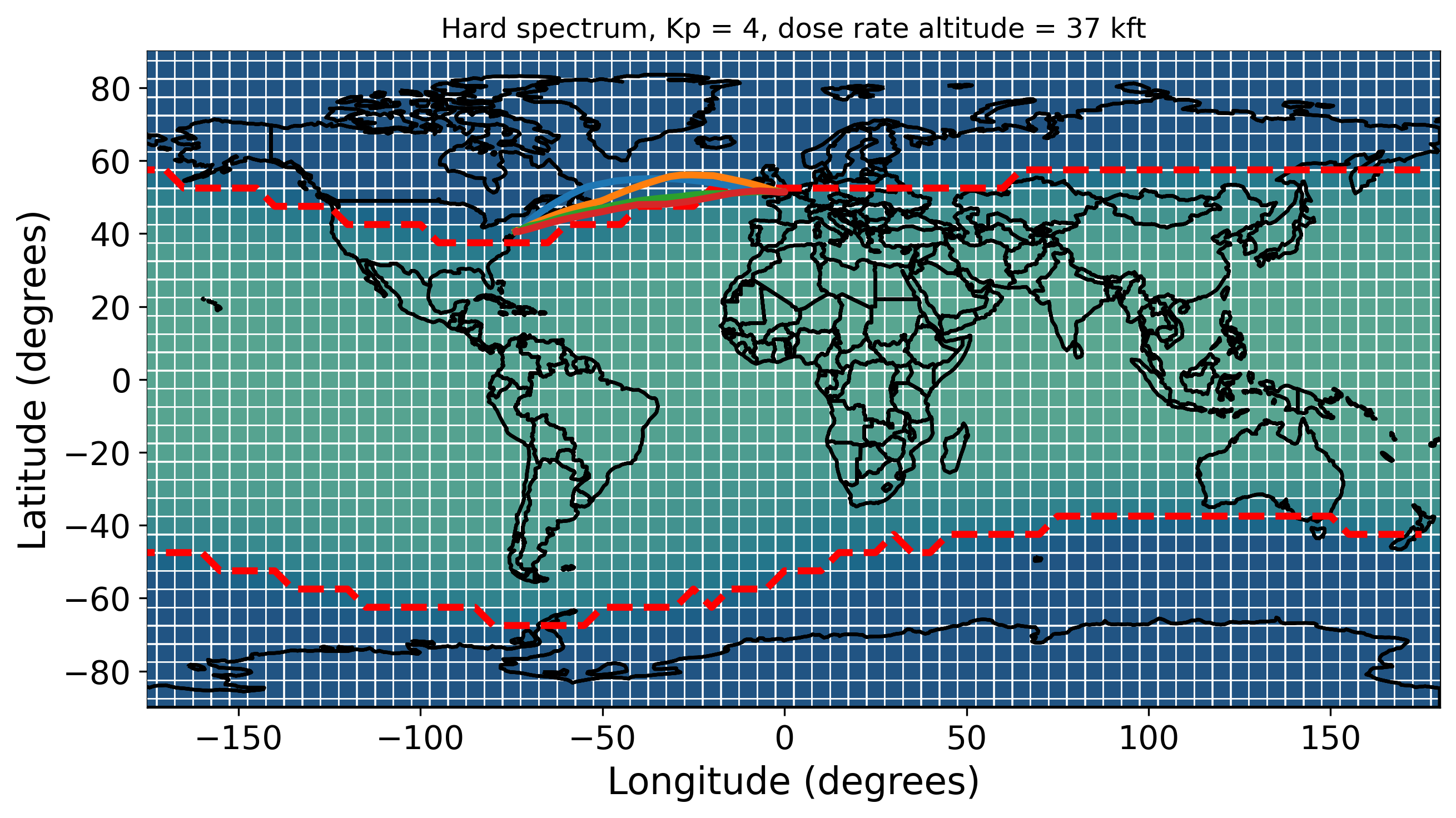}}
   \subfigure{\includegraphics[trim={0 0 0 0},clip,width=0.4\columnwidth]{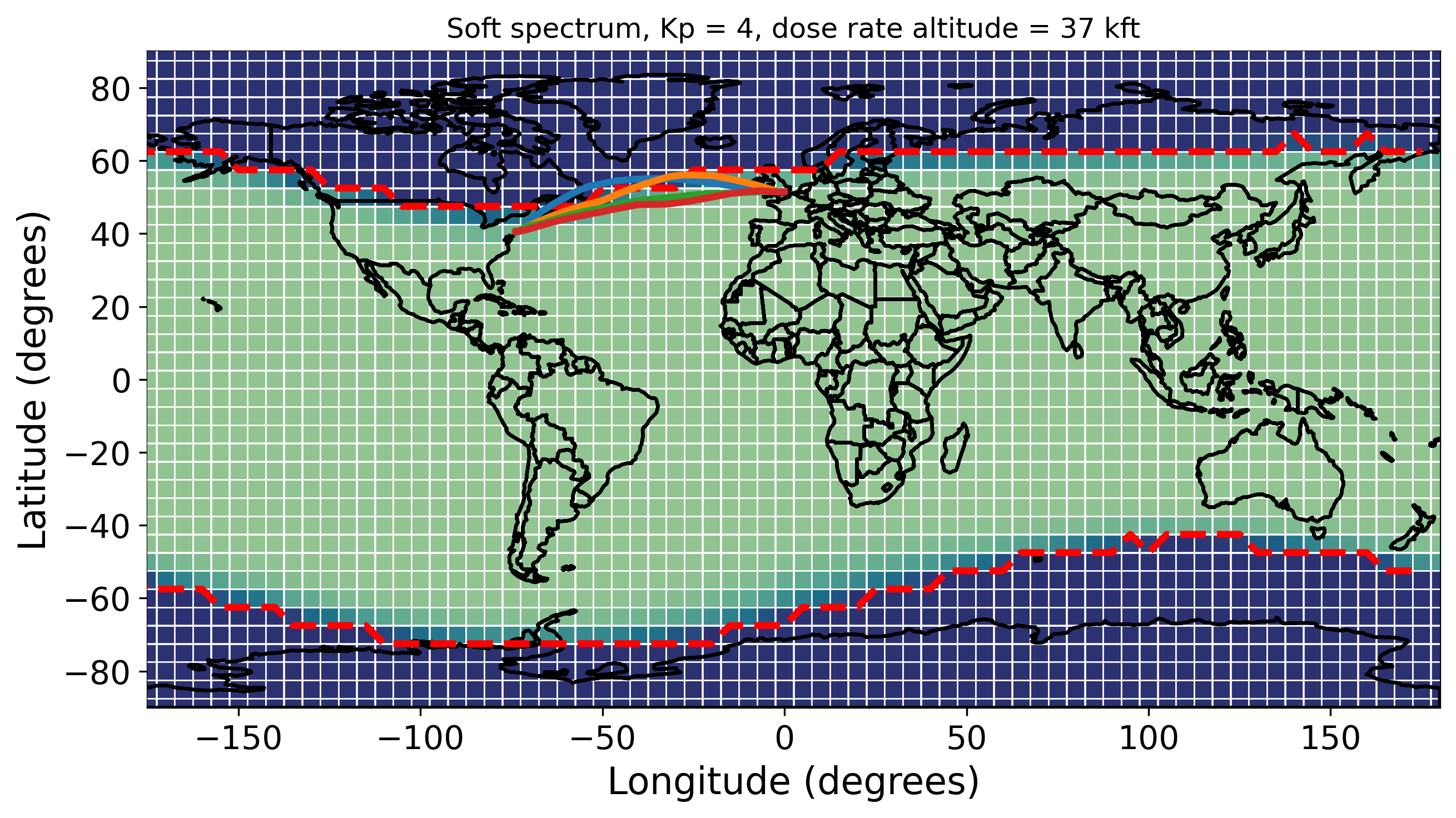}}
   \subfigure{\includegraphics[trim={0 0 0 0},clip,width=0.4\columnwidth]{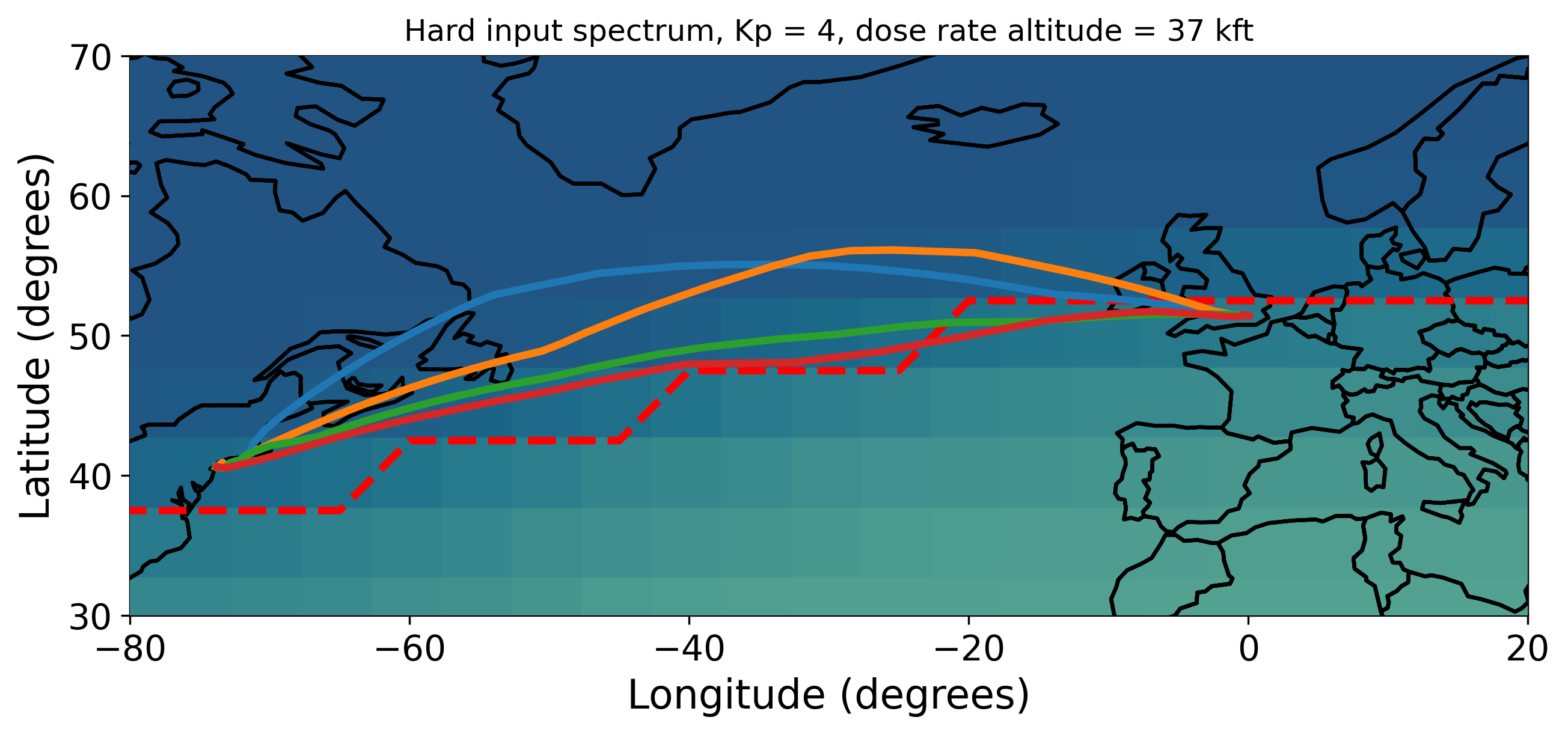}}
   \subfigure{\includegraphics[trim={0 0 0 0},clip,width=0.4\columnwidth]{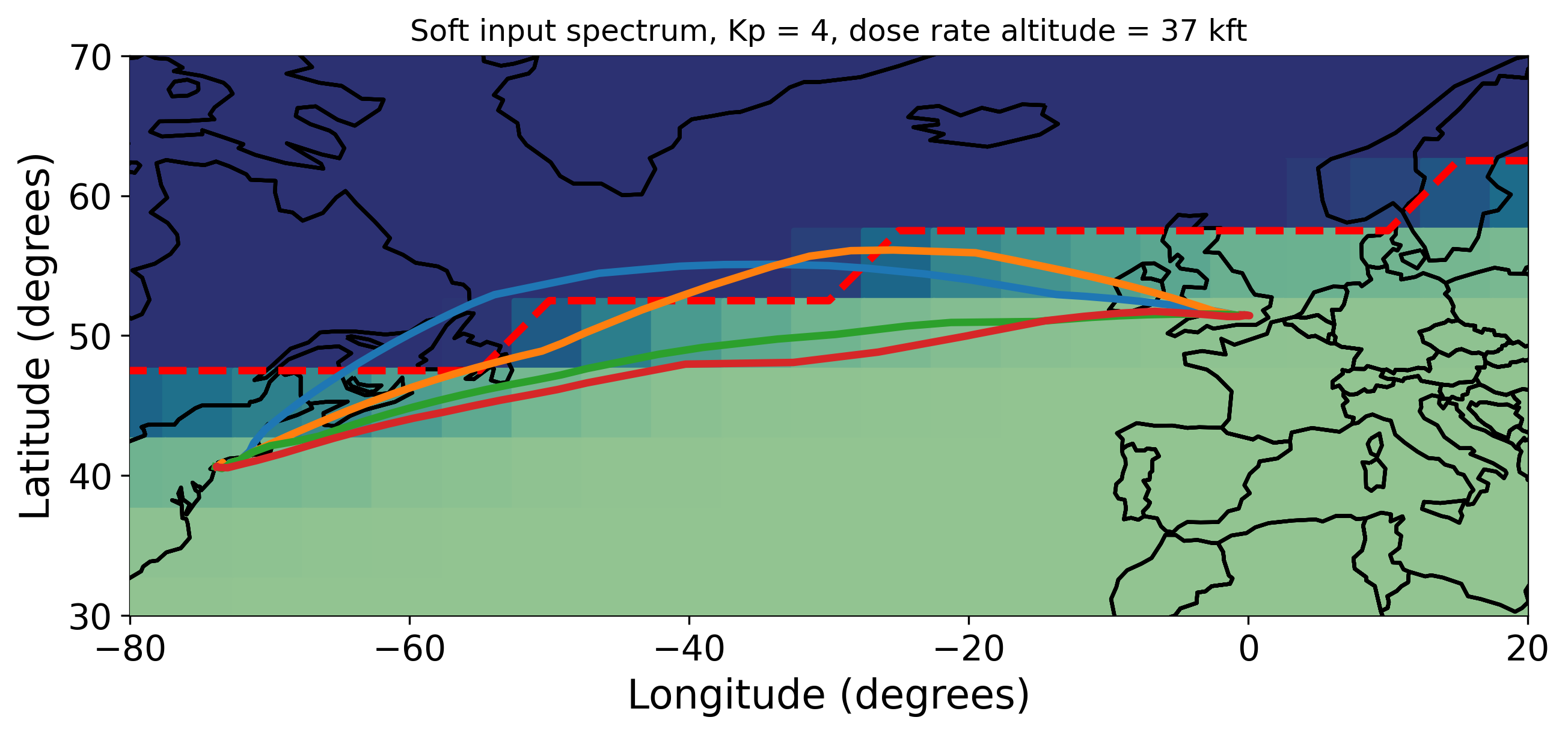}}
   
   \subfigure{\includegraphics[trim={22cm 3cm 0 3cm},clip,width=0.4\columnwidth]{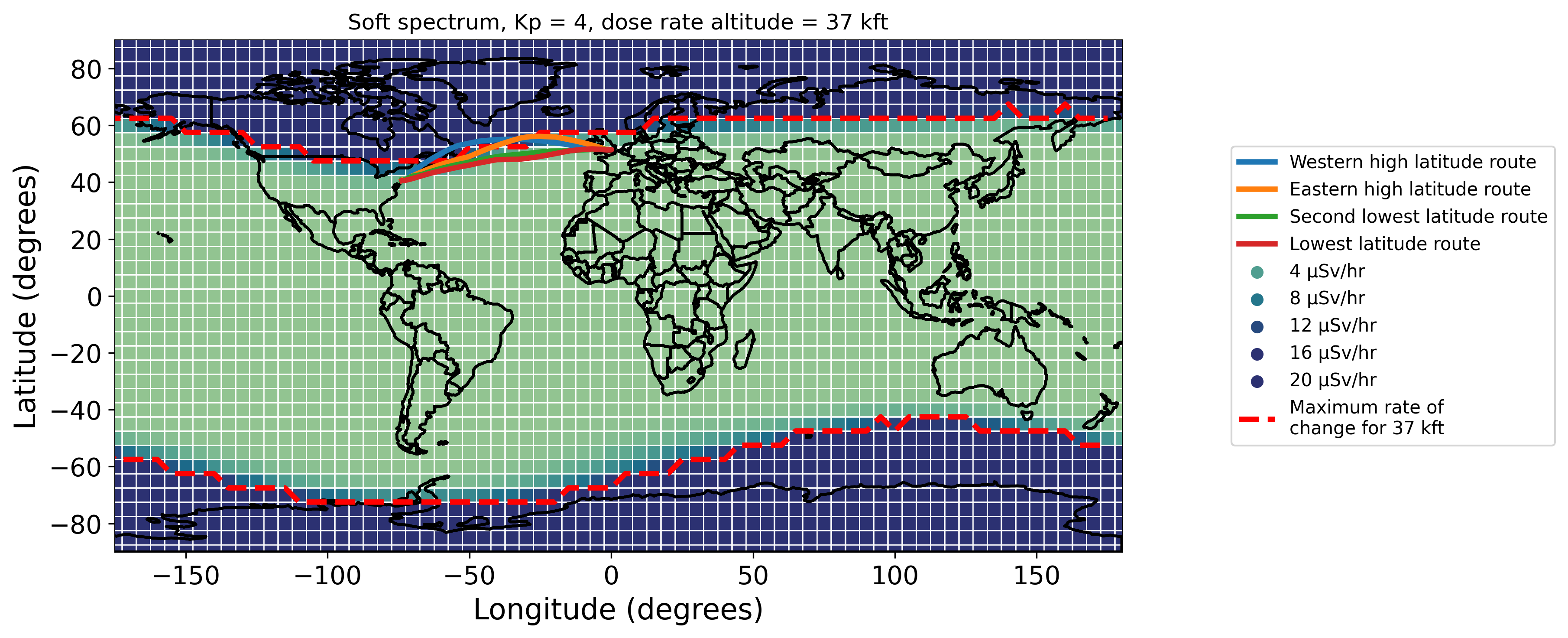}}

   \caption{Dose rate maps of Earth at 34,000 ft and the paths of the flight routes that were investigated in this paper. The maximum rate of change of dose rate with respect to latitude is also plotted to indicate the location of the transition region between the low and high dose rate regions. In the high spectral index case, the transition region between the low dose rate `equatorial' region and the high dose rate region lies between the flight routes.}
   \label{fig:WorldMapsAndFlightRoutes}
\end{figure}

\begin{figure}
   \centering
   \subfigure{\includegraphics[trim={0 0 0 0},clip,width=0.4\columnwidth]{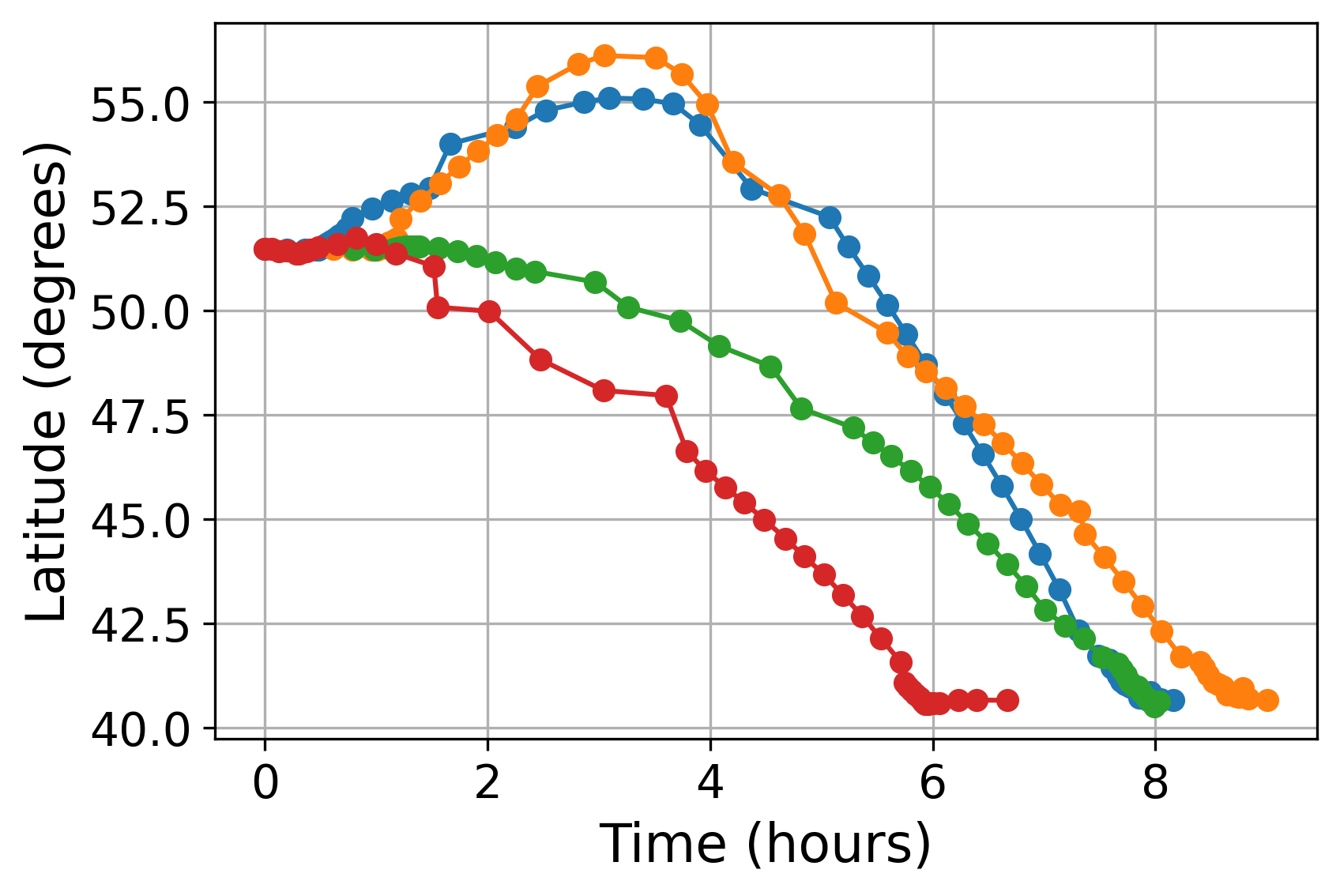}}
   \subfigure{\includegraphics[trim={0 0 0 0},clip,width=0.4\columnwidth]{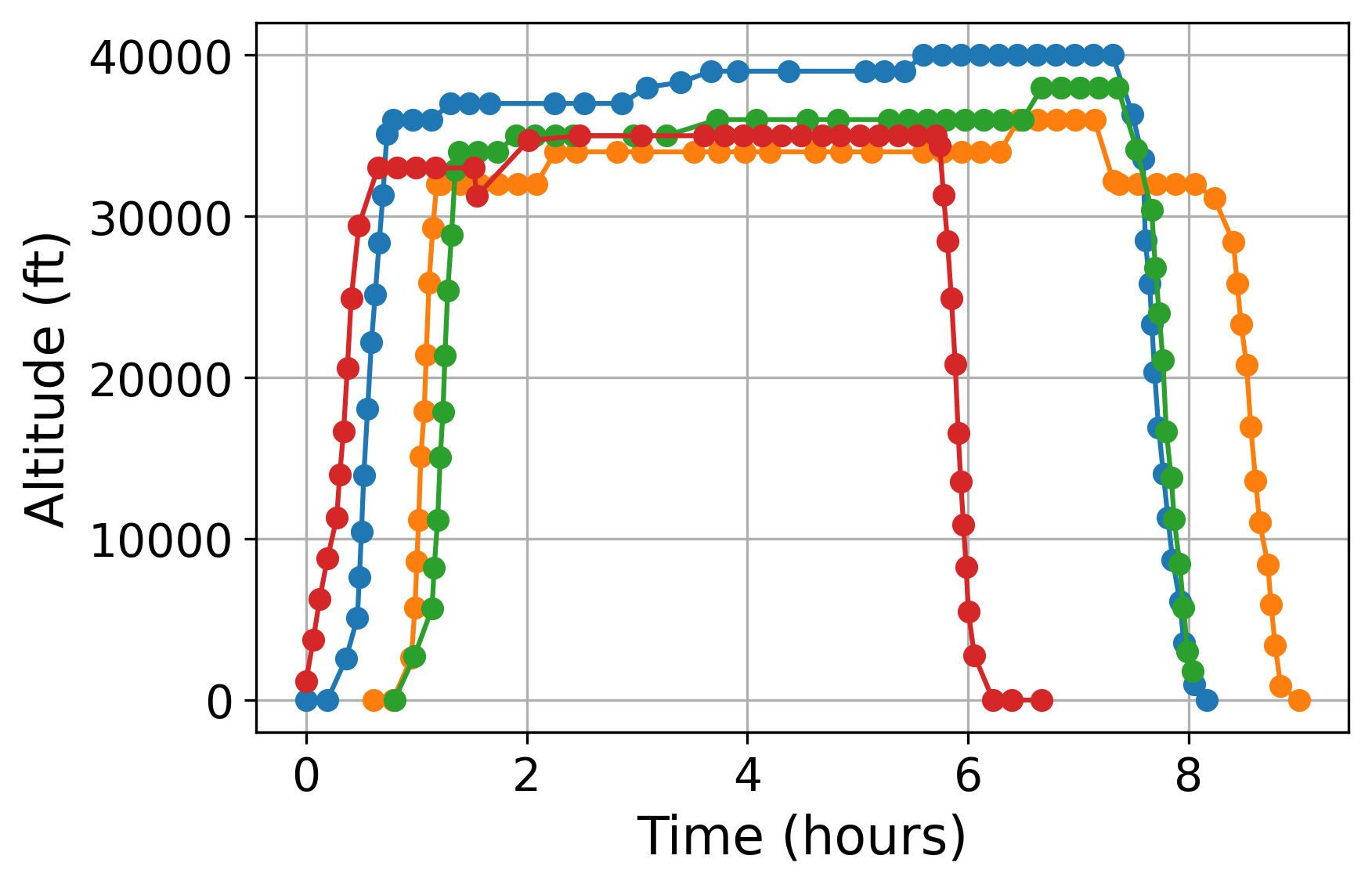}}
   
   \subfigure{\includegraphics[trim={0 2cm 0 2cm},clip,width=0.4\columnwidth]{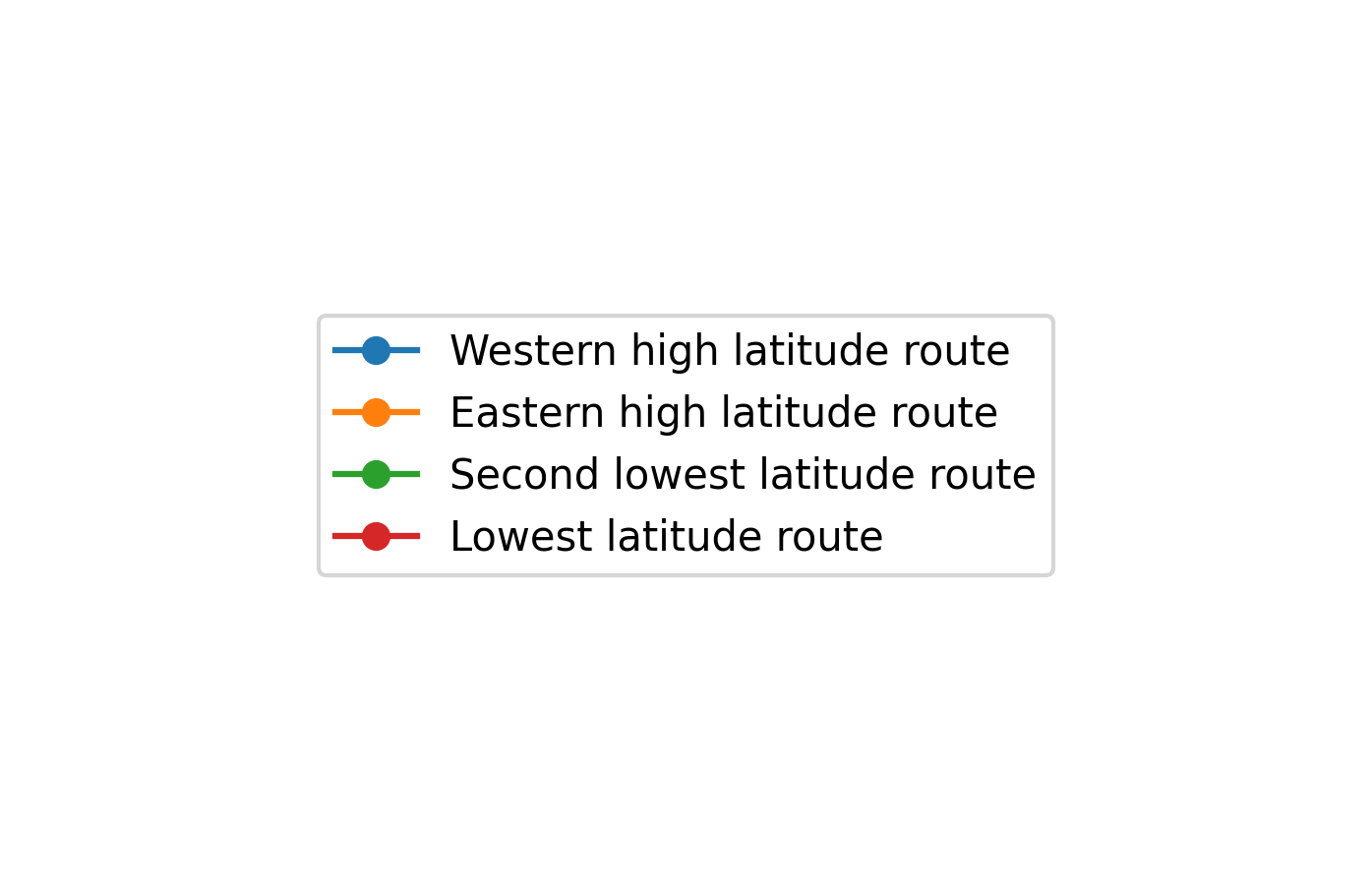}}
   
   \caption{The latitude and altitude time profiles for the flights shown in figure~\ref{fig:WorldMapsAndFlightRoutes}. The lowest latitude flight was a flight from New York JFK to London Heathrow LHR, in contrast to the other flights which went from LHR to JFK, and the flight route profile has therefore been reversed for the lowest latitude flight for better comparison.}
   \label{fig:flightRouteProfiles}
\end{figure}

This is illustrated by 
figure~\ref{fig:WorldMapsAndFlightRoutes}, which shows a map of 
four flight routes heading between Heathrow airport (London, UK) 
and John F. Kennedy airport (New York City, USA), as well as the 
corresponding dose rates at 37,000 ft from the simulations 
described in this paper. The latitude and altitude profiles can 
also be seen in figure~\ref{fig:flightRouteProfiles}. These are 
real flight routes collected from FlightRadar24 
\citep{FlightRadar24}, which flew between 2020 and 2022. It 
should however be noted that the `time' on the x-axis does not 
necessarily correspond with the actual take-off time of the 
aircraft, rather the time that Flightradar24 began recording the 
aircraft's location. In addition, the Eastern high latitude 
flight and the second lowest latitude flight's time coordinates 
have been translated backwards to be 2 hours less than the value 
given by Flightradar24 for easier comparison with the other 
flights on time plots, as the flights took several hours to 
take-off from the beginning of Flightradar24's recording. The 
lowest latitude flight was actually a flight from New York JFK 
to London Heathrow, in contrast to the other flights, which 
travelled in the opposite direction. The flight route was 
therefore reversed in this analysis so that coordinates roughly 
correspond with those of the other flight routes (i.e. in the 
lowest latitude flight case, time represents approximately the 
time until landing, whereas in all the other flights, time 
represents approximately the time from the flight beginning).

The routes encompass a range of different latitudes, and it can 
be seen in figure~\ref{fig:WorldMapsAndFlightRoutes} that in 
the low spectral index/soft spectrum case at Kp = 4, some of the 
flights cross through the transition region into and out of the 
high dose regions, while others stay in the low dose rate 
region for the entirety of the flight. During the hard spectrum 
conditions, all of the flights investigated happen to be within 
the high dose rate region for Kp = 4.

\begin{figure}
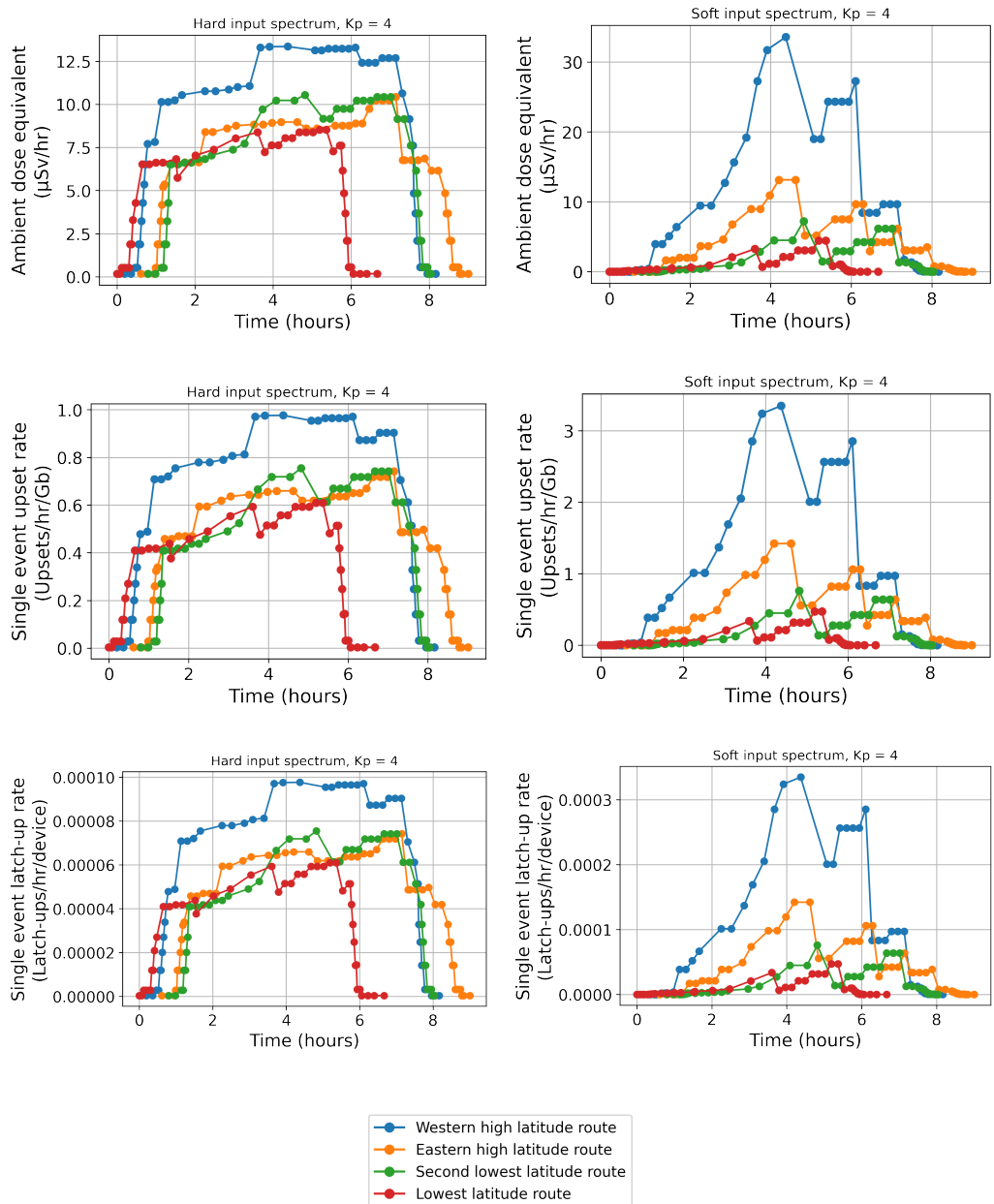

   \centering
   \foreach \doseType in {adose,SEU,SEL}{
         \foreach \spectrum in {Hard,Soft}{
            \subfigure{\includegraphics[trim={0 0 0 0},clip,width=0.4\columnwidth]{Spec\spectrum inputspectrum_KpIndex4_\doseType_no_legend.png}}
            }
         }
         \subfigure{\includegraphics[trim={0 2cm 0 2cm},clip,width=0.4\columnwidth]{flightRouteAltitudeProfiles_legend.png}}
   \caption{Dose rate characteristics of the flight routes 
   investigated. In both spectral index cases, the higher latitude 
   flights experience generally higher dose rates, however in the 
   high spectral index case the high latitude flights experience 
   significantly higher doses than the low latitude flight.}
   \label{fig:AllDosesVsTimeKp4}
\end{figure}

\begin{figure}
   \centering
   \subfigure{\includegraphics[trim={0 0 0 0},clip,width=0.4\columnwidth]{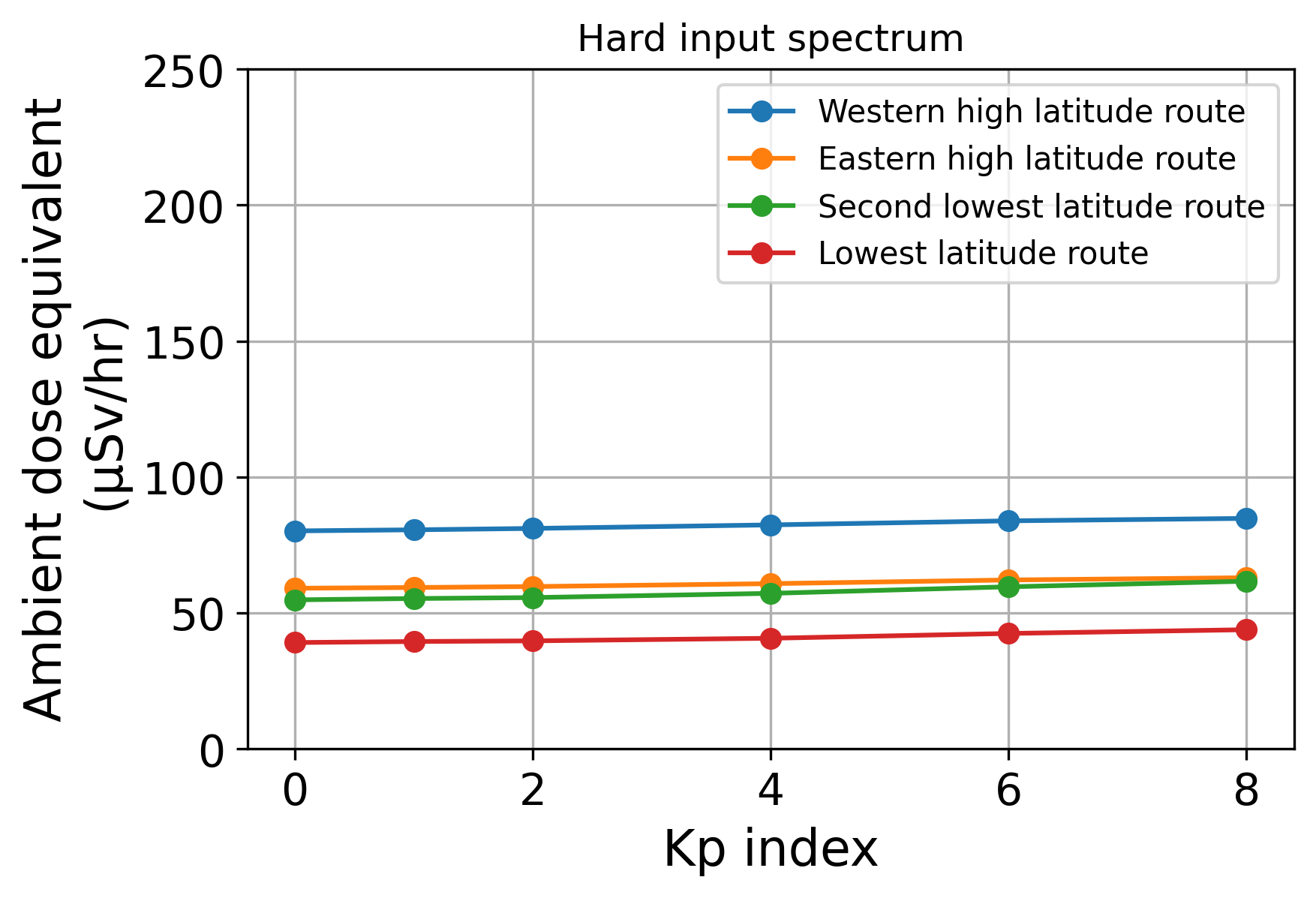}}
   \subfigure{\includegraphics[trim={0 0 0 0},clip,width=0.4\columnwidth]{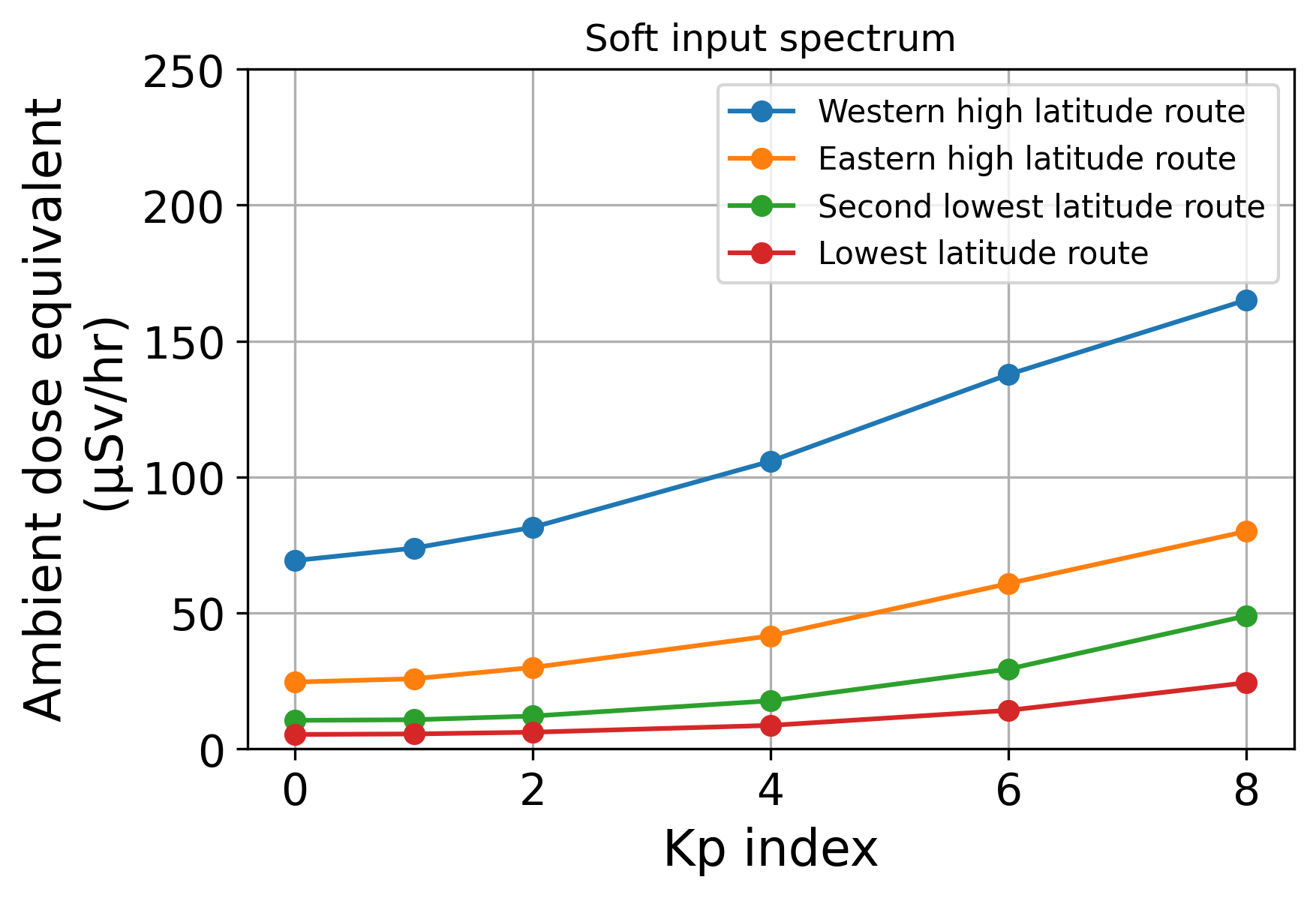}}
   \caption{The total accumulated/integrated dose for each of the 
   flights as a function of Kp index. There are only slight 
   variations for each of the selected flights in dose with Kp 
   index in the hard spectrum case, however there are large 
   variations in dose in the soft spectrum case for all of the 
   flights.}
   \label{fig:AccDoseVsKp}
\end{figure}

\begin{figure}
   \centering
   \foreach \flight in {Westernhighlatitude,Easternhighlatitude,Secondlowestlatitude,Lowestlatitude}{
         \foreach \spectrum in {Hard,Soft}{
            \subfigure{\includegraphics[trim={0 0 0 0},clip,width=0.4\columnwidth]{FlightVariationWithKp\spectrum Spec\flight route_no_legend.png}}
            }
         }
   
   \subfigure{\includegraphics[trim={0 0 0 0},clip,width=0.4\columnwidth]{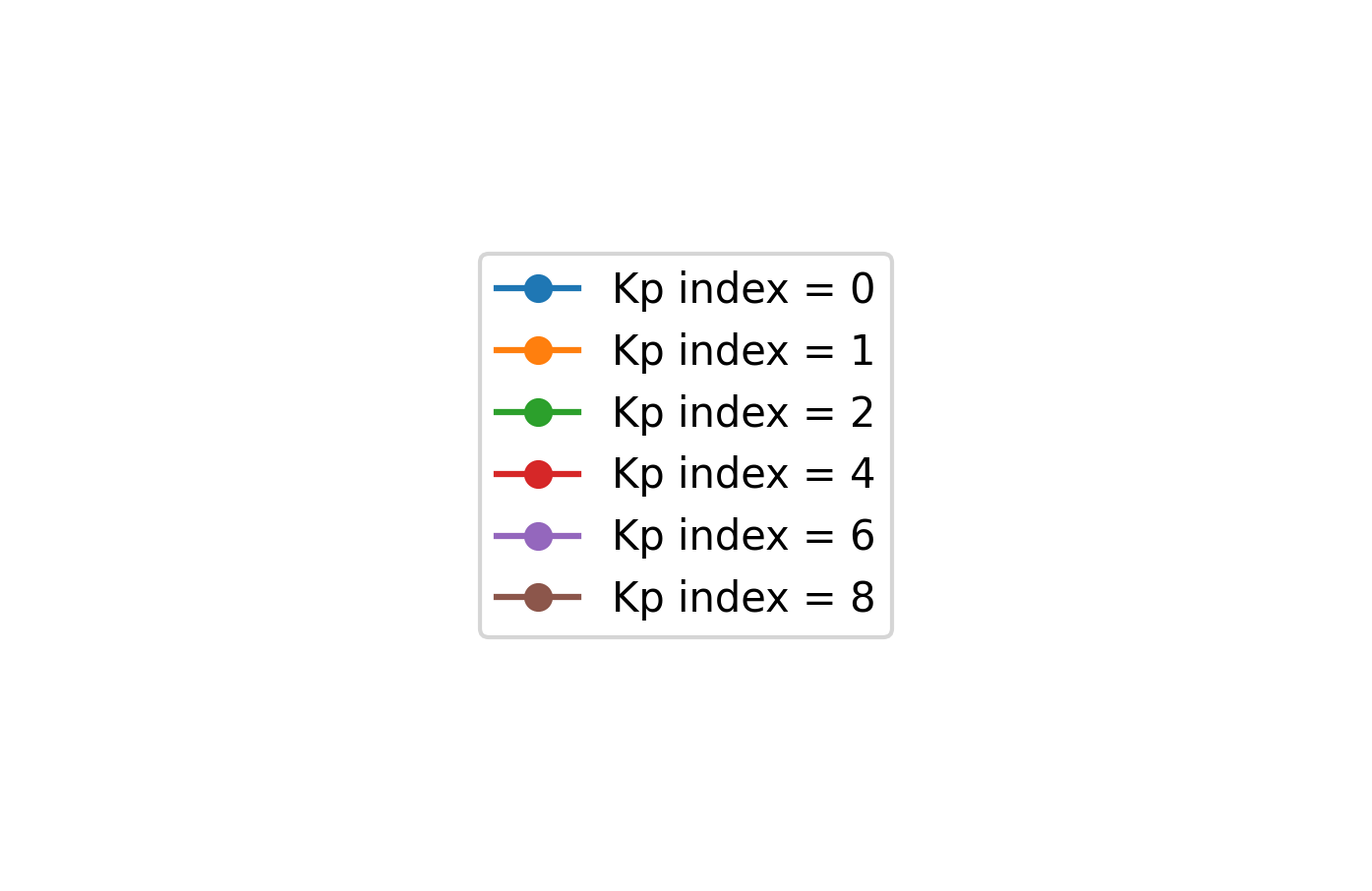}}
   \caption{The full dose rates over time for each flight route and each simulated Kp index. As shown previously in figure~\ref{fig:AccDoseVsKp}, dose rates are significantly higher for higher Kp indices in the soft spectrum case.}
   \label{fig:FlightRouteDoseRates}
\end{figure}

The dose rates and accumulated doses for these flight routes are 
displayed in figure~\ref{fig:AllDosesVsTimeKp4}, 
table~\ref{tab:PeakAndAccDoseTable}, 
figure~\ref{fig:AccDoseVsKp} and 
figure~\ref{fig:FlightRouteDoseRates}, 
where it can be seen that in the soft input spectral index case, 
the high latitude flights are subject to significantly higher 
dose rates than the lower latitude flights. This is not the case 
in the low spectral index/hard spectrum case, which is likely 
caused by the fact that the transition region boundary is 
consistently lower than all the flight routes in the hard 
spectrum case, while the high latitude flights cross the 
transition region in the soft spectrum case. 

Accumulated doses were calculated through integrating the dose 
rate over time for each flight route. Accumulated SEU and SEL 
count rates were also found using this method, giving 
accumulated latch-up probabilities for each SRAM device on the 
western high latitude flight route of about 0.06\% in the hard 
spectrum case and 0.11\% in the soft spectrum case. These 
accumulated probabilities assume a static event with constant 
spectral conditions taken from two arbitrary snapshots within 
GLE42 in 1989. Taking into the account the fact that aircraft 
may have many SRAM devices, that many devices aboard an aircraft 
may have significantly higher single event damage cross-sections 
than SRAMs \citep{dyer2020single}, and most importantly that 
GLEs can be significantly larger than the event used here (The 
peak GLNM percentage increase in GLE05, February 1956 was 
approximately 12 times higher than the peak GLNM percentage 
increase in GLE42, 1989 \citep{dyer2017extreme}) means that the 
probability of many high latitude aircraft experiencing numerous 
devices becoming damaged simultaneously may become high during a 
large ground-level enhancement. 

\begin{table}[]
   \centering
   \caption{Peak dose rates and accumulated doses for each of 
   the investigated flight routes and incoming spectra. The 
   integrated values were determined through taking the integral 
   over the curves displayed in figure~\ref{fig:AllDosesVsTimeKp4}}
   \scalebox{0.5}{
      \begin{tabular}{|c|c|c|c|c|c|c|c|c|c|}
         \cline{3-10}
         \multicolumn{2}{c|}{} & \multicolumn{2}{|c|}{Western high latitude route} & \multicolumn{2}{c|}{Eastern high latitude route} & \multicolumn{2}{c|}{Second lowest latitude route} & \multicolumn{2}{c|}{Lowest latitude route} \\
         \cline{3-10}
         \multicolumn{2}{c|}{} & Hard spectrum & Soft spectrum & Hard spectrum & Soft spectrum & Hard spectrum & Soft spectrum & Hard spectrum & Soft spectrum \\
         \hline
         \multirow{2}{*}{Ambient dose equivalent} & Peak dose rate ($\mu Sv / hr$) & 13.35 & 33.58 & 10.43 & 13.14 & 10.54 & 7.22 & 8.52 & 4.44 \\
                                                   & Accumulated dose ($\mu Sv$) & 82.37 & 105.8 & 60.76 & 41.57 & 57.19 & 17.68 & 40.67 & 8.62 \\
         \hline
         \multirow{2}{*}{Single event upset rate} & Peak upset rate (Upsets/hr/Gb) & 0.98 & 3.35 & 0.74 & 1.42 & 0.75 & 0.76 & 0.61 & 0.47 \\
                                                   & Accumulated upset count (Upsets/Gb) & 5.89 & 10.98 & 4.34 & 4.47 & 3.91 & 1.77 & 2.74 & 0.87 \\
         \hline
         \multirow{2}{*}{Single event latch-up rate} & Peak latch-up rate ($\times10^{-5}$ Latch-ups/hr/device) & $9.76$ & $33.47$ & $7.41$ & $14.21$ & $7.55$ & $7.59$ & $6.09$ & $4.72$ \\
                                                      & Accumulated latch-up count ($\times10^{-5}$ Latch-ups/device) & $58.94$ & $109.8$ & $43.42$ & $44.73$ & $39.05$ & $17.67$ & $27.44$ & $8.71$ \\
         \hline
      \end{tabular}
   }
   \label{tab:PeakAndAccDoseTable}
\end{table}

\begin{figure}
   \centering
   \foreach \KpIndex in {0,4,8}{
         \foreach \spectrum in {Hard,Soft}{
            \subfigure{\includegraphics[trim={0 0 0 0},clip,width=0.4\columnwidth]{ZoomedInWorldMapAndFlights_Spec\spectrum inputspectrum_KpIndex\KpIndex_altitude37_no_legend.png}}
            }
         }
   \subfigure{\includegraphics[trim={22cm 1cm 0 1cm},clip,width=0.4\columnwidth]{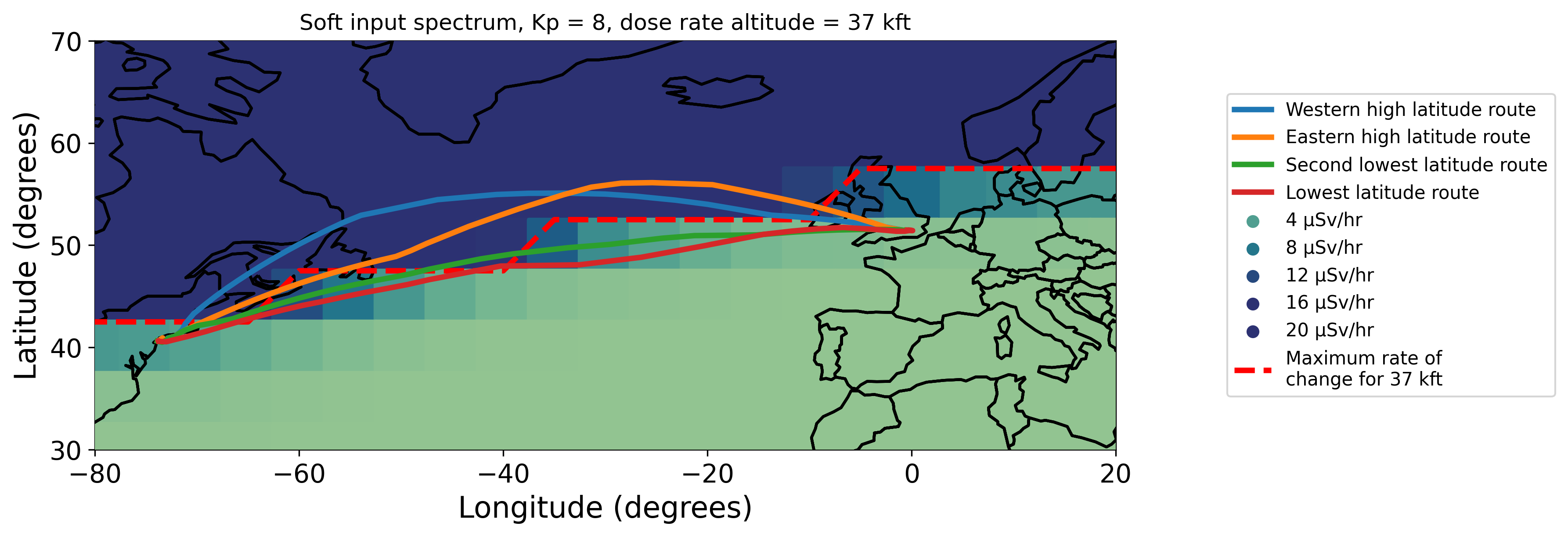}}
   \caption{Dose rate maps around the investigated flight routes 
   for several Kp indices. In the high spectral index case, the 
   transition dose rate region moves towards lower latitudes 
   with increasing Kp index, as was previously shown in 
   figure~\ref{fig:xsectionKpAll}, and crosses over the flight 
   routes.}
   \label{fig:FlightsDoseRatesAndKpIndex}
\end{figure}

It can also be seen that all flight routes experience 
significantly higher dose rates at higher Kp indices in the soft 
spectrum/high spectral index case. This does not seem to be true 
in the hard spectrum/low spectral index case, where high Kp 
indices only slightly increase dose rates. This is likely due to 
the transition region between moving southwards with increasing 
Kp index in the soft spectrum case, such that it encompasses a 
higher percentage of each flight route, an effect that was 
previously shown in figure~\ref{fig:xsectionKpAll} and 
figure~\ref{fig:xsectionPercentIncreases}. This effect is also 
shown in figure~\ref{fig:FlightsDoseRatesAndKpIndex}, which 
shows how dose rate maps at 37 kft change with varying Kp index 
for both spectral cases.

These results for the high spectral index case in 
figure~\ref{fig:FlightRouteDoseRates} and in 
figure~\ref{fig:xsectionPercentIncreases} previously complicate 
the possible strategy of flying to lower latitudes during a 
large ground-level enhancement. During high spectral index 
conditions, the actual location of the boundary between the low 
and high dose rate regions will be subject to a large amount of 
uncertainty, as it will be highly susceptible to the current 
conditions of Earth's magnetic field and of the spectral index 
of the solar particle spectrum bombarding Earth - parameters 
which currently have a large degree of uncertainty. Previous 
work using the QARM model has also made this point by showing 
the deviation flight route dose rates can have from those of a 
great circle route \citep{dyer2007solar}. The effect of Earth's 
magnetic field on dose rates that an aircraft might experience 
is displayed in figure~\ref{fig:FlightsDoseRatesAndKpIndex}.

It should be noted that not every model uses Kp index as an 
input parameter. However, all models use complex methods of 
representing Earth's magnetosphere and its effect on radiation 
dose rates in Earth's atmosphere. Given how quickly dose rate 
varies with latitude at the transition region under the softer 
spectral conditions in this work, small deviations in 
simulations of Earth's magnetosphere from reality may lead to 
large errors in dose rate predictions near the transition region 
and for transatlantic flights. 

This means that at least for the current level of technology and 
scientific knowledge of Earth's magnetosphere, a large buffer 
zone might have to be assumed a number of degrees south of the 
boundary zone location as predicted by current dose rate 
calculation software. Aircraft attempting to fly out of the high 
dose rate polar region would have to ensure they crossed the 
whole buffer zone to be reasonably confident of no longer 
experiencing high radiation dose rates.

Alternatively, the only way for an aircraft to know if and when 
it has truly crossed into the lower dose region would be if they 
are equipped with a multi-particle onboard radiation detector, 
and the possibility of equipping all aircraft with radiation 
dose monitors has been discussed in previous work 
\citep{clewerCitizen}. Monitors for this purpose were previously 
used on Concorde aircraft 
\citep{dyer2003calculations,bagshaw2000british}, and emergency 
descent procedures were put in place in case of high radiation 
dose rate levels.


\section{Conclusions}

The results create an overall picture that during a ground-level 
enhancement, the Earth's atmosphere can be divided into three 
approximate zones; a low dose rate equatorial and low altitude 
region, a high dose rate polar and high altitude region, and a 
transition region between the two.

The properties of the transition region define many of the 
geometric features of the GLE. The transition region varies in 
shape and structure with both the incoming particle spectra and 
Earth's magnetospheric conditions. This transition region is 
therefore the region where dose rates exhibit the highest levels 
of systematic uncertainty and any aircraft flying through this 
region (such as transatlantic flights) will experience a highly 
uncertain radiation dose. This is shown in this paper, where 
several flight routes between London Heathrow and New York JFK 
airports were simulated to experience significantly different 
dose rates during the same spectral and magnetospheric 
conditions. Dose rates for these flights were also shown to vary 
significantly with Kp index during relatively soft spectral 
conditions.

The results presented in this paper indicate that the high dose 
rate polar region stretches further south during highly 
disturbed solar conditions. During extreme ground-level 
enhancements that have an intensity on a scale that hasn't been 
so far recorded in modern history, Earth's magnetosphere could 
be significantly more disturbed than has been measured since the 
dawn of the space age and the high dose rate polar region in 
such events may therefore move significantly further south. It 
would likely pass closer to the equator than 45\textdegree in 
latitude, particularly over North America and Australia. This 
would breach well into the ICAO latitude bands of MNH and MSH 
(Middle latitudes Northern Hemisphere and Middle latitudes 
Southern Hemisphere), but specifically at these longitude 
regions. 

\section{Open Research Statement}
All of the data used to generate the results and plots in this 
paper can be found at https://doi.org/10.5281/zenodo.7410799 
along with relevant usage instructions.

\section{Funding Information}
This research was funded as part of the UKRI SWIMMR SWARM project.


\bibliography{agusample}
   

\end{document}